\documentclass[a4paper,fleqn,usenatbib]{mnras}

\usepackage{graphicx}
\usepackage{amsmath}
\usepackage{amssymb}
\usepackage[normalem]{ulem}

\newcommand*{\factorthree}{0.33}
\newcommand*{\factorsix}{0.39}

\newcommand{\fable}{{\sc fable}}
\newcommand{\MACSIS}{{\sc macsis}}

\title[Evolution of cluster scaling relations]{The redshift evolution of X-ray and Sunyaev-Zel'dovich scaling relations in the FABLE simulations}

\author[Henden et al.]{Nicholas A. Henden,$^{1}$\thanks{E-mail: n.henden@ast.cam.ac.uk}
Ewald Puchwein$^{1,2,3}$
and Debora Sijacki$^{1,2}$
\\
$^{1}$Institute of Astronomy, University of Cambridge, Madingley Road, Cambridge, CB3 0HA, UK\\
$^{2}$Kavli Institute for Cosmology, University of Cambridge, Madingley Road, Cambridge CB3 0HA, UK\\
$^{3}$Leibniz-Institut f\"{u}r Astrophysik Potsdam (AIP), An der Sternwarte 16, D-14482 Potsdam, Germany\\
}

\date{Accepted XXX. Received YYY; in original form ZZZ}

\pubyear{2019}

\begin{document}
\label{firstpage}
\pagerange{\pageref{firstpage}--\pageref{lastpage}}
\maketitle

\begin{abstract}
  We study the redshift evolution of the X-ray and Sunyaev-Zel'dovich (SZ) scaling relations for galaxy groups and clusters in the \fable\ suite of cosmological hydrodynamical simulations. Using an expanded sample of $27$ high-resolution zoom-in simulations, together with a uniformly-sampled cosmological volume to sample low-mass systems, we find very good agreement with the majority of observational constraints up to $z \sim 1$. We predict significant deviations of all examined scaling relations from the simple self-similar expectations. While the slopes are approximately independent of redshift, the normalisations evolve positively with respect to self-similarity, even for commonly-used mass proxies such as the $Y_{\mathrm{X}}$ parameter. These deviations are due to a combination of factors, including more effective AGN feedback in lower mass haloes, larger binding energy of gas at a given halo mass at higher redshifts and larger non-thermal pressure support from kinetic motions at higher redshifts. Our results have important implications for cluster cosmology from upcoming SZ surveys such as SPT-3G, ACTpol and CMB-S4, as relatively small changes in the observable--mass scaling relations (within theoretical uncertainties) have a large impact on the predicted number of high-redshift clusters and hence on our ability to constrain cosmology using cluster abundances. In addition, we find that the intrinsic scatter of the relations, which agrees well with most observational constraints, increases at lower redshifts and for lower mass systems. This calls for a more complex parametrization than adopted in current observational studies to be able to accurately account for selection biases.
\end{abstract}


\section{Introduction}\label{sec:intro}
The abundance and spatial distribution of galaxy groups and clusters holds the potential to provide precise constraints on cosmological parameters, such as the matter density of the Universe or the amplitude and slope of the matter power spectrum (see \citealt{Allen2011, Kravtsov2012, Planelles2015} for recent reviews). Ongoing and future surveys such as the Dark Energy Survey \citep{DES2005}, Euclid \citep{Laureijs2011}, the Large Synoptic Survey Telescope \citep{LSST2012}, eROSITA \citep{Merloni2012, Pillepich2012, Pillepich2018}, \textit{Athena} \citep{Nandra2013}, SPT-3G \citep{Benson2014} and Advanced ACTpol \citep{Henderson2016} will enable us to take full advantage of this potential by vastly expanding the number and mass range of known groups and clusters.
Yet the potential of such surveys to probe the underlying cosmology relies on our ability to relate the abundance of observed clusters, as a function of some observable, to theoretical predictions for the abundance of collapsed objects, as a function of their mass \citep{Kravtsov2012}.

Cluster masses are typically inferred via the relationship between the total mass and an observable calibrated to a sample of clusters with more direct mass measurements, for example from gravitational lensing or an X-ray hydrostatic analysis.
These mass--observable scaling relations are required to relate the theoretical mass function to the observed number counts and to understand the selection function of the survey, which describes how the observed cluster sample relates to the underlying population.
Despite recent progress in the calibration of the mass--observable relations, the uncertainty in their slope and normalisation continues to dominate the error budget of current cosmological studies of clusters (e.g. \citealt{Rozo2010, Sehgal2011, Mantz2014a, Bocquet2015, PlanckXXIV2015}).
Other aspects of the mass--observable relations are also not yet fully understood. For example, the origin of intrinsic scatter in the relations or their extension to low mass clusters or groups.
Moreover, the mass-observable relations may vary with redshift beyond that expected in simple hierarchical models of cluster formation. As the increased size and depth of future surveys push cluster detections to increasingly high redshift, constraints on the redshift evolution of the cluster scaling relations will become increasingly important to exploit the full potential of these new samples.

The paucity of well-defined cluster samples at high redshift, and the lack of low mass clusters and galaxy groups in existing samples, strongly limits current constraints on the redshift evolution of the scaling relations.
Existing observational studies (e.g. \citealt{Vikhlinin2002, Maughan2006, Reichert2011, Hilton2012, Maughan2012, Sereno2015d}) also find seemingly contradictory results. For example, some studies (e.g. \citealt{Ettori2004, Reichert2011, Hilton2012}) measure a negative evolution of the X-ray luminosity--temperature relation with respect to self-similarity, while others find zero or even positive evolution (e.g. \citealt{Vikhlinin2002, Kotov2005, Maughan2006, Pacaud2007}).
One of the dominant causes for this lack of consensus is the difficulty in accounting for selection bias in small, often heterogeneous samples of high-redshift clusters drawn from different surveys, which can mimic evolution (e.g. \citealt{Pacaud2007, Short2010}).

Theoretical modelling of cluster formation can aid in understanding these issues by studying the evolution of cluster scaling relations for the same set of objects, or a well-defined subsample, over cosmic time.
In the past decade, semi-analytic prescriptions have made significant progress in the modelling of realistic galaxy clusters (e.g. \citealt{DeLucia2007, Bower2008, Somerville2008, Guo2011}), however these methods do not fully capture the effects of baryonic processes during cluster formation, which can have a significant impact on cluster masses (e.g. \citealt{VanDaalen2011, Cui2014, Cusworth2014, Velliscig2014}).

An alternative approach is to use cosmological hydrodynamical simulations. These follow the highly non-linear, dark matter-dominated growth of large-scale structure while, at the same time, self-consistently evolving the baryon component to make predictions for cluster observables. Over the past two decades, cosmological hydrodynamical simulations have rapidly progressed our understanding of the complex interplay between gravitational collapse and astrophysical processes such as feedback from massive stars and active galactic nuclei (AGN), resulting in rapid improvements in the realism of simulated galaxy clusters.
In particular, AGN feedback has proven vital for reproducing a range of cluster observables (e.g. \citealt{Sijacki2007, Puchwein2008, Puchwein2010, Fabjan2010, McCarthy2010, McCarthy2011}).
Recent progress by several independent groups has yielded a range of simulations that reproduce various cluster observables, such as the X-ray and Sunyaev-Zel'dovich (SZ) scaling relations (e.g. \citealt{LeBrun2014, Pike2014, McCarthy2017, Truong2018}), the density, temperature and metallicity profiles of the intracluster medium (ICM; e.g. \citealt{Planelles2014, MACSIS, CEAGLE, Biffi2017, Vogelsberger2018}), and the apparent dichotomy between cool-core and non-cool-core clusters (e.g. \citealt{Rasia2015, Hahn2017, Barnes2017, Barnes2018}).

A number of these simulations have been utilised to study the redshift evolution of the cluster scaling relations, with occasionally dissimilar results (e.g. \citealt{Fabjan2011, LeBrun2017, MACSIS, Truong2018}). Yet some variation in their predictions is to be expected given that the simulations use different sets of physical models with different parametrizations. For this reason it is important to explore a range of plausible models in order to constrain the dependence of the theoretical predictions on the physical modelling. In addition, observational constraints from future surveys have the potential to distinguish between models and thus to constrain the non-gravitational physics important to the formation and evolution of galaxy clusters.

In the present study we explore the redshift evolution of X-ray and SZ scaling relations from the \textit{Feedback Acting on Baryons in Large-scale Environments} (\fable) simulations, a suite of cosmological hydrodynamical simulations incorporating an updated version of the Illustris model for galaxy formation \citep{Genel2014, Vogelsberger2014, Sijacki2015} with improved agreement with observations over a much wider mass scale, from galaxies to massive clusters. Originally presented in \cite{Henden2018} (hereafter Paper~I), the \fable\ suite employed in this study consists of a uniformly-sampled cosmological volume together with a series of 27 high-resolution zoom-in simulations of galaxy groups and clusters. This constitutes a much larger simulated sample compared with Paper I where just six zoom-in simulations were employed. Here we extend the analysis of the $z = 0$ scaling relations shown in Paper~I out to $z \approx 2$ using our expanded sample.

This paper is organised as follows. Section~\ref{sec:methods} briefly introduces the \fable\ simulations and describes our methods for calculating observable quantities and our choice of sample selection.
In Section~\ref{sec:xray} we compare the X-ray scaling relations with observations at intermediate to high redshifts and investigate the evolution of the relations out to $z = 1.8$ with comparison to the recent simulation studies of \cite{MACSIS} and \cite{Truong2018}.
In Section~\ref{sec:SZ} we explore the redshift evolution of the scaling between the SZ signal and total mass, including a comparison to observed clusters at $z \lesssim 1$. Lastly we investigate how different predictions for the relation can affect the predicted cluster counts for future SZ surveys.

Throughout this paper, spherical overdensity masses and radii use the critical density of the Universe as a reference point. Hence, $M_{500}$ is the total mass inside a sphere of radius $r_{500}$ within which the average density is 500 times the critical density of the Universe.

\section{Methods}\label{sec:methods}
\subsection{The \fable\ simulations}
The \fable\ simulations use the cosmological hydrodynamic moving-mesh code \textsc{arepo} \citep{Springel2010} with a suite of physical models relevant for galaxy formation based on the Illustris simulation \citep{Vogelsberger2013, Vogelsberger2014, Genel2014}, including radiative cooling \citep{Katz1996, Wiersma2009}, star and black hole formation \citep{Springel2003, Springel2005, Vogelsberger2013} and associated supernovae and AGN feedback \citep{Vogelsberger2013, DiMatteo2005, Springel2005, Sijacki2007, Sijacki2015}.

As detailed in Paper~I, we have revisited the modelling of feedback from supernovae and AGN in order to improve on some of the shortcoming of Illustris, in particular the gas content of massive haloes. The strength of the AGN feedback in Illustris was calibrated to reproduce the massive end of the galaxy stellar mass function, however the feedback acted too violently on the gas component of massive haloes ($M_{500} \approx 10^{13}-10^{14} M_{\odot}$), leaving them almost devoid of gas at $z=0$ \citep{Genel2014}. In \fable\ we have calibrated our AGN feedback model to reproduce both the galaxy stellar mass function and the gas mass fractions of galaxy groups ($M_{500} \lesssim 10^{14} M_{\odot}$).
Cluster-sized haloes were not present in the calibration volume due to computational constraints, although in Paper~I we applied our calibrated model to more massive haloes using the ``zoom-in'' technique to show that the gas mass fractions of \fable\ clusters remain in good agreement with observations for clusters as massive as $M_{500} \approx 10^{15} M_{\odot}$.

In Paper~I we analysed a full-volume simulation approximately 60 Mpc on a side together with a small set of zoom-in simulations of groups and clusters performed with the \fable\ model. We demonstrated good agreement with $z \approx 0$ observations for a number of key observables, such as the total stellar mass, the X-ray luminosity--total mass and gas mass relations and the Sunyaev-Zel'dovich (SZ) signal--mass relation.
For the present study we have greatly expanded our sample of zoom-in simulations from just \textit{six} in Paper~I to \textit{twenty-seven}. Following Paper~I, the resimulated haloes were selected from the dark matter-only (3 $h^{-1}$ Gpc)$^3$ Millennium-XXL simulation \citep{Angulo2012} to be approximately logarithmically spaced over the mass range $10^{13} M_{\odot} \lesssim M_{500} \lesssim 3 \times 10^{15} \, M_{\odot}$ at $z=0$.
The high-resolution region of the simulations extends to approximately $5 \, r_{500}$ at $z=0$ and the high-resolution dark matter particles have a mass of $m_{\mathrm{DM}}=5.5\times$10$^7$ $h^{-1}$~M$_{\odot}$. The gravitational softening length was fixed to $2.8125$~$h^{-1}$~kpc in physical coordinates at redshift $z \leq 5$ and fixed in comoving coordinates for $z > 5$.
Mode amplitudes in the initial conditions were scaled to a Planck cosmology \citep{PlanckXII2015} with cosmological parameters $\Omega_{\Lambda}=$~0.6911, $\Omega_{\mathrm{M}}=$~0.3089, $\Omega_{\mathrm{b}}=$~0.0486, $\sigma_8=$~0.8159, $n_s=$~0.9667 and $H_0=67.74$~km~s$^{-1}$~Mpc$^{-1}$. We assume this as our fiducial cosmology throughout the rest of this paper.

In addition to the main halo of each zoom-in simulation we also consider ``secondary'' friends-of-friends (FoF; \citealt{Davis1985}) haloes within the high-resolution region. We include in our sample any FoF halo that is not contaminated by low-resolution dark matter particles within $5 \, r_{500}$ at the given redshift. This ensures that the halo properties are unaffected by the zoom-in technique. Indeed, we do not find any evidence that the X-ray or SZ properties of these secondary haloes depend systematically on their distance from the main halo or from the edge of the high-resolution region, which can be non-spherical.

In addition, the galaxy group population ($M_{500} \lesssim 10^{14} M_{\odot}$) is supplemented with haloes from a periodic volume of side length 40~$h^{-1}$ comoving Mpc.
The box contains 512$^3$ dark matter particles with a mass of $m_{\mathrm{DM}}=3.4\times$10$^7$ $h^{-1}$~M$_{\odot}$ and approximately 512$^3$ baryonic resolution elements (gas cells and stars), which have a target mass of $\overline{m}_{\mathrm{b}}=6.4\times$10$^6$~$h^{-1}$~M$_{\odot}$.
The gravitational softening length was fixed to $2.393$~$h^{-1}$~kpc in physical coordinates below $z=5$ and fixed in comoving coordinates at higher redshifts.
The simulation assumes a Planck cosmology \citep{PlanckXII2015} with cosmological parameters $\Omega_{\Lambda}=$~0.6935, $\Omega_{\mathrm{M}}=$~0.3065, $\Omega_{\mathrm{b}}=$~0.0483, $\sigma_8=$~0.8154, $n_s=$~0.9681 and $H_0=67.9$~km~s$^{-1}$~Mpc$^{-1}=h \times$100~km~s$^{-1}$~Mpc$^{-1}$. We rescale the appropriate quantities to the cosmology used in the zoom-in simulations, although this has a negligible effect on the results given the similarity between the parameters.

\subsection{Calculating X-ray properties}\label{subsec:xray_props}
We estimate bolometric X-ray luminosities and spectroscopic temperatures for our simulated haloes via the method described in Paper~I with some minor alterations.

First we define a series of temperature and metallicity bins and estimate the total emission measure of gas associated with each bin. Using the {\sc XSPEC} package (\citealt{Arnaud1996}, version 12.8.0) we then generate an APEC emission model \citep{Smith2001} for each bin with the calculated emission measure. The sum of the spectra of all bins determines the mock X-ray spectrum.
We include the effects of Galactic HI absorption on the spectrum via a {\sc wabs} model in {\sc XSPEC} with a column density of $5 \times 10^{20}$ cm$^{-2}$.
We mimic observations with either the \textit{Chandra} or \textit{Athena} X-ray observatories by convolving the mock spectrum with an appropriate response function as described below (Section~\ref{subsubsec:response}).
We follow the standard practice of fitting a single-temperature APEC model to the spectrum in the energy range $0.5$--$10$ keV for \textit{Chandra} and $0.2$--$12$ keV for \textit{Athena}. We fix the redshift to the input value but leave the temperature, metallicity and normalisation free to vary during the fit. The spectroscopic temperature is thus the temperature of the best-fitting model and the bolometric X-ray luminosity is calculated from the model in the range $0.01$--$100$ keV.

The two changes to our procedure relative to Paper~I are the addition of Galactic HI absorption to the spectra and the inclusion of the metallicity information of the gas.
Adding absorption improves the realism of our spectra, although its effect on the derived X-ray luminosity and temperature is small ($\lesssim 2$ per cent).
Whereas in Paper~I we assumed a constant metallicity of $0.3$ times the solar value for simplicity, here we utilise the metallicity of the gas tracked by the simulations. The conclusions of Paper~I are unchanged by using this updated method, however in the present study we are concerned with the exact slope of the X-ray scaling relations, which can be sensitive to low temperature systems ($\lesssim 2$ keV) where metal line emission is a significant contributor to the X-ray luminosity.

For each halo we calculate an X-ray spectrum for the gas within a circular aperture of radius $r_{500}$ centred on the minimum of the gravitational potential, integrated along the length of the simulation volume along one of the coordinate axes. The same projection axis is used for all haloes. We use a projected rather than a spherical aperture as this is more akin to observations. However, we caution that the resultant spectrum can be biased by hot gas along the line of sight. We find that the X-ray luminosity can be boosted by as much as $\sim 15$ per cent by gas that lies in projection, although the effect on the spectroscopic temperature is small ($\lesssim 2$ per cent). We shall comment on the effect of switching to a spherical aperture in situations where projection has an appreciable effect on the derived X-ray scaling relation. We note that other quantities, such as the total mass, gas mass and mass-weighted temperature, are measured directly from the simulation within a \textit{spherical} aperture centred on the same location.

As in Paper~I we exclude cold gas with a temperature below $3 \times 10^4$~K and gas above the density threshold for star formation. The thermal properties of such gas are not reliably predicted by the simulation due to the lack of molecular cooling and the simple multiphase model for star-forming gas, respectively. We exclude this gas from the mass-weighted temperature calculation for the same reasons.

We make one further temperature cut on the gas that excludes very high temperature bubbles created by our relatively simple model for radio-mode AGN feedback. Excessively hot AGN-heated bubbles bias the derived X-ray luminosity and spectroscopic temperature in a few per cent of systems and occur when a particularly strong feedback event has been very recently triggered. In reality, AGN-driven bubbles are thought to be supported by non-thermal pressure and should only contribute to the X-ray temperature once thermalisation has occurred.
We find that such gas can be reliably excluded by applying a temperature threshold of $4.0$ times the virial temperature, $k_B T_{200} = G M_{200} \mu m_p / 2 r_{200}$. This greatly reduces the presence of outliers, which can otherwise bias the scatter inferred from the X-ray scaling relations. We note that for the vast majority of systems no gas is excluded by this choice of threshold.

In addition we calculate the X-ray analogue of the integrated SZ effect, known as $Y_{\mathrm{X}}$. First introduced by \cite{Kravtsov2006}, $Y_{\mathrm{X}}$ is equal to the product of the core-excised spectroscopic temperature and total gas mass and is considered a low-scatter mass proxy that is especially robust to the cluster dynamical state (e.g. \citealt{Arnaud2007, Maughan2007, Nagai2007c}) and to the baryonic physics included in simulations (e.g. \citealt{Short2010, Fabjan2011, Planelles2014} but see also \citealt{LeBrun2014}).
To obtain $Y_{\mathrm{X}}$ we calculate the core-excised spectroscopic temperature within a projected annulus of inner radius $0.15 \, r_{500}$ and outer radius of $r_{500}$.

\subsubsection{Choice of response function}\label{subsubsec:response}
In Section~\ref{subsubsec:obs_comp} we compare the X-ray properties of our simulations to observational data. For this we employ the response function and effective area energy curve of the \textit{Chandra} ACIS-I detector, which is commonly used in cluster X-ray studies.
We adopt a very large exposure time of $10^7$ seconds so that we are not limited by photon noise even in low-mass galaxy groups at $z = 1$. The fits are performed in the energy range $0.5$--$10$ keV.

When investigating the redshift evolution of the X-ray scaling relations (Section~\ref{subsec:evol}) we find that current X-ray observatories such as \textit{Chandra} possess insufficient effective area at low energies to reliably measure group and cluster temperatures out to high redshift ($z \approx 2$).
In particular, our tests have shown that the spectroscopic temperature can be biased high for situations in which a significant proportion of the X-ray emission is redshifted below the energy range used for the spectral fit. For example, using the \textit{Chandra} response and fitting the spectra in the $0.5$--$10$ keV range, we find that the spectroscopic temperature is biased high compared with the mass-weighted temperature at $z \gtrsim 1$. This bias increases with increasing redshift and is larger for lower temperature systems as a larger fraction of the X-ray emission is redshifted below the minimum energy of the fit. For the sample described in Section~\ref{subsec:sample} this effect dominates the redshift evolution of the slope and normalisation of the spectroscopic temperature-based relations at $z \gtrsim 0.6$.
Lower (higher) values for the minimum energy of the fit show decreased (increased) bias, however the effective area of \textit{Chandra} ACIS-I becomes negligible at $\lesssim 0.5$ keV so that values lower than $0.5$ keV have little effect.

For this reason we generate spectra out to $z \approx 2$ using the X-ray Integral Field Unit (X-IFU) on board the future \textit{Athena} X-ray observatory \citep{Barret2018}, which will possess an order of magnitude larger effective area than \textit{Chandra} over a wider $0.2$--$12$ keV bandpass.
For this we use response matrices and effective area energy curves produced for the so-called cost-constrained configuration of \textit{Athena} as described in \cite{Barret2018}. For the mock \textit{Athena} observations we adopt a very long exposure of $10^8$ seconds, which ensures that the derived spectroscopic temperature is converged with respect to the total photon count for the lowest temperature objects in our sample at $z \approx 2$ ($\sim 1$ keV). Such an exposure time is clearly impractical however it allows us to present predictions of the X-ray scaling relations over an extended halo mass range out to high redshift.

\subsection{Fitting of cluster scaling relations}\label{subsec:fitting}
For all scaling relations we relate the property $Y$ to the property $X$ with a best-fitting power-law of the form
\begin{equation}\label{eq:powerlaw}
Y = E(z)^{\gamma} \; 10^A \, \left(\frac{X}{X_0}\right)^{\beta},
\end{equation}
where $A$ and $\beta$ describe the normalisation and slope of the relation, respectively, and $E(z)^{\gamma}$ corresponds to the expected self-similar evolution of the normalisation, where $E(z) = \sqrt{\Omega_m (1+z)^3 + \Omega_{\Lambda}}$ and the exponent $\gamma$ is derived in Appendix~\ref{A:SS}. $X_0$ is the pivot point, which we set to $M_{500} = 2 \times 10^{14} \, M_{\odot}$, $T_{500} = 3$ keV or $M_{\mathrm{gas, 500}} = 2 \times 10^{13} \, M_{\odot}$ for the total mass, temperature or gas mass, respectively.
These are close to the average values of our sample (Section~\ref{subsec:sample}) across all redshifts analysed ($z \leq 1.8$).

We perform the fitting in log-space using the orthogonal BCES method described in \cite{Akritas1996}, which is commonly used in observational studies (e.g. \citealt{Pratt2009, Zhang2011a, Maughan2012, Giles2016}). We note that in our case of no measurement errors this method reduces to orthogonal regression (e.g. \citealt{Isobe1990}).
We have repeated our analyses using two other common choices for the fitting procedure, namely the BCES(Y$|$X) method \citep{Akritas1996} and the Bayesian approach described in \cite{Kelly2007}.
We confirm that these two methods yield identical values for the best-fitting parameters. The orthogonal BCES method yields marginally higher values for the slope, although the difference is less than $5$ per cent and comparable to the uncertainties. The offset is systematic across redshift bins and has a negligible effect on the redshift evolution of the slope and normalisation of the relations.

We compute the intrinsic scatter about the best-fitting relation following \cite{Tremaine2002} for which
\begin{equation}\label{eq:scat}
\sigma = \sqrt{\frac{1}{N-2} \sum_{i=1}^{N} [\mathrm{log}_{10}(Y_i) - \mathrm{log}_{10}(F(X_i))]^2 },
\end{equation}
where $N$ is the number of data points, as described by their position ($X_i, Y_i$) in the space of observables $X$ and $Y$, and $F$ is the best-fitting power-law relation.
We have confirmed that the scatter calculated in this way matches the best-fitting value found via the method of \cite{Kelly2007}.

We estimate confidence intervals on the best-fitting parameters via bootstrap resampling of the data. Specifically, we generate $10^4$ resamples with replacement and obtain the best-fitting parameters for each resample. The confidence interval for each parameter is defined as the empirical quantiles of the bootstrap distribution of the parameter following the ``basic'' bootstrap method described in \cite{Davison1997}.
Quoted uncertainties on the best-fitting parameters correspond to the 68 per cent confidence interval.

\subsection{Sample selection}\label{subsec:sample}
Whilst a single power law adequately describes the scaling of massive clusters, at low masses the slope of the relation can change due to the influence of non-gravitational processes such as feedback, which have a larger impact on lower mass haloes due to their shallower gravitational potential wells. In \fable\ we find that certain scaling relations -- particularly those involving X-ray luminosity or temperature -- show signs of a steepening in the regime of low-mass galaxy groups with masses $M_{500} \lesssim 3 \times 10^{13} \: M_{\odot}$ and average temperatures $T_{500} \lesssim 1$~keV, consistent with previous simulation results (e.g. \citealt{Dave2008, Puchwein2008, Gaspari2014, Planelles2014}) as well as some observational studies (e.g. \citealt{Helsdon2000, Mulchaey2000a, Sanderson2003, Eckmiller2011, Bharadwaj2015, Kettula2015}).

To ensure that the best-fitting power law relation is not biased by low-mass groups we include only haloes above a mass threshold of $M_{500} > 3 \times 10^{13} \, E(z)^{-0.5} \: M_{\odot}$ in our sample.
The redshift evolution of the lower mass threshold is parametrised by the factor $E(z)^{-0.5}$, which was chosen to be similar to that of an SZ-selected sample based on the results of the 2500 deg$^2$ SPT-SZ survey (see Appendix~\ref{A:mass_evol_dependence}).
Applying an SZ-like selection allows us to maximise the size of our sample at high redshift so that we are able to robustly derive the best-fitting scaling relations in single redshift bins.
We have tested that our results are not systematically affected by this choice, which we discuss in Appendix~\ref{A:mass_evol_dependence}.

This choice corresponds to a sample of $39$ haloes at $z=0$, $68$ at $z=1$ and $44$ at $z=1.8$. Note that we do not extend our analyses beyond $z=1.8$ as the number of cluster-scale haloes with $M_{500} > 10^{14} M_{\odot}$ falls to one.
Our sample is not as large as some recent simulation studies (e.g. \citealt{LeBrun2017, MACSIS}) because our simulations are run at comparatively high resolution, which limits the number of simulations we are able to realistically perform. Even so, our high-redshift samples are still comparable in size to recent observational studies of local scaling relations (e.g. \citealt{Zou2016, Giles2017, Ge2018, Nagarajan2018}).

Some simulation studies choose to limit their sample to $M_{500} \gtrsim 10^{14} M_{\odot}$ (e.g. \citealt{MACSIS, Truong2018}) to avoid a possible break in the cluster scaling relations. Indeed, there is some observational evidence that the scaling relation between X-ray luminosity and total mass or temperature experiences a break at $M_{500} \sim 10^{14} M_{\odot}$ ($\sim 3$ keV; e.g. \citealt{Hilton2012, Maughan2012, Lovisari2015}), although others find a more gradual shift in slope (e.g. \citealt{Eckmiller2011, Bharadwaj2015, Kettula2015}) or no change at all (e.g. \citealt{Anderson2014a, Zou2016, Babyk2018a}).
In Appendix~\ref{A:mass_dependence} we discuss the changes to our best-fitting scaling relations when restricting our sample to more massive haloes with $M_{500} > 6 \times 10^{13} \, E(z)^{-0.5} \; M_{\odot}$ or $M_{500} > 10^{14} \, E(z)^{-0.5} \; M_{\odot}$.
In general we find that higher mass thresholds bring the best-fitting slope and the evolution of the normalisation slightly closer to self-similarity, including somewhat smaller intrinsic scatter. We opt for a relatively low mass threshold ($M_{500} > 3 \times 10^{13} \, E(z)^{-0.5} \: M_{\odot}$) to obtain robust statistics but comment on those aspects of the scaling relations that are affected by choosing a more massive sample.

The wide halo mass range of our sample means that the best-fitting power law may be biased towards low-mass haloes, which are more abundant due to the shape of the halo mass function.
To assess this effect we have repeated our analyses fitting to the median relation in bins of width $0.1$ dex rather than to the individual points. We find that the change to the best-fitting slope and normalisation is qualitatively similar to that shown in Appendix~\ref{A:mass_dependence} when raising the lower mass threshold of the sample, though with a negligible change to the intrinsic scatter.
Quantitatively, the size of the effect is similar to, or smaller than, switching from the fiducial sample to the sample of haloes with $M_{500} > 10^{14} \, E(z)^{-0.5} \; M_{\odot}$.
At higher masses we do not expect a significant bias of this type as the sample is limited to the central haloes of the zoom-in simulations in this mass range, which were selected to be approximately equally-spaced in logarithmic halo mass at $z=0$.

\section{X-ray Scaling Relations}\label{sec:xray}
In this section we investigate five X-ray scaling relations: gas mass, $Y_{\mathrm{X}}$ and X-ray luminosity as a function of total mass and total mass and X-ray luminosity as a function of temperature.
In Section~\ref{subsec:comp} only we also consider the relation between X-ray luminosity and gas mass.
The form of each scaling relation is arbitrary as our choice of fitting procedure does not distinguish between dependent and independent variables.
The expected self-similar scalings for these relations are derived in Appendix~\ref{A:SS}.
In Section~\ref{subsec:comp} we compare to observations of intermediate- and high-redshift clusters and in Section~\ref{subsec:evol} we investigate the redshift evolution of the X-ray scaling relations in relation to other recent simulation results.

\subsection{Comparison to observations at intermediate and high redshift}\label{subsec:comp}
\begin{figure*}
\begin{center}
{\includegraphics[width=\factorsix\textwidth]{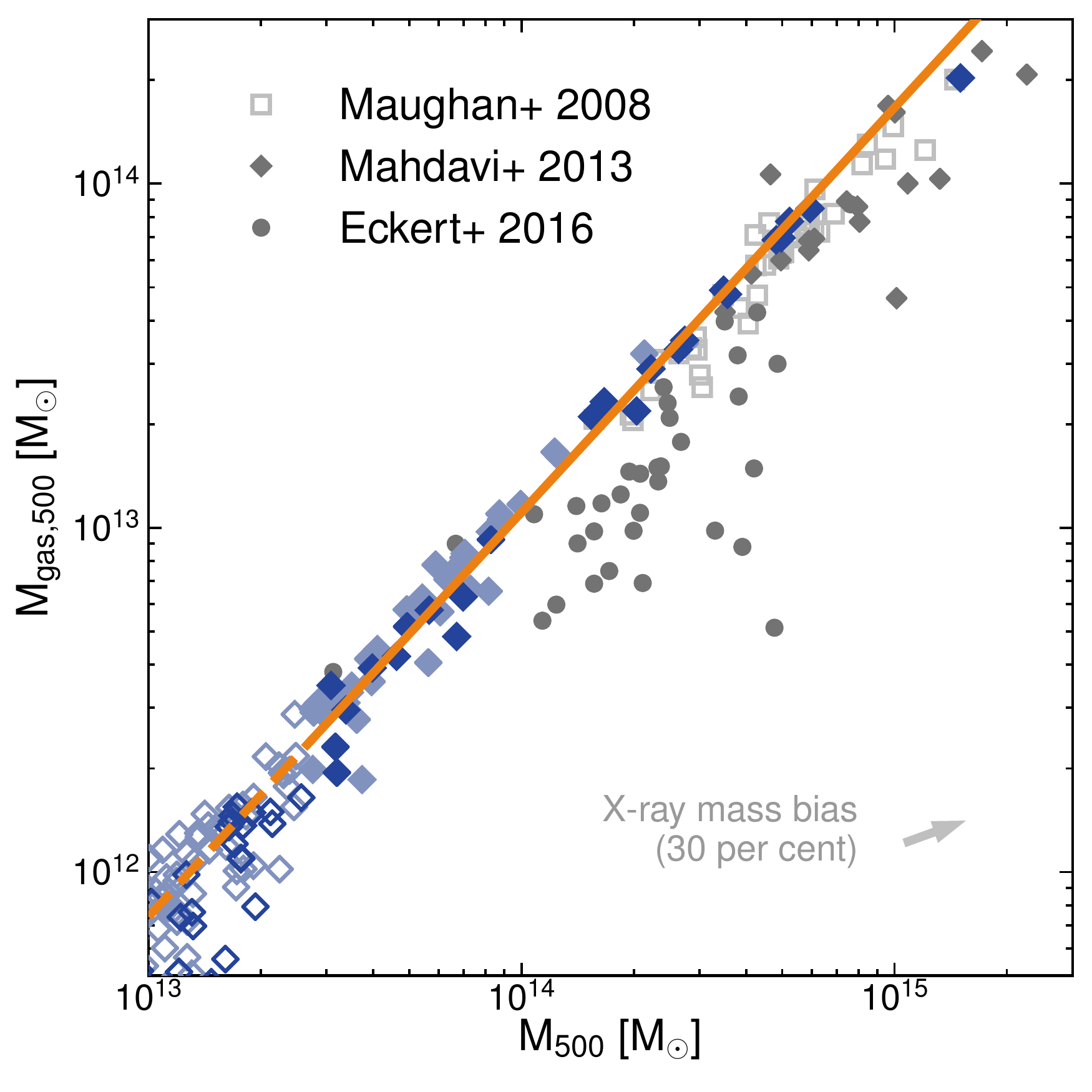}}
{\includegraphics[width=\factorsix\textwidth]{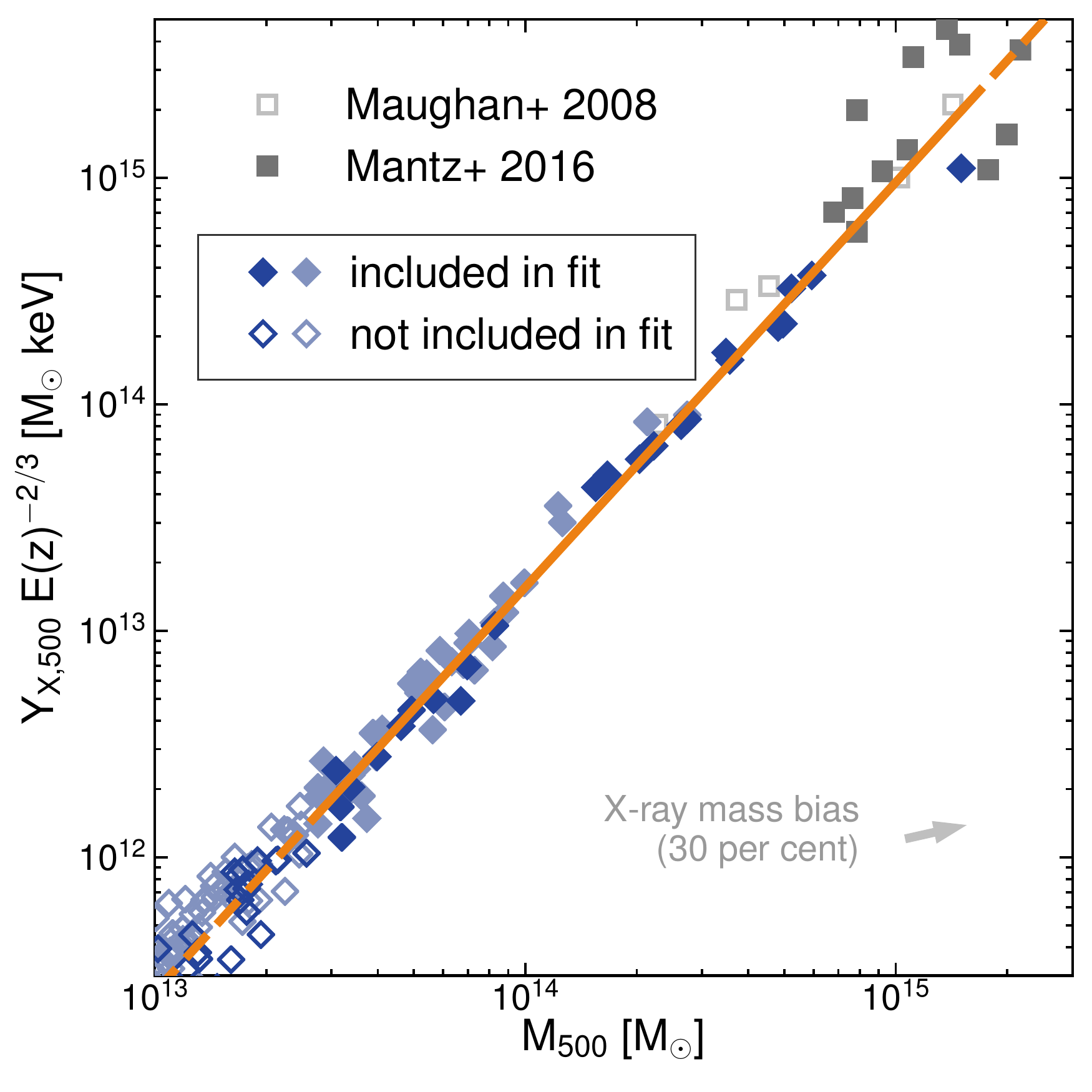}}
{\includegraphics[width=\factorsix\textwidth]{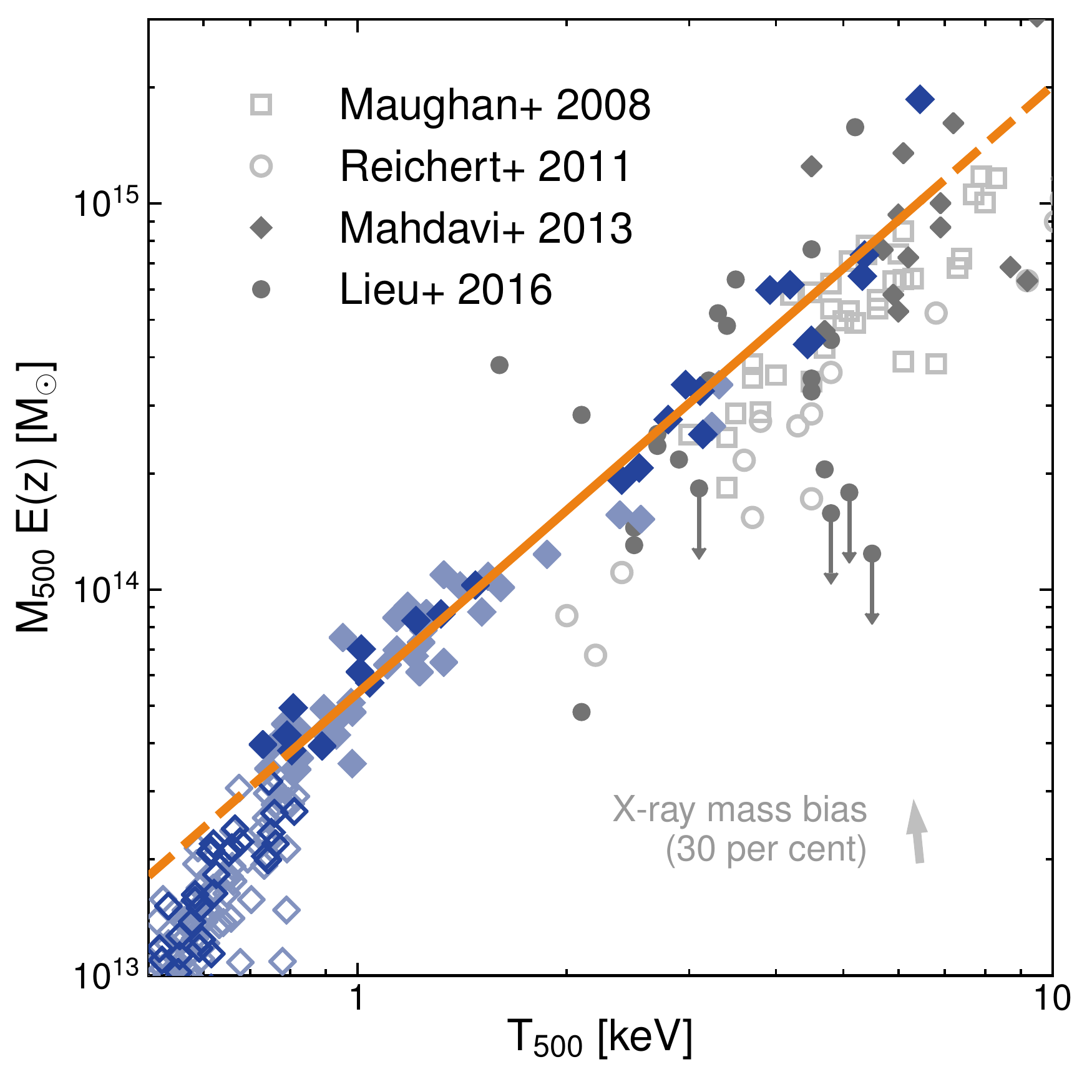}}
{\includegraphics[width=\factorsix\textwidth]{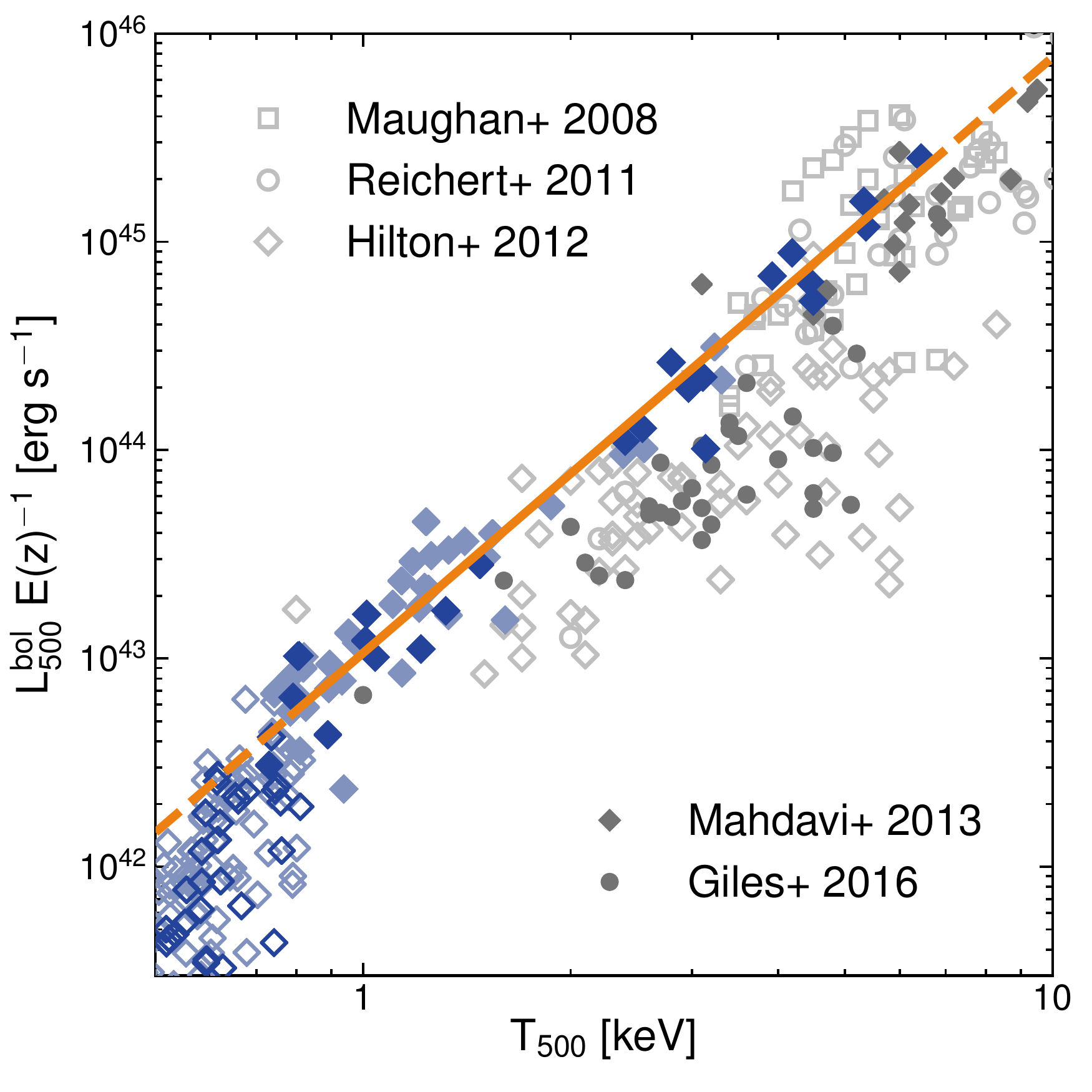}}
{\includegraphics[width=\factorsix\textwidth]{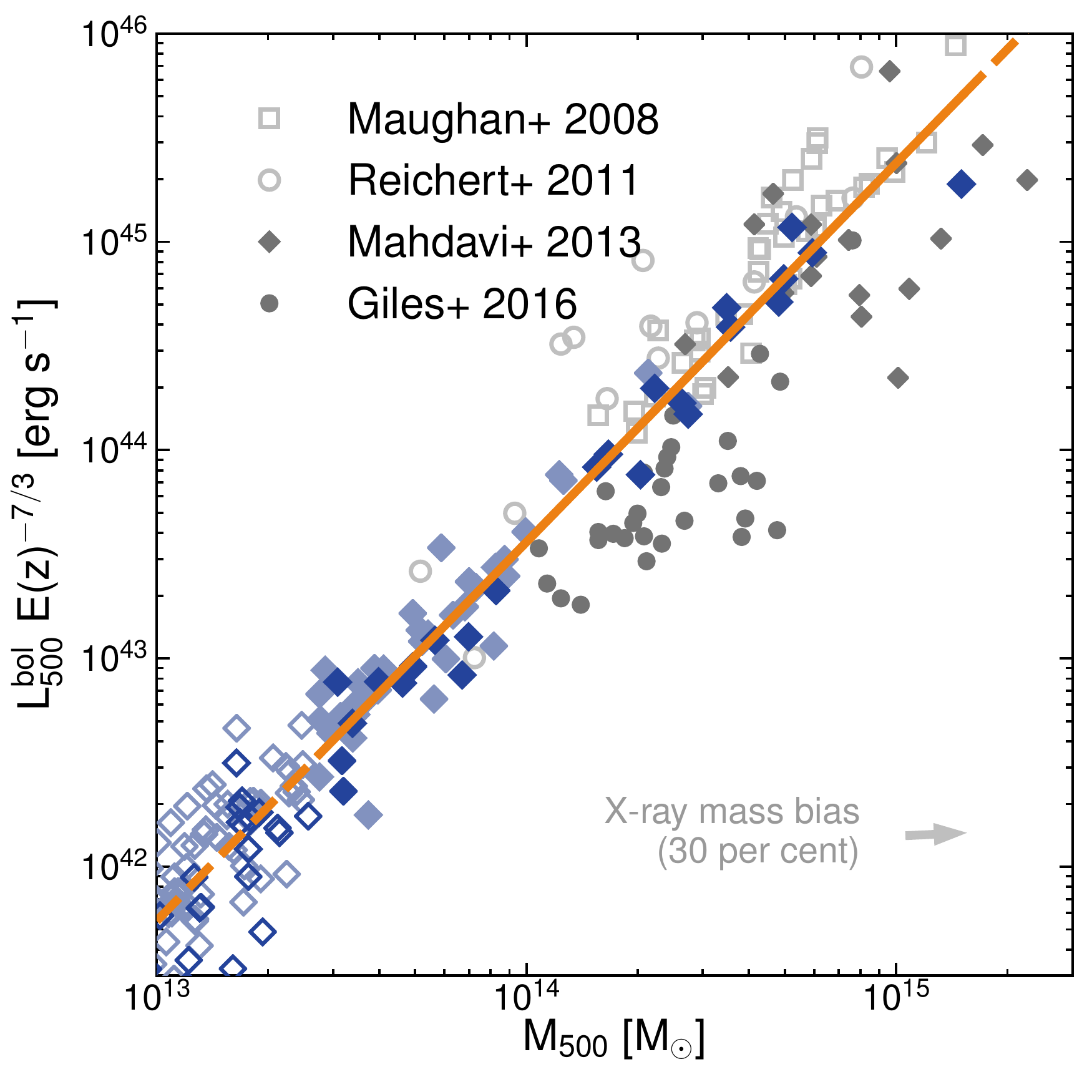}}
{\includegraphics[width=\factorsix\textwidth]{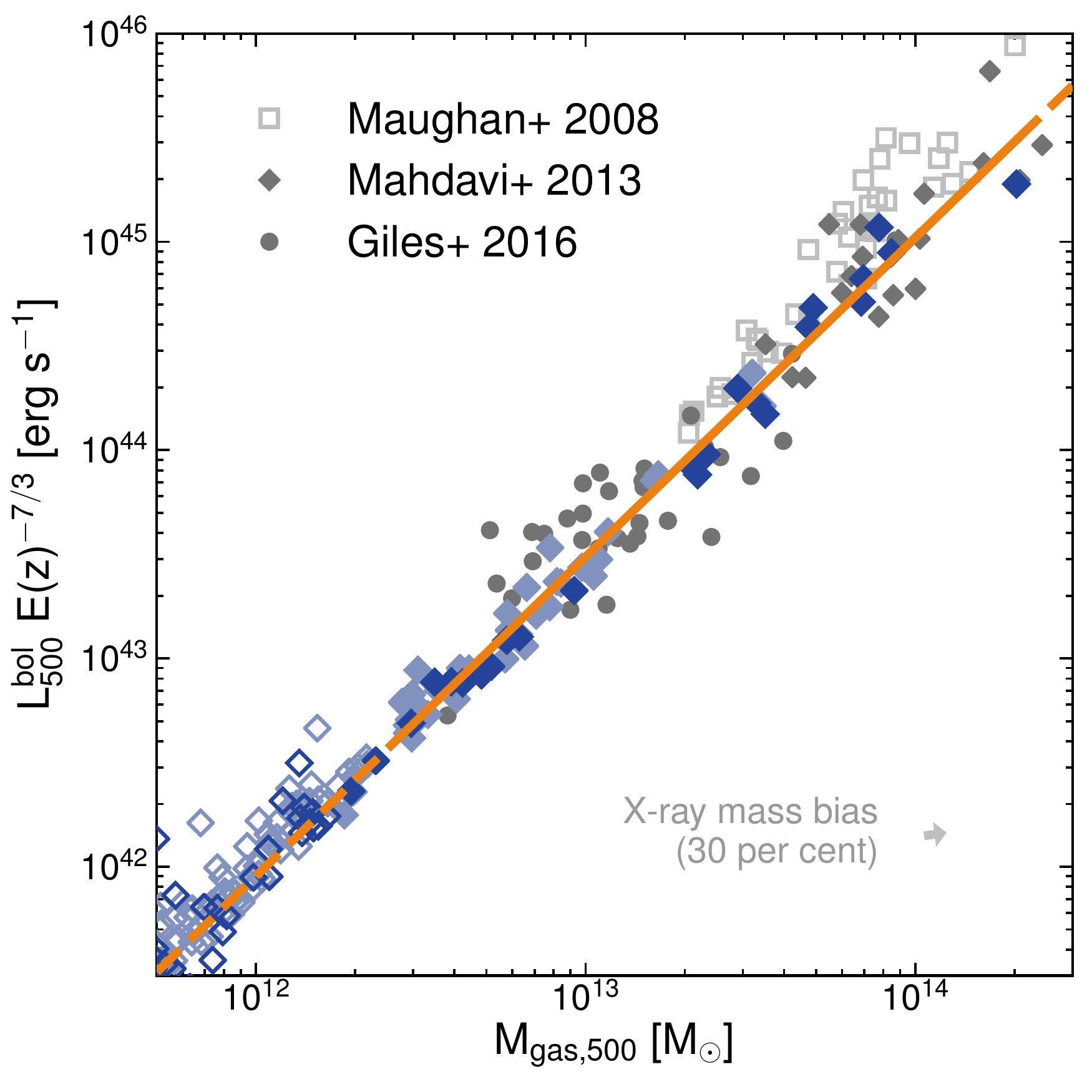}}
\end{center}
\caption{X-ray scaling relations at $z=0.4$ compared to observed groups and clusters at similar redshift.
  Dark blue symbols show haloes in the full-volume simulation and the most massive halo in each zoom-in simulation, while light blue symbols highlight lower mass companion clusters within the high-resolution region of the zoom-in simulations.
  Solid lines show the best-fitting power law relations of the mass-limited sample described in Section~\ref{subsec:sample} (filled diamonds) and become dashed in the approximate region of extrapolation.
  Dark grey, solid symbols indicate observational data based on weak lensing mass estimates while light grey, open symbols show data based on X-ray hydrostatic masses. The grey arrow in the bottom-right of these panels shows how the latter are expected to change when correcting for a potential X-ray hydrostatic mass bias under the assumption that their mass is currently biased low by $30$ per cent. The total masses of the simulated objects are measured directly from the simulation and should be compared with the less biased weak lensing masses where possible.
  }
    \label{fig:xray_z04}
\end{figure*}

\begin{figure*}
\begin{center}
{\includegraphics[width=\factorsix\textwidth]{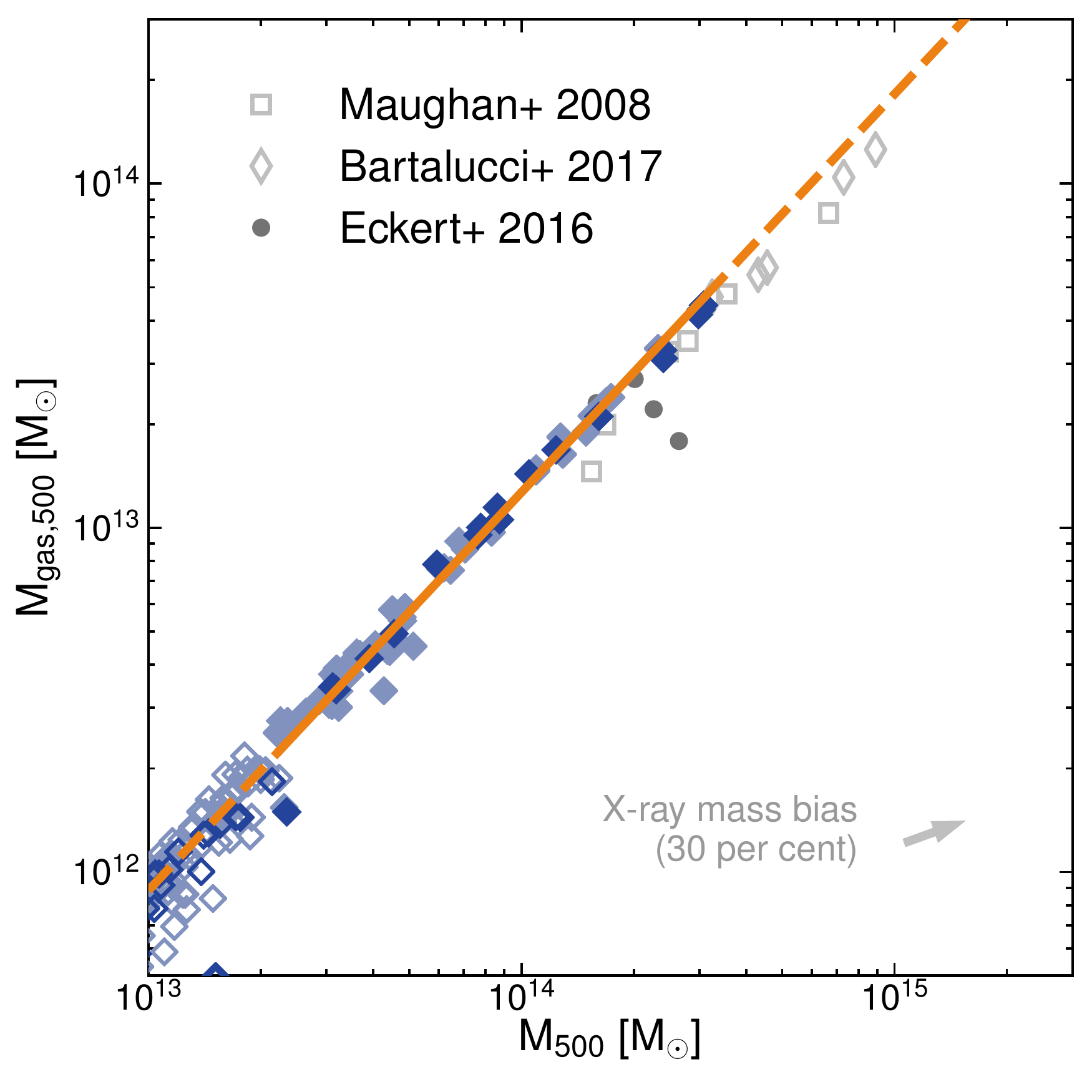}}
{\includegraphics[width=\factorsix\textwidth]{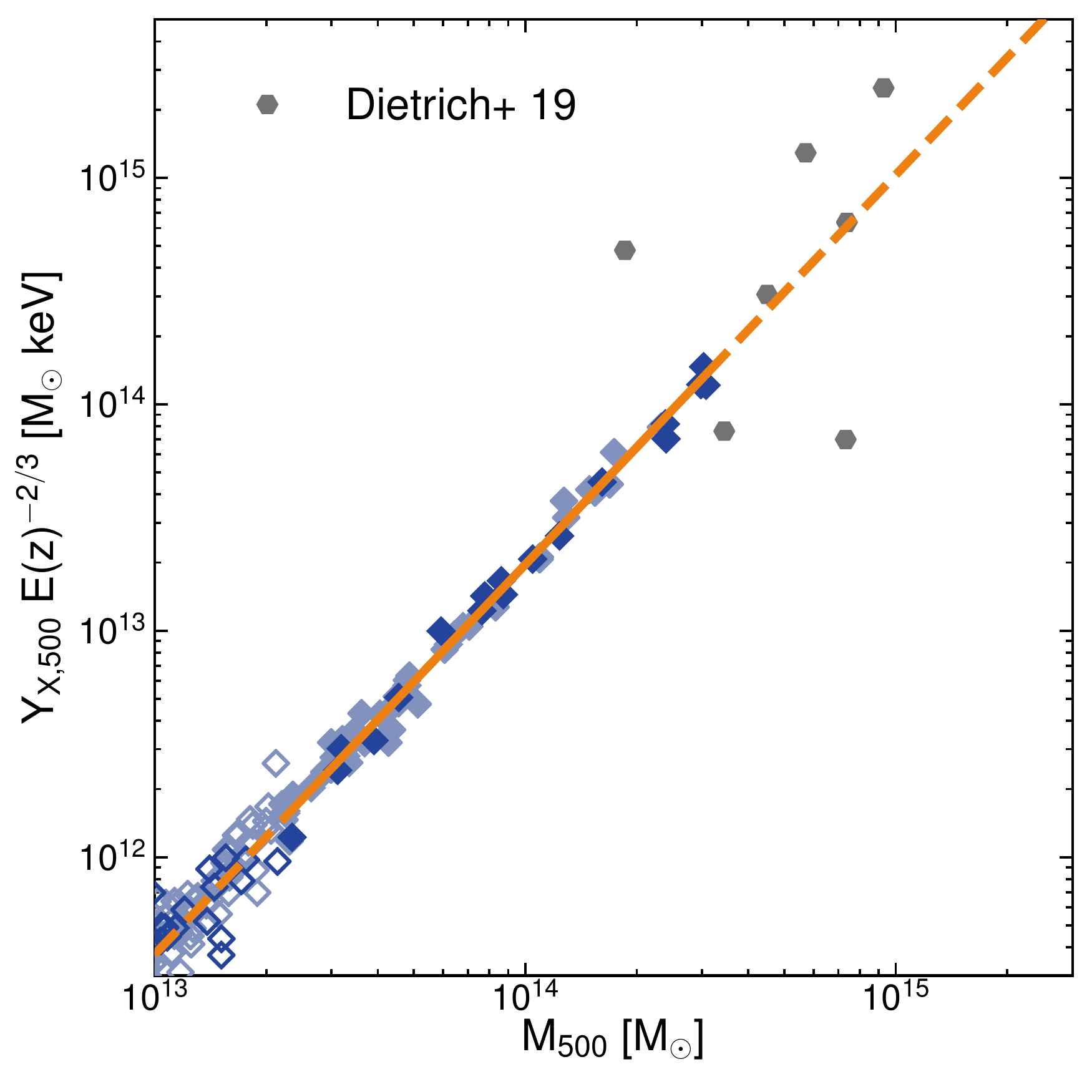}}
{\includegraphics[width=\factorsix\textwidth]{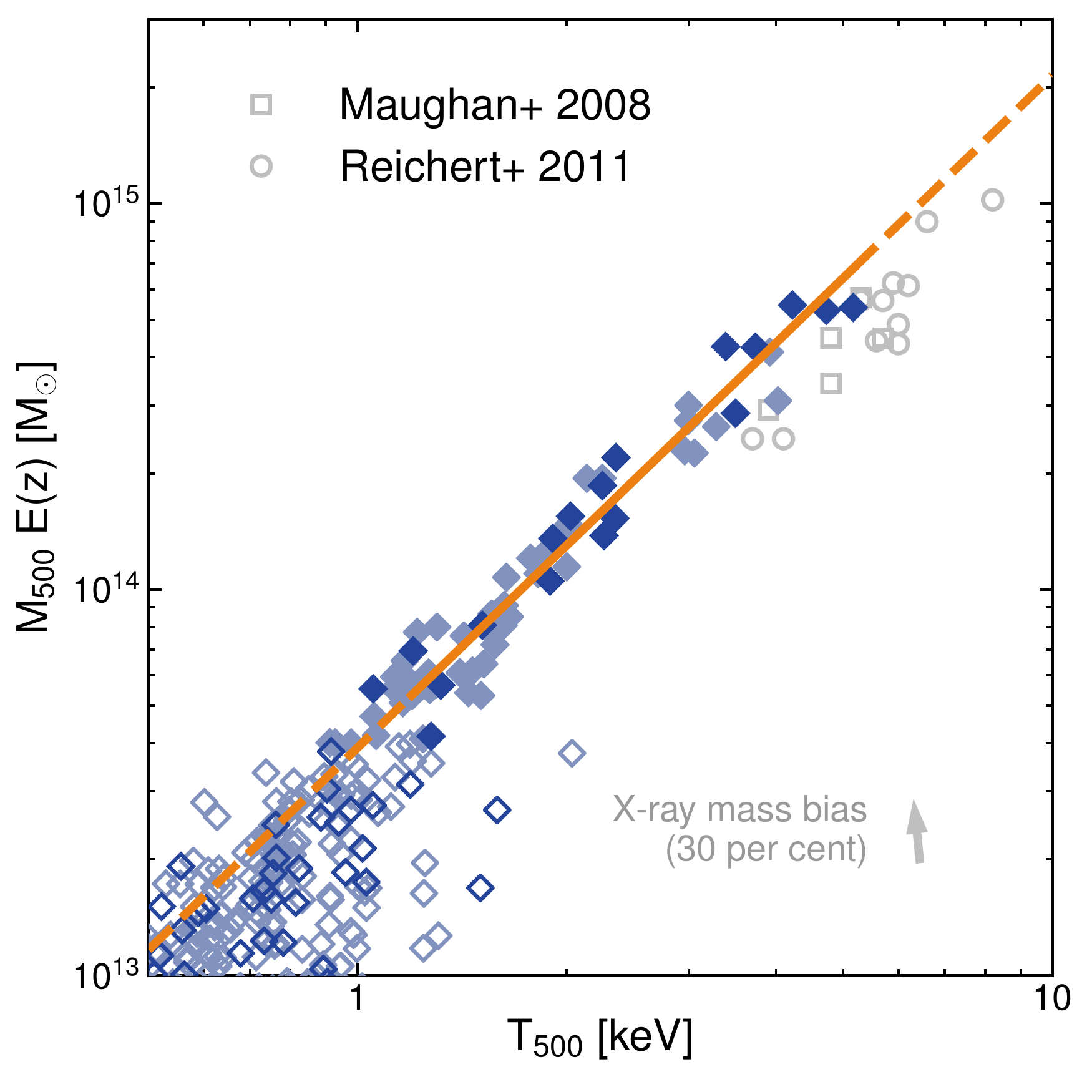}}
{\includegraphics[width=\factorsix\textwidth]{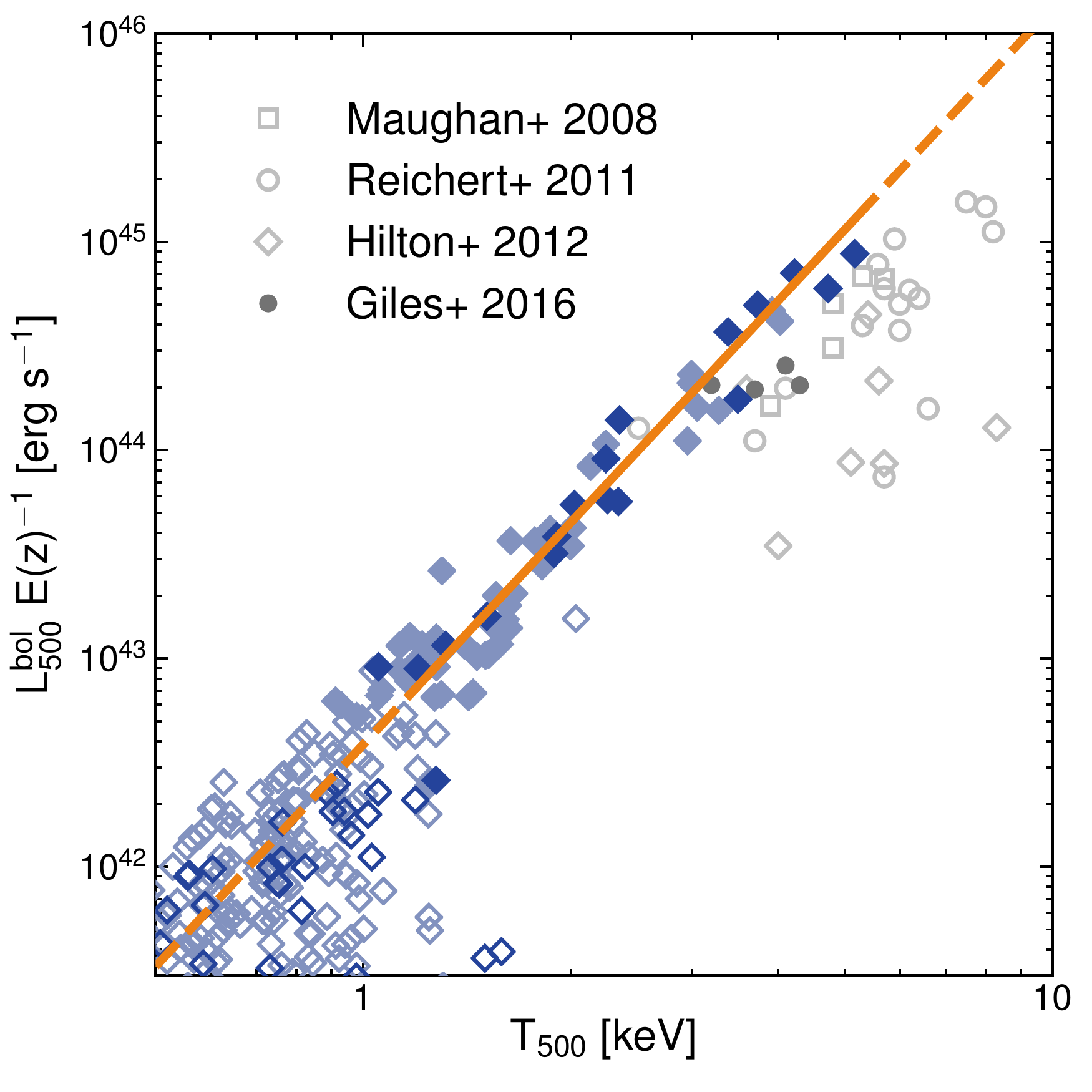}}
{\includegraphics[width=\factorsix\textwidth]{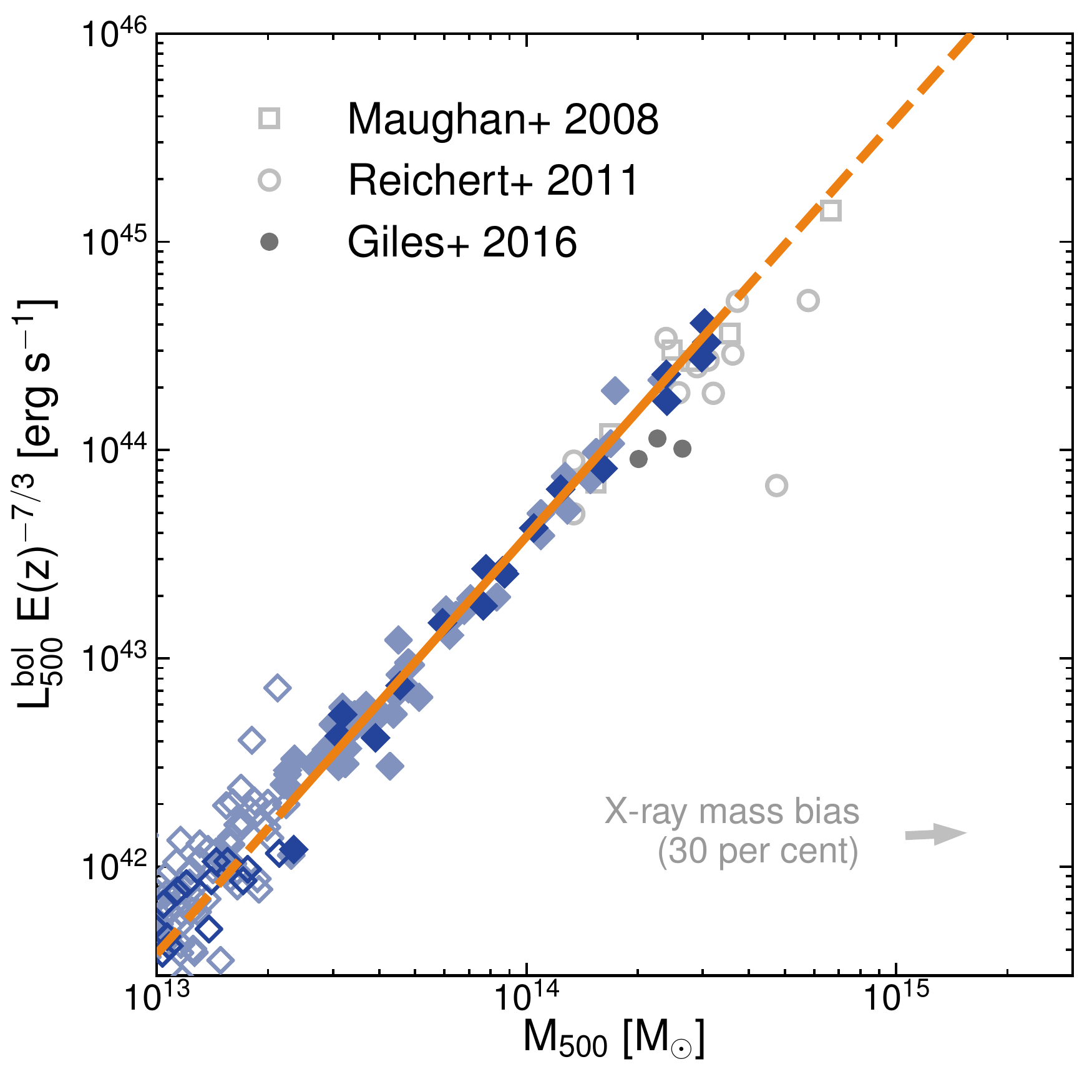}}
{\includegraphics[width=\factorsix\textwidth]{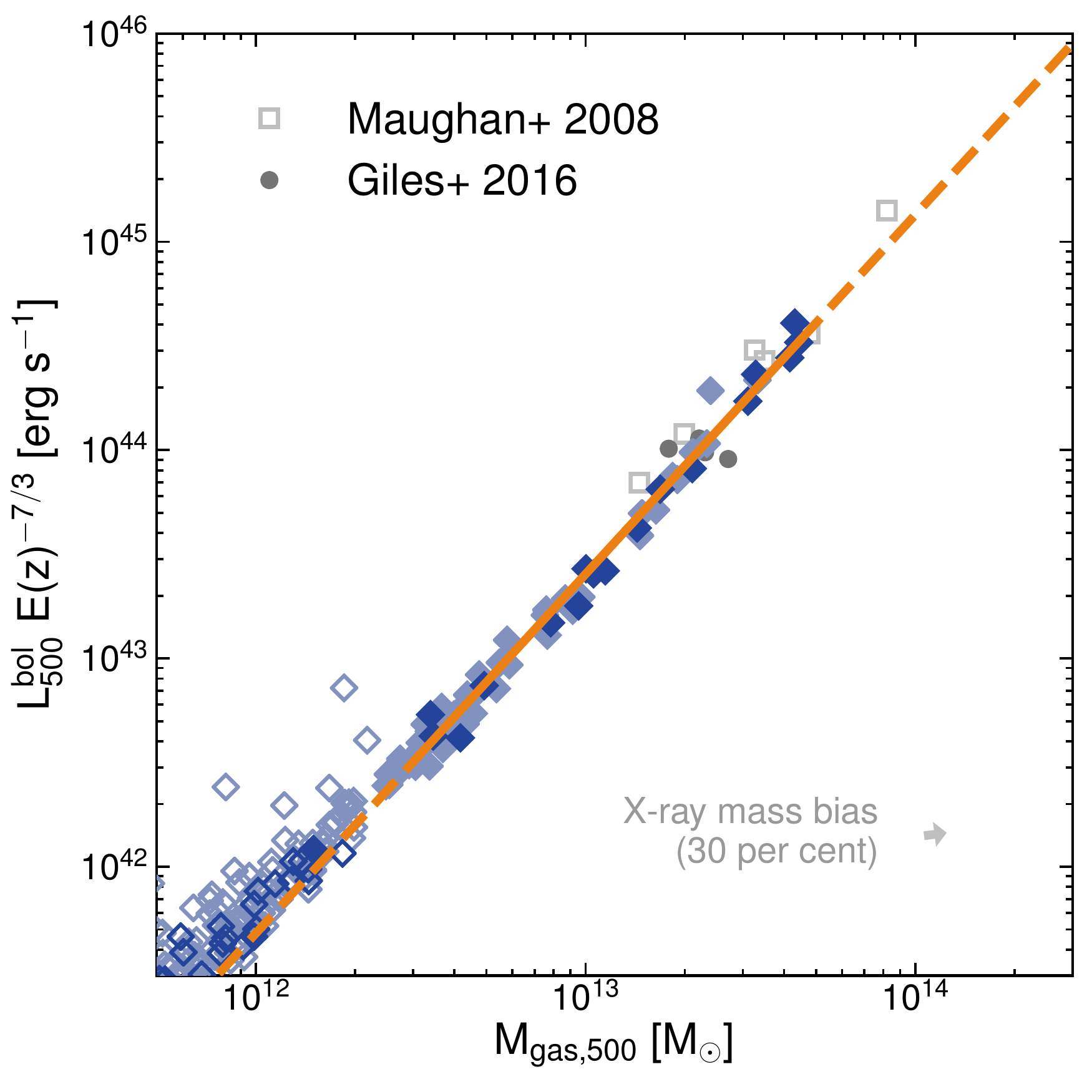}}
\end{center}
\caption{X-ray scaling relations at $z=1.0$ compared to observations at similar redshift. Symbol styles are identical to those of Fig.~\ref{fig:xray_z04}.
  }
    \label{fig:xray_z1}
\end{figure*}

In Fig.~\ref{fig:xray_z04} and \ref{fig:xray_z1} we compare the X-ray scaling relations at $z=0.4$ and $z=1.0$ with observed samples of clusters of similar median redshift (grey symbols).
Solid lines show the best-fitting power law relation for the mass-limited sample defined in Section~\ref{subsec:sample}, as indicated by filled diamonds. Observational data based on weak lensing mass measurements are distinguished from X-ray hydrostatic mass estimates with filled and open symbols, respectively.
In this section we mimic \textit{Chandra} ACIS-I observations (see Section~\ref{subsec:xray_props}).

\subsubsection{Observational data}\label{subsubsec:obs_data}
We compare extensively to results from the XXL-100-GC sample, which consists of the 100 brightest clusters in the XXL survey over the redshift range $0.05 < z < 1.1$ \citep{Pacaud2016}.
For the total mass--temperature relation ($z=0.4$ only) we compare to XXL-100-GC clusters with direct weak lensing mass estimates from \cite{Lieu2016}, restricting the sample to clusters at $0.259 \leq z \leq 0.52$ with a median redshift of $0.41$.
For the X-ray luminosity and gas mass-based relation we compare to data from \cite{Giles2016} and \cite{Eckert2016} for which masses are estimated from the weak lensing calibrated total mass--temperature relation derived in \cite{Lieu2016}. We choose XXL-100-GC clusters at $0.3 < z < 0.5$ and $0.91 < z < 1.05$ with median redshifts $0.39$ and $0.99$, respectively.
We note that the spectroscopic temperatures of the XXL-100-GC were measured within a circular aperture of fixed radius $300$ kpc, however, \cite{Giles2016} find no systematic difference between this temperature and the temperature measured within $r_{500}$.

Additional weak lensing-based data come from \cite{Mahdavi2013} for a sample of clusters in the redshift range $0.15 < z < 0.55$ with combined \textit{Chandra} and \textit{XMM-Newton} X-ray data. We restrict our comparison to clusters at $0.28 < z < 0.55$ with a median redshift of $0.40$. We caution that the weak lensing mass estimates for this sample were revised upwards in \cite{Hoekstra2015} by approximately 20 per cent on average. However, updated values for the X-ray quantities (due to the associated increase in $r_{500}$) are not available. We therefore compare to the published data from \cite{Mahdavi2013} and describe in the text how an increase in the mass estimates may affect our comparison.

For the $Y_{\mathrm{X}}$--total mass relation at $z=0.4$ we also compare to \textit{Chandra} and \textit{ROSAT} data from \cite{Mantz2016}. We restrict their sample to clusters with weak lensing mass measurements at $0.35 < z < 0.45$ with a median redshift of $0.40$.

\cite{Maughan2008}, \cite{Reichert2011} and \cite{Hilton2012} study the X-ray scaling relations for large samples of clusters out to high redshift using X-ray hydrostatic mass estimates.
\cite{Maughan2008} analyse 115 clusters at $0.1 < z < 1.3$ with archived \textit{Chandra} data. For the $z=0.4$ and $z=1$ comparisons we limit their sample to $0.30 < z < 0.50$ (median $0.40$) and $0.83 < z < 1.24$ (median $0.96$), respectively. For the $Y_{\mathrm{X}}$--total mass comparison at $z=0.4$ we use a subset of their clusters at $0.4 \lesssim z \lesssim 0.46$ with direct X-ray hydrostatic mass estimates given in \cite{Maughan2007}.
\cite{Reichert2011} combine numerous published data sets to study the evolution of the X-ray scaling relations out to $z \sim 1.5$. For the $z=0.4$ and $z=1$ samples we use clusters at $0.30 < z < 0.50$ and $0.90 \leq z < 1.11$ with median redshifts of $0.41$ and $0.99$, respectively.
\cite{Hilton2012} measure the evolution of the X-ray luminosity--temperature relation out to $z \sim 1.5$ using 211 clusters from the \textit{XMM} Cluster Survey \citep{Mehrtens2012}. For the $z=0.4$ and $z=1$ comparisons we restrict their sample to $0.30 \leq z \leq 0.50$ and $0.91 \leq z \leq 1.13$ with median values of $0.41$ and $1.00$, respectively.

At $z = 1$ we supplement our comparison with data from \cite{Bartalucci2017c} and \cite{Dietrich2019}.
\cite{Bartalucci2017c} study five clusters at $0.93 < z < 1.13$ detected via the SZ effect with gas mass estimates derived from combined \textit{XMM-Newton} and \textit{Chandra} data. Halo masses are estimated using the total mass--$Y_{\mathrm{X}}$ relation of \cite{Arnaud2010} assuming self-similar evolution.
We compare our $Y_{\mathrm{X}}$--total mass relation at $z=1$ to the high-redshift cluster sample studied in \cite{Dietrich2019} with weak lensing mass estimates from \textit{Hubble Space Telescope} data \citep{Schrabback2018} and X-ray data from \textit{Chandra}. We use seven of their clusters at $0.87 \leq z \leq 1.13$ and extract individual values of $Y_{\mathrm{X}}$ and $M_{500}$ as presented in their fig. 11.

\subsubsection{Comparison to observations}\label{subsubsec:obs_comp}

Fig.~\ref{fig:xray_z04} and \ref{fig:xray_z1} reveal many of the same trends as our comparison to local cluster data presented in Paper~I. In particular, the \fable\ clusters lie on the upper end of the observed scatter in the X-ray luminosity--temperature relation. This implies that the X-ray luminosities or spectroscopic temperatures of the simulated clusters may be over- or underestimated, respectively. Which of these interpretations dominates depends on whether observations based on X-ray hydrostatic or weak lensing masses are used for the comparison.

It is clear from Fig.~\ref{fig:xray_z04} and \ref{fig:xray_z1} that several of the observed scaling relations change significantly when using weak lensing mass estimates (filled symbols) as opposed to X-ray hydrostatic masses (open symbols).
From Fig.~\ref{fig:xray_z04} we see that the $z=0.4$ total mass--temperature and X-ray luminosity--total mass relations based on weak lensing are offset in normalisation compared with data based on X-ray masses, particularly if the \cite{Mahdavi2013} weak lensing masses are revised upwards by $\sim 20$ per cent as suggested by \cite{Hoekstra2015}. It is unclear whether this difference continues to $z=1$ due to the lack of weak lensing data at high redshift, although a similar offset in normalisation has been shown at $z \approx 0$ (e.g. \citealt{Lieu2016} and Paper~I).

These discrepancies imply a systematic bias between masses measured from X-rays and masses measured via weak lensing. The size of such an offset -- commonly referred to as an X-ray mass bias -- is currently under debate.
Some observational studies find that the two methods yield equivalent results (e.g. \citealt{Gruen2014, Israel2014, Smith2015, Applegate2016}) while others find that X-ray hydrostatic masses are biased low compared to weak lensing masses by $\sim 25$--$30$ per cent within $r_{500}$ (e.g. \citealt{Donahue2014, VonderLinden2014, Hoekstra2015, Sereno2017, Simet2017, Hurier2017}). In addition, cosmological constraints from cluster abundance studies seem to require an even larger X-ray mass bias of $\sim 40$ per cent in order to reconcile their results with \textit{Planck} measurements of the primary CMB \citep{PlanckXXIV2015, Salvati2018}.

Weak lensing is expected to give a less biased estimate of the true mass than the X-ray hydrostatic method because it is relatively insensitive to the equilibrium state of the cluster (see \citealt{Hoekstra2013} and \citealt{Mandelbaum2017} for recent reviews).
This is generally confirmed by numerical studies, which find significantly smaller biases in weak lensing mass measurements ($\lesssim 10$ per cent; e.g. \citealt{Meneghetti2010, Becker2011, Rasia2012, Henson2017}) compared with X-ray hydrostatic masses ($\sim 10$--$35$ per cent; e.g. \citealt{Nagai2007a, Kay2012, Rasia2012, LeBrun2014, Biffi2016}).

Given that we do not model an X-ray hydrostatic mass bias in our analyses, care must be taken when comparing the simulated relations to data based on X-ray masses (open symbols).
To aid this comparison, in Fig.~\ref{fig:xray_z04} and \ref{fig:xray_z1} we indicate how the X-ray data are expected to change when correcting for a possible X-ray mass bias in which the X-ray masses are biased low by $30$ per cent. We have calculated the expected change in the observable quantities due to the change in the aperture radius, $r_{500}$, from our simulated clusters with $M_{500} > 10^{14} M_{\odot}$ at $z=0.4$ and $z=1$ and find that, on average, the gas mass, spectroscopic temperature, bolometric X-ray luminosity, and $Y_{\mathrm{X}}$ increase by $15$, $-2$, $3$ and $14$ per cent, respectively.
  These changes are indicated with an arrow in the bottom-right of each panel, with the exception of the X-ray luminosity--temperature relation for which the change is negligible.
  We note that these shifts in the observable quantities are due only to the change in aperture and do not explicitly take into account additional sources of bias such as gas clumping or instrument calibration, which may bias the inferred gas mass, X-ray luminosity or temperature (e.g. \citealt{Nagai2007a, Rasia2012, Schellenberger2015}).

Taking a potential X-ray mass bias into consideration, our comparison to the observed total mass--temperature and X-ray luminosity--total mass relations (middle- and bottom-left panels) implies that the simulated clusters possess realistic global temperatures but are somewhat over-luminous at fixed mass.
Given that the relation between X-ray luminosity and gas mass is a good match to the observations (bottom-right panel), the discrepancy in X-ray luminosity at fixed total mass must largely be driven by an overestimate in the gas mass.
Indeed, the \fable\ systems lie on the upper end of the scatter in the gas mass--total mass relation compared with the weak lensing-based studies of \cite{Mahdavi2013} and \cite{Eckert2016}, as well as with the data based on X-ray hydrostatic masses if these are biased low.
As we discuss in Paper~I, an increase in the efficiency of our AGN feedback model could reduce this discrepancy by ejecting larger gas masses from massive haloes. However, it is likely that a more sophisticated modelling of AGN feedback is required to simultaneously reproduce the cluster thermodynamic profiles.

Although the \fable\ model tends to overpredict the gas masses and X-ray luminosities at fixed mass or temperature, the size of the offset does not change dramatically between $z=0$ and $z=1$.
This gives us confidence that the \fable\ model makes reliable predictions for the redshift evolution of the cluster scaling relations, which is the main topic of this paper.
Indeed, in the following section we investigate the redshift evolution of the X-ray scaling relations and show that the \fable\ predictions are often bracketed by the results of two other recent simulation works that show different levels of agreement with observations.

\subsection{Evolution of the X-ray scaling relations}\label{subsec:evol}
In the following sections we assess the redshift evolution of the best-fitting parameters (slope, normalisation and intrinsic scatter) of each X-ray scaling relation.
Here we mimic X-ray observations with the planned \textit{Athena} X-IFU instrument (see Section~\ref{subsec:xray_props}).
We compare with recent results from the \MACSIS\ (\citealt{MACSIS}; hereafter B17) and \cite{Truong2018} (hereafter T18) galaxy cluster simulations, which we describe below.

The \MACSIS\ suite of zoom-in simulations consists of 390 galaxy clusters simulated with the baryonic physics model of the {\sc bahamas} simulation \citep{McCarthy2017}. The model was calibrated to reproduce the present-day galaxy stellar mass function and the hot gas mass fractions of galaxy groups and clusters. The calibrated simulations reproduce a broad range of group and cluster properties at $z \approx 0$, including the X-ray and SZ scaling relations and the thermodynamic profiles of the ICM \citep{McCarthy2017}.
B17 investigate the redshift evolution of a combined sample of clusters from the \MACSIS\ and {\sc bahamas} simulations out to $z=1.5$. Their sample consists of haloes above a redshift-independent mass limit of $M_{500} > 10^{14} M_{\odot}$, yielding 1294 clusters at $z=0$ and 225 at $z=1$.
B17 calculate bolometric X-ray luminosities and spectroscopic temperatures from synthetic X-ray observations following \cite{LeBrun2014}.
We note that B17 use X-ray hydrostatic masses estimated from their mock X-ray data, which are biased low compared with the mass measured directly from the simulation \citep{Henson2017}.
We caution that this may bias the slope of the observable--mass relations and their intrinsic scatter somewhat high compared with \fable\ and T18.
On the other hand, we do not expect a significant change to the redshift evolution of the slope, normalisation or intrinsic scatter of the \MACSIS\ relations given that \cite{Henson2017} find no evidence for a redshift dependence of the X-ray mass bias.

T18 analyse a suite of 29 zoom-in simulations including AGN feedback that reproduce a wide range of cluster properties, most notably the observed dichotomy between cool-core and non-cool-core clusters \citep{Rasia2015} and their thermodynamic profiles \citep{Rasia2015, Planelles2017}.
Their zoom-in simulations are centred on clusters drawn from a $1 \, h^{-3}$ Gpc$^3$ parent simulation \citep{Bonafede2011} with a high-resolution Lagrangian region extending to at least five times the virial radius.
Their cluster sample comprises all objects in the high-resolution regions above an evolving halo mass threshold of $M_{500} > 10^{14} E(z)^{-1} M_{\odot}$.
T18 compute bolometric X-ray luminosities from pre-calculated cooling-function tables assuming the \textsc{APEC} model and approximate the spectroscopic temperature using the ``spectroscopic-like'' temperature \citep{Mazzotta2004}.
Both T18 and B17 calculate their X-ray luminosities and temperatures within a spherical, core-excised aperture ($0.15 < r / r_{500} < 1$). We have repeated our analyses for the same aperture and comment on the change to the best-fitting parameters in the text.

We calculate the normalisation of the relations at the pivot points used in our own fitting procedure, which are $M_{500} = 2 \times 10^{14} M_{\odot}$ and $T_{500} = 3$ keV.
These are very close to the pivot points used in T18 ($M_{500} \approx 1.5 \times 10^{14} M_{\odot}$ and $T_{500} = 3$ keV) although somewhat lower than \MACSIS\ ($M_{500} = 4 \times 10^{14} M_{\odot}$ and $T_{500} = 6$ keV; B17).
We propagate the uncertainty in the normalisation assuming that the uncertainties on the best-fitting slope are independent of the normalisation and normally distributed.
These assumptions are not necessarily valid for \MACSIS\ and T18, however we find no systematic bias in the uncertainties when applying the same procedure to our own uncertainties for a range of different pivot points.
To enable a comparison of the redshift evolution of the normalisation we plot the normalisation at each redshift relative to the $z=0$ value. Because the best-fitting scaling relations include a term for self-similar evolution of the normalisation (Section~\ref{subsec:fitting}), any deviation of the curves from horizontal indicates departure from self-similarity. Positive (negative) evolution refers to a normalisation that increases (decreases) with increasing redshift relative to the self-similar expectation.

B17 and T18 calculate the intrinsic scatter as in equation~\ref{eq:scat} except that B17 take $N$ rather than $N-2$ in the denominator. Their sample is large enough however that the difference is negligible.
When quoting the intrinsic scatter measured in other studies we convert to units of dex to be consistent with the definition in equation~\ref{eq:scat}.

We also make frequent reference to the simulation study of \cite{LeBrun2017} although we do not plot their best-fitting parameters. \cite{LeBrun2017} analyse the cosmo-OWLS suite of simulations \citep{LeBrun2014}, which employ four different galaxy formation models. Unless otherwise stated we refer to their fiducial AGN 8.0 simulation, which \cite{LeBrun2014} have shown produces the best match to observations. The AGN 8.0 model is similar to that of {\sc bahamas} and \MACSIS\ except for slight adjustments to the parameters of the stellar and AGN feedback models (see table 1 in \citealt{McCarthy2017}). \cite{LeBrun2017} construct a sample of all haloes with $M_{500} > 10^{13} M_{\odot}$ at each redshift up to $z=1.5$. They fit their scaling relations using both a single power law and a broken power law with low- and high-mass slopes below and above the pivot point ($M_{500} = 10^{14} M_{\odot}$). We refer to their single power law fit unless stated otherwise.

In the following sections we discuss the X-ray scaling relations in turn. We focus much of our attention on interpreting the gas mass--total mass and total mass--temperature relations (Sections \ref{subsubsec:Mgas-M} and \ref{subsubsec:M-T}) as the other relations are closely related to these.

\subsubsection{Gas mass--total mass scaling relation}\label{subsubsec:Mgas-M}

\begin{figure*}
\begin{center}
{\includegraphics[width=\factorthree\textwidth]{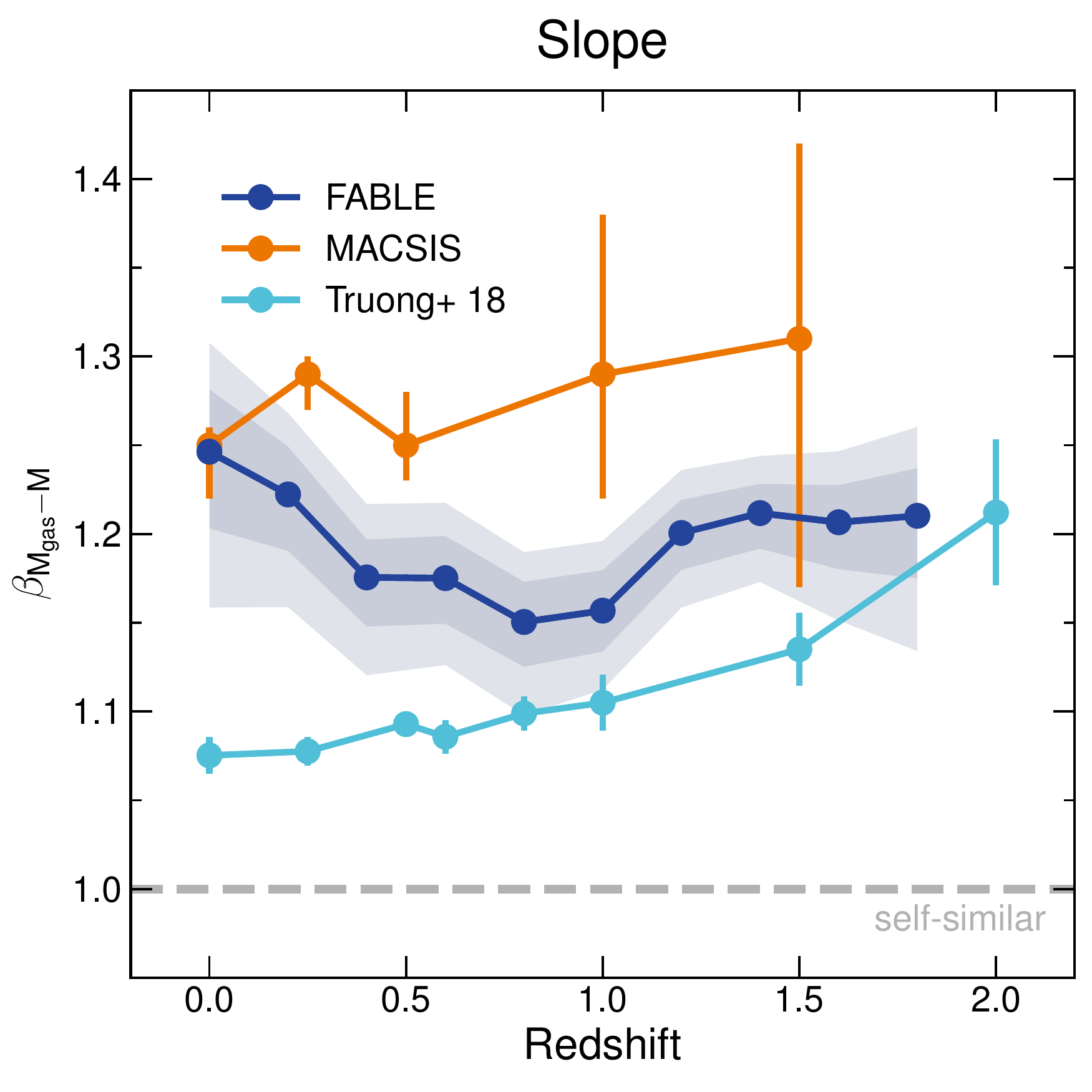}}
{\includegraphics[width=\factorthree\textwidth]{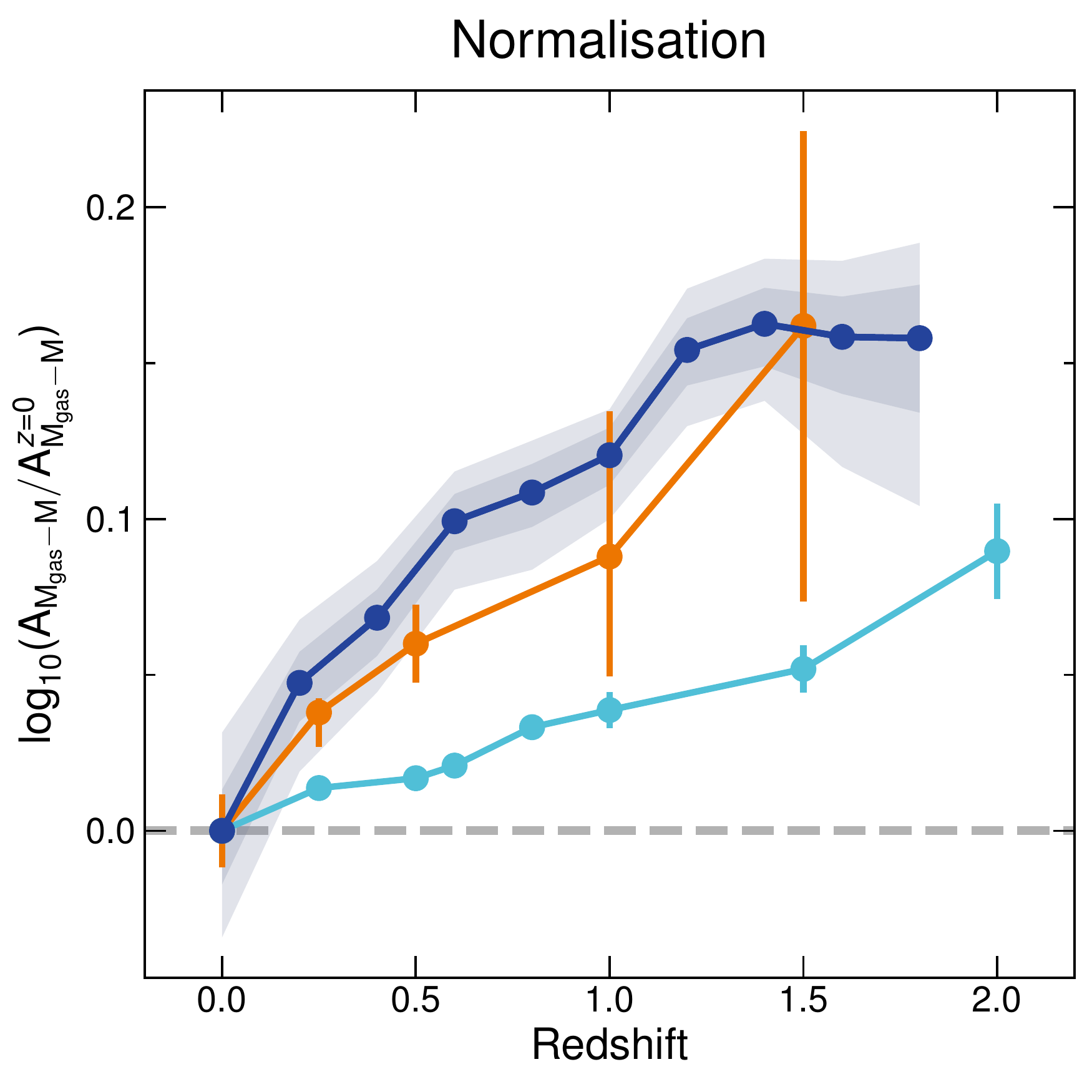}}
{\includegraphics[width=\factorthree\textwidth]{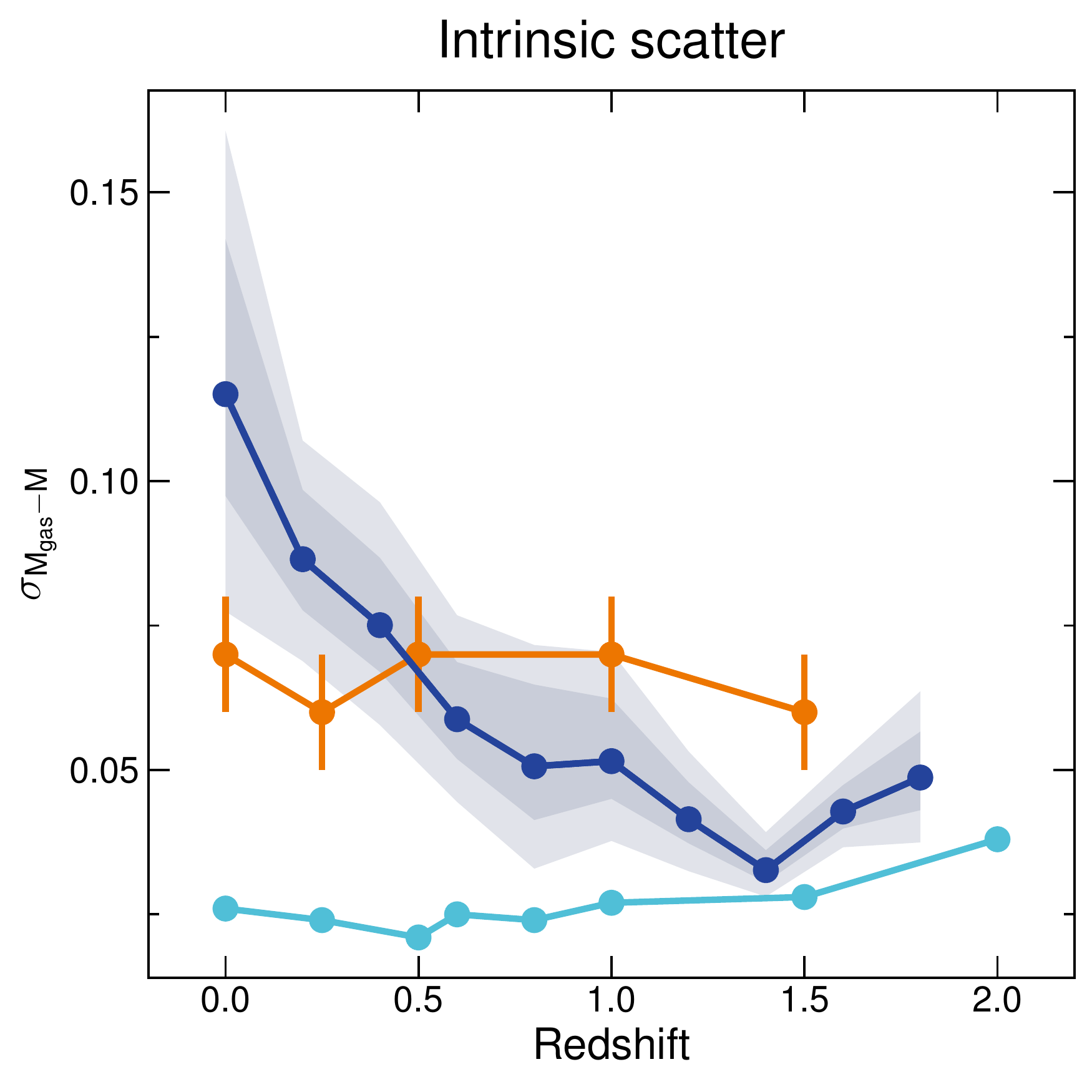}}
\end{center}
\caption{The redshift evolution of the gas mass--total mass relation for the \fable\ (dark blue), \MACSIS\ (orange) and T18 (light blue) simulations. The panels show, from left to right, the slope, normalisation and intrinsic scatter of the best-fitting power law relation as a function of redshift. Shaded regions about the \fable\ relation represent the 68 and 95 per cent confidence intervals on the best-fitting parameters estimated from bootstrap resampling.
The uncertainties on the intrinsic scatter are unknown for the T18 relation. Grey dashed lines indicate the self-similar expectation.
  }
    \label{fig:Mgas-M}
\end{figure*}

Fig.~\ref{fig:Mgas-M} shows, from left to right, the redshift evolution of the best-fitting slope, normalisation and intrinsic scatter of the gas mass--total mass relation, respectively.

\paragraph{Gas mass--total mass slope}
The self-similar expectation for the slope of the gas mass--total mass relation is unity (i.e. a constant gas mass fraction). This is indicated by a horizontal dashed line in the left-hand panel of Fig.~\ref{fig:Mgas-M}.
\fable, \MACSIS\ and T18 predict a gas mass--total mass slope significantly greater than unity out to $z \sim 2$. This is consistent with the results of \cite{Chiu2016, Chiu2018} who find a steeper than self-similar mass trend in the gas content of galaxy groups and clusters out to $z \approx 1.3$ with no significant redshift dependence.
Previous numerical studies have shown that simulations incorporating radiative cooling and star formation yield a steeper than self-similar gas mass--total mass relation due to the conversion of gas into stars, which occurs more efficiently in lower mass systems (e.g. \citealt{Stanek2010, Battaglia2013, Planelles2014, LeBrun2017}). In addition, simulations that include some form of AGN feedback produce an even steeper slope since feedback is able to expel gas more efficiently from lower mass haloes due to their shallower potential wells, leading to a tilt in the relation (e.g. \citealt{Puchwein2008, Fabjan2011, McCarthy2011, Gaspari2014}).

At $z=0$, \fable\ and \MACSIS\ predict a slope of $1.25^{+0.04}_{-0.04}$ and $1.25^{+0.01}_{-0.03}$, respectively. These values are in good agreement with a number of observational constraints, including \cite{Arnaud2007} ($1.25 \pm 0.06$), \cite{Gonzalez2013} ($1.26 \pm 0.03$), \cite{Eckert2016} ($1.21^{+0.11}_{-0.10}$) and \cite{Chiu2018} ($1.32 \pm 0.07$). T18 derive a significantly shallower slope of $1.08 \pm 0.01$ at $z=0$, although still steeper than self-similar. Some observational studies measure a similarly shallow slope, for example \cite{Lin2012} ($1.13 \pm 0.03$), \cite{Mahdavi2013} ($1.04 \pm 0.10$) and \cite{Mantz2016a} ($1.04 \pm 0.05$), the latter two results being consistent with self-similarity.
Differences between studies can be attributed to a number of factors. One of the most influential of these is the sample selection, such as the distribution of mass, redshift and dynamical state of the clusters. For instance, the sample of \cite{Mantz2016a} is comprised of massive relaxed clusters that, due to their deep potential wells, have lost a comparatively small fraction of their gas through AGN feedback.
Indeed, we find a mass dependence in the gas mass--total mass slope in our simulations (see Appendix~\ref{A:mass_dependence}). For example, if we restrict our $z=0$ sample to haloes with $M_{500} > 10^{14} M_{\odot}$ as for the \MACSIS\ and T18 samples then the best-fitting slope drops to $1.16^{+0.05}_{-0.05}$, which lies in between their results.

The simulation results additionally depend on the model implementation of AGN feedback.
The \fable, \MACSIS\ and T18 models all inject AGN feedback energy thermally, however they differ as to when and how much energy is input. In both \fable\ and \MACSIS\ the frequency of energy injection is controlled by a duty cycle, although in \MACSIS\ the duty cycle depends on the ability of the AGN to heat the surrounding gas \citep{Booth2009, McCarthy2017}, while in \fable\ it is controlled either by a fixed accumulation time or by the mass growth of the black hole, depending on whether the AGN is in the quasar- or radio-mode (see Paper~I and \citealt{Sijacki2007}).
Conversely, the T18 model for AGN feedback inputs feedback energy continuously \citep{Steinborn2015}. Continuous AGN feedback may have a gentler impact on the ICM compared with the duty cycle models of \fable\ and \MACSIS, which store feedback energy before injecting it into the surrounding gas in a single energetic event. For example, a number of studies have found that more energetic but less frequent AGN feedback events are more effective at reducing the gas content of galaxy groups and clusters (e.g. \citealt{LeBrun2014}, \citealt{Schaye2015} and Appendix A of Paper~I). Since AGN feedback is more effective in lower mass haloes, this may explain why the \fable\ and \MACSIS\ models predict a slightly steeper gas mass--total mass relation compared with T18.

Although the AGN feedback models used in \fable\ and \MACSIS\ are similar in the sense that the feedback energy is input thermally and on a duty cycle, the slope of the gas mass--total relation in \fable\ is somewhat shallower than \MACSIS\ at most redshifts.
This suggests that the removal of gas via AGN feedback is less efficient in \fable\ than in \MACSIS. Indeed, in Section~\ref{subsubsec:obs_comp} we showed that \fable\ haloes are seemingly too gas-rich at fixed halo mass, implying the need for more efficient feedback.
Some of the difference in slope can also be attributed to X-ray mass bias in the \MACSIS\ results, which \cite{Henson2017} show increases mildly with mass, thereby steepening the gas mass--total mass relation.

At $z \lesssim 1$ the gas mass--total mass slope decreases mildly with increasing redshift.
This is consistent with \cite{LeBrun2017} who predict a decrease in slope with increasing redshift out to $z=1.5$ for their simulations with AGN feedback.
This evolution can partly be attributed to the reduced efficiency of gas expulsion by AGN feedback with increasing redshift, which we discuss in the following section.
In addition, the change in slope at $z \lesssim 1$ corresponds to an increase in the proportion of central black holes operating in the radio-mode of AGN feedback (from a constant $\sim 90$ per cent at $z \gtrsim 1$ to almost $100$ per cent at $z = 0$ in our sample), which occurs when the black hole accretion rate drops below one per cent of the maximal Eddington rate. In Appendix A of Paper~I we show that the radio-mode of feedback is more efficient than the quasar-mode at reducing the gas mass fractions of group-scale haloes. As a result, the increasing prevalence of the radio-mode with decreasing redshift may partially explain the increase in the gas mass--total mass slope.

At $z \gtrsim 1$ the slope is approximately independent of redshift. In fact, within the uncertainties the evolution in the slope is consistent with zero over a much wider redshift range ($0.4 \lesssim z \leq 1.8$), which is consistent with \MACSIS\ and with the simulations with AGN feedback in \cite{Fabjan2011} and \cite{Battaglia2013}.
T18 find a positive evolution in the slope at $z \gtrsim 1$, however they attribute this to the decreasing mass of their sample with increasing redshift coupled with gas mass fractions that decline towards lower mass haloes due to the aforementioned effects of AGN feedback. This effect is likely accentuated for their sample compared with \fable\ and \MACSIS\ as their minimum mass threshold falls more rapidly with redshift. These results suggest that, whilst differences in the theoretical modelling lead to slightly different predictions for the slope of the gas mass--total mass relation, the change in slope with redshift is typically small in comparison.

\paragraph{Gas mass--total mass normalisation}
In the middle panel of Fig.~\ref{fig:Mgas-M} we plot the normalisation of the gas mass--total mass relation as a function of redshift. All three simulations yield a positive evolution in normalisation relative to the self-similar model, which predicts no evolution (i.e. constant gas fraction with redshift).
This implies either that AGN feedback is less effective at expelling gas at high redshift or that the efficiency with which gas is cooled and converted into stars decreases with increasing redshift. In \fable\ it does not seem that the latter explanation holds since the total stellar mass within $r_{500}$ at fixed total mass shows little evolution at $z<2$ (not shown), consistent with the results of an SZ-selected sample at $0.2 < z < 1.25$ \citep{Chiu2018}.
This implies that the evolution of the normalisation is driven largely by AGN feedback. At high redshift, clusters of a given mass are denser and have deeper gravitational potential wells. The increased binding energy of a halo means that AGN feedback must supply more energy in order to eject gas beyond $r_{500}$. Furthermore, AGN at high redshift have had less time in which to affect the ICM of their host clusters and their progenitors. These effects combine to increase the normalisation of the gas mass--total mass relation with increasing redshift.
In addition, part of the evolution in the normalisation at $z \lesssim 1$ may be driven by the increasing prevalence of radio-mode feedback, which is able to more efficiently eject gas from massive haloes.

T18 predict a slightly weaker evolution than \fable\ and \MACSIS. This, combined with the difference in the slope of the relations, suggests that hot gas removal by cooling, star formation and AGN feedback is slightly less efficient in their simulations, at least at $z \lesssim 2$. This may be driven by differences in the AGN feedback implementation as discussed in the previous section.
Contrary to these results, simulations works such as \cite{Planelles2013} and \cite{Battaglia2013} find a constant gas and baryon mass fraction with redshift up to $z=1$, which suggests that feedback is not removing gas at all in these simulations at $z < 1$.

\paragraph{Gas mass--total mass intrinsic scatter}
The three simulations predict quite different levels of intrinsic scatter at low redshift. For instance, at $z=0$ the measured scatter is $0.12^{+0.03}_{-0.02}$ in \fable, $0.07 \pm 0.01$ in \MACSIS\ and $0.026$ in T18.
Observations typically measure an intrinsic scatter close to $\sigma_{\mathrm{M_{gas}}} \approx 0.05$, for example, \cite{Arnaud2007} ($0.044$), \cite{Mahdavi2013} ($0.07 \pm 0.03$), \cite{Mantz2016a} ($0.04 \pm 0.01$) and \cite{Chiu2018} ($0.05 \pm 0.01$).
These constraints are lower than the \fable\ prediction at low redshift but slightly higher than T18.

Some of the variation can be attributed to sample differences, in particular the mass range. Indeed, less massive objects tend to exhibit larger scatter in their X-ray properties, as has been shown for the cosmo-OWLS, \textsc{bahamas} and \MACSIS\ simulations \citep{LeBrun2017, Farahi2017}.
Similarly, \cite{Eckmiller2011} measure significantly increased intrinsic scatter for a sample of 26 local galaxy groups compared with a more massive sample of 64 clusters from the HIFLUGCS survey \citep{Hudson2010}.
As our sample is significantly less massive than that of \MACSIS, T18 and the aforementioned observational studies, our scatter measurement is likely biased high by low-mass objects. Indeed, in Appendix~\ref{A:mass_dependence} we show that increasing the mass of our sample significantly lowers the intrinsic scatter at all redshifts. For instance, at $z=0$ the intrinsic scatter drops from $0.12^{+0.03}_{-0.02}$ to $0.07^{+0.03}_{-0.01}$ when restricting our sample to the same mass range as \MACSIS\ and T18 at $z=0$ ($M_{500} \geq 10^{14} M_{\odot}$). This brings the scatter into agreement with \MACSIS\ and most observational constraints.

The increase in intrinsic scatter towards lower masses is likely associated with the increasing influence of non-gravitational processes such as stellar and AGN feedback in less massive haloes.
Indeed, numerical studies such as \cite{Stanek2010} and \cite{LeBrun2017} have shown that preheating or AGN feedback not only increases the intrinsic scatter of the gas mass at fixed halo mass but also strengthens the trend of increasing scatter with decreasing halo mass (see e.g. fig. 9 in \citealt{Stanek2010}). To confirm this result we have re-simulated our ($40 \, h^{-1}$ Mpc)$^3$ volume without AGN feedback. We find that the \fable\ model has increased intrinsic scatter compared with the non-AGN run for all of the scaling relations presented here and, although the volume is limited to haloes with $M_{500} \lesssim 10^{14} M_{\odot}$, there is a mild trend of increasing scatter towards lower halo masses in all scaling relations.

T18 predict an intrinsic scatter two to three times smaller than \fable\ and \MACSIS\ at $z=0$ for the same mass threshold, $M_{500} \geq 10^{14} M_{\odot}$. T18 expect their instrinsic scatter to be biased low due to their small sample size, which limits the number of outliers. However, their sample is three times larger than ours for the same mass range.
Part of this offset may be explained by differences in the AGN feedback modelling, in particular the magnitude and frequency of the thermal energy injections. As mentioned above, the AGN feedback models of \fable\ and \MACSIS\ operate on a duty cycle while the T18 model inputs energy continuously. With a duty cycle, a large amount of feedback energy can be input to the ICM in one event. This can have a sudden and significant impact on ICM properties such as the gas mass and temperature, which we confirm in our model. The current ICM properties can therefore vary depending on the time that has passed since the last feedback event. As a result we might expect AGN feedback models with a duty cycle to produce stronger or more numerous outliers than a continuous feedback model, yielding a larger intrinsic scatter.

We find that the intrinsic scatter decreases with increasing redshift in \fable, in contrast to T18 and \MACSIS\ which predict little to no evolution. This may be driven by the increasing binding energy at fixed mass or the decreasing fraction of radio-mode AGN with increasing redshift discussed above.
We point out that significant redshift evolution in the intrinsic scatter could have important implications for cluster cosmology, which requires knowledge of the intrinsic scatter in order to properly account for selection biases (e.g. \citealt{Maughan2012}).

\subsubsection{Total mass--temperature scaling relation}\label{subsubsec:M-T}

\begin{figure*}
\begin{center}
{\includegraphics[width=\factorthree\textwidth]{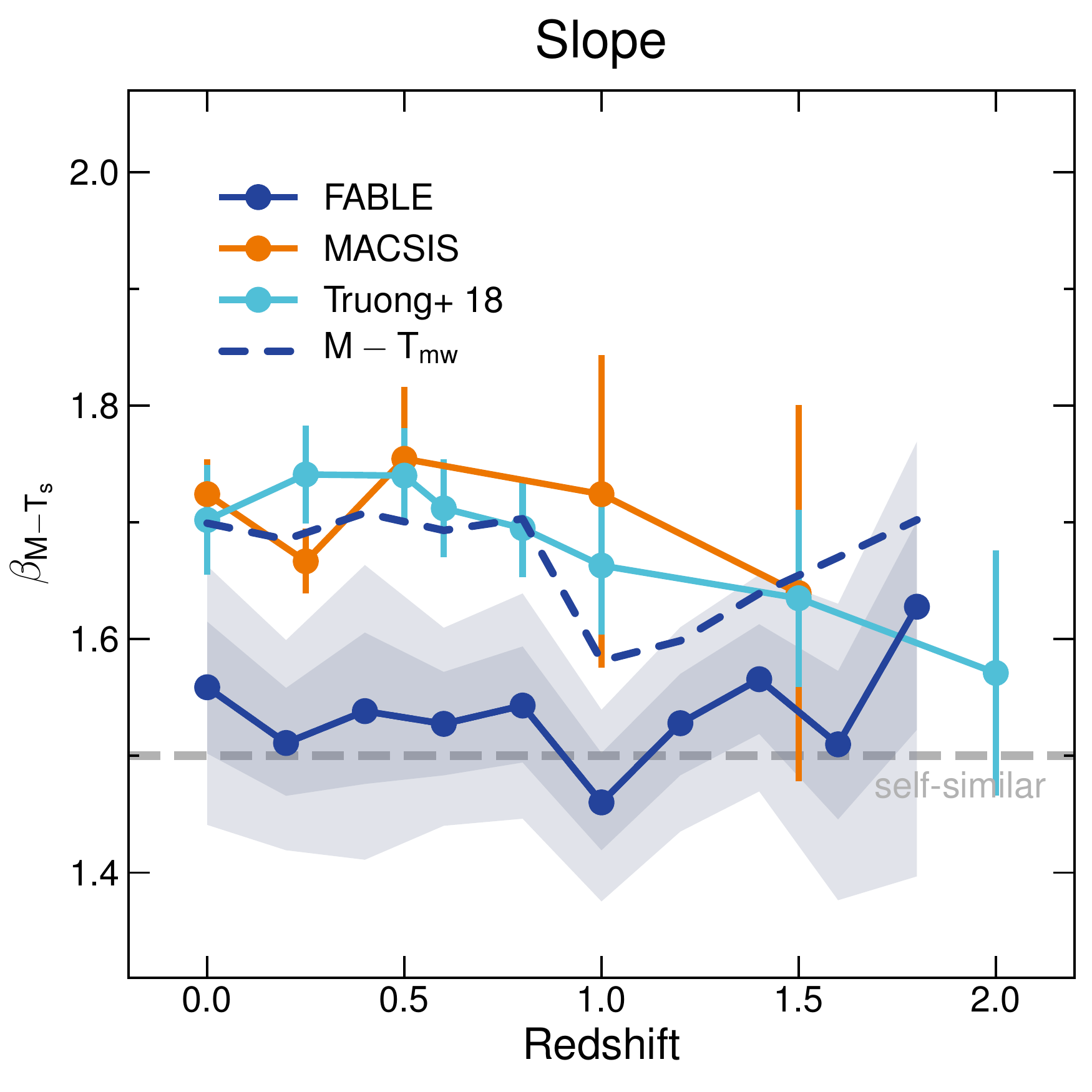}}
{\includegraphics[width=\factorthree\textwidth]{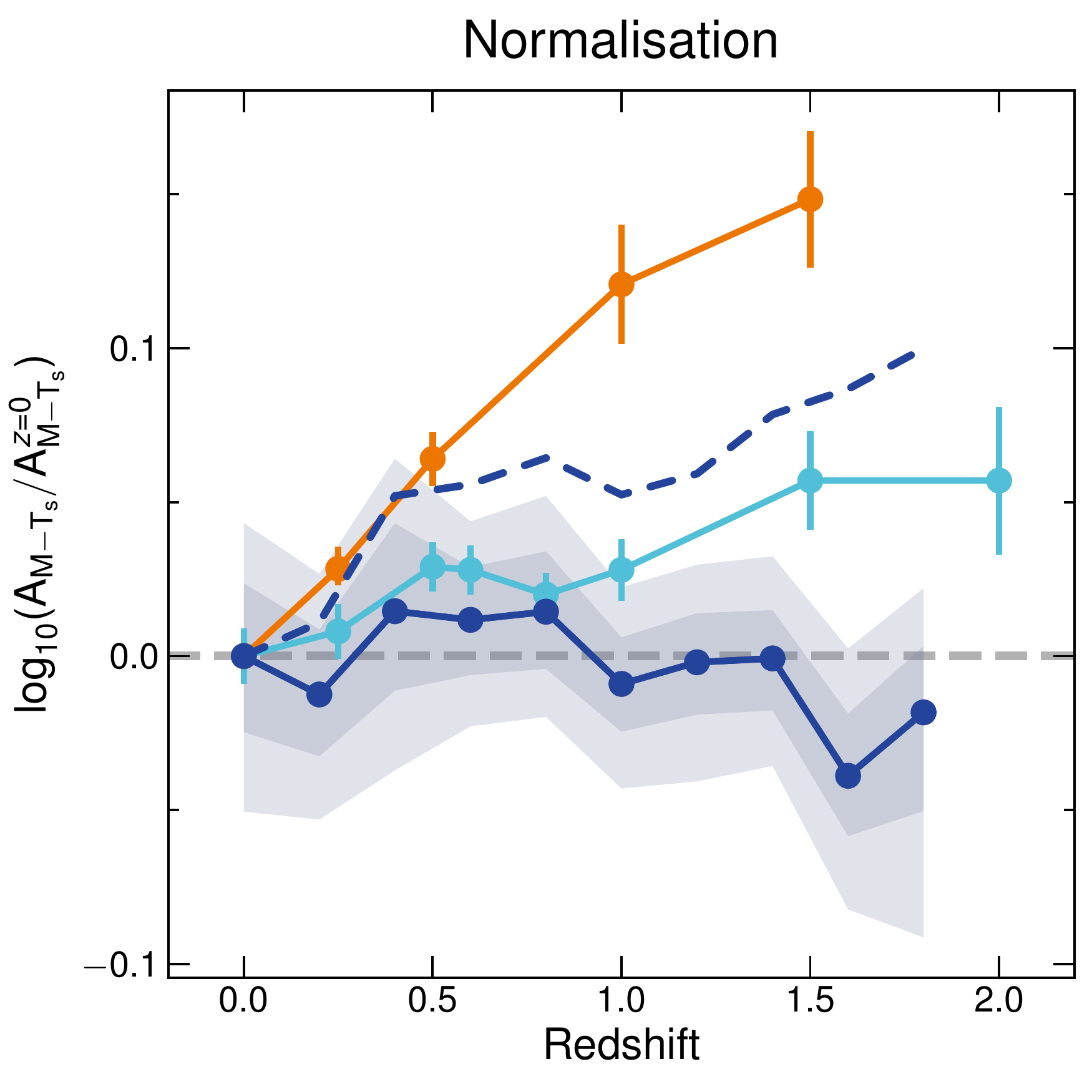}}
{\includegraphics[width=\factorthree\textwidth]{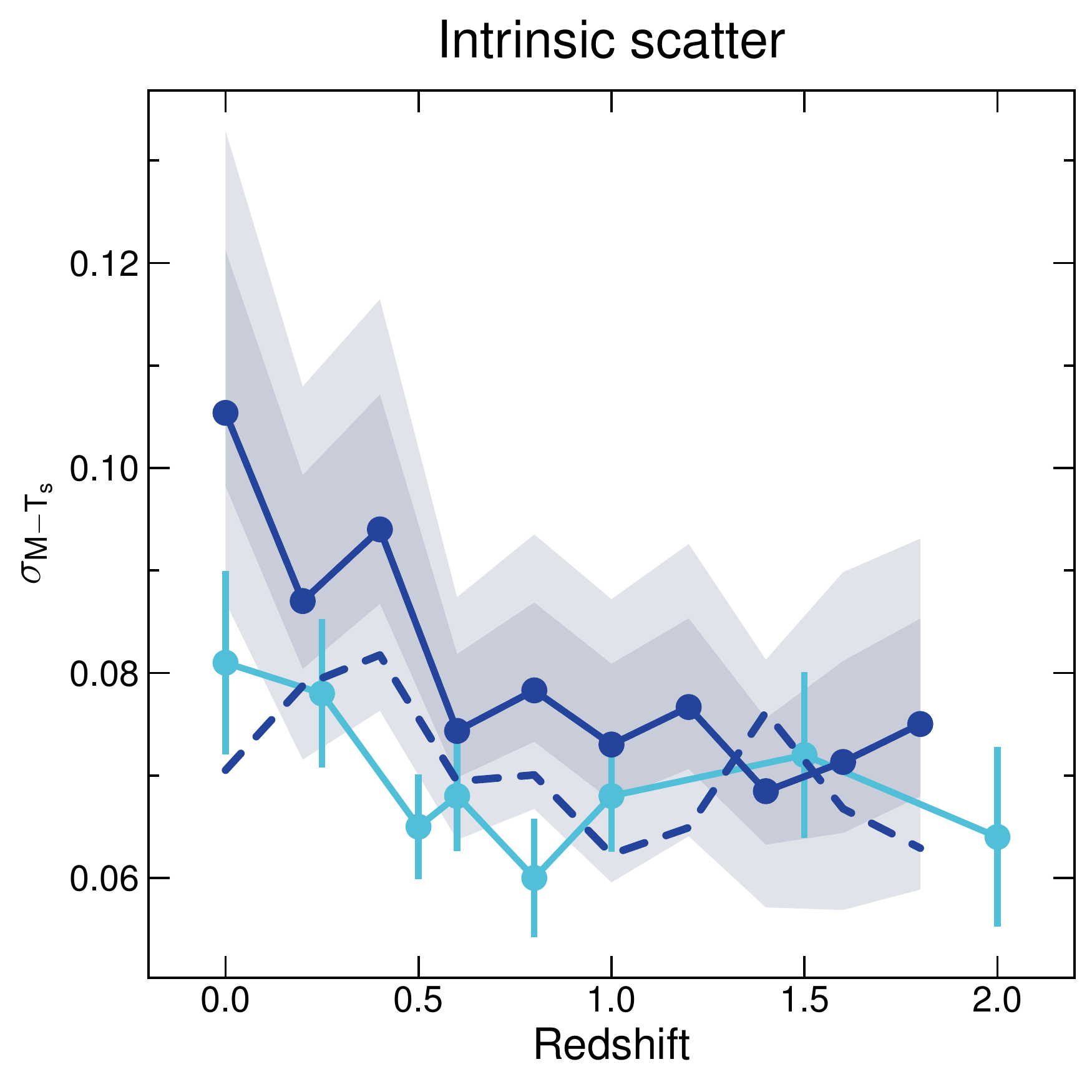}}
\end{center}
\caption{As Fig.~\ref{fig:Mgas-M} showing the redshift evolution of the total mass--temperature relation. The left, middle and right-hand panels show the slope, normalisation and intrinsic scatter, respectively. Blue solid and dashed lines correspond to the \fable\ relation calculated using the spectroscopic temperature and mass-weighted temperature, respectively. The intrinsic scatter in the \MACSIS\ relation is not shown as B17 compute their scatter at fixed mass rather than fixed temperature.
  }
    \label{fig:M-T}
\end{figure*}

In Fig.~\ref{fig:M-T} we plot the redshift evolution of the best-fitting parameters of the total mass--spectroscopic temperature relation. We also show with a dashed line the parameters of the total mass--temperature relation based on the mass-weighted temperature. We do not plot the uncertainties on these parameters but they are comparable to those of the spectroscopic temperature relation. The spectroscopic temperature is closer to that measured in X-ray observations, however it can be biased with respect to the mass-weighted temperature, which is a direct measure of the total thermal energy of the ICM.

\paragraph{Total mass--temperature slope}
The slope of the total mass--temperature relation based on the mass-weighted temperature (dashed line) is steeper than self-similar at all redshifts (i.e., $\beta_{\mathrm{M-T}} > 1.5$), in good agreement with \MACSIS\ and T18. The same departure from self-similarity was also found in the numerical studies of \cite{Stanek2010}, \cite{Fabjan2011}, \cite{Pike2014} and \cite{LeBrun2017} up to $z \sim 1$.

There are several ways in which the physical processes included in these simulations can affect the average temperature of the ICM, which in the self-similar scenario is determined solely by the depth of the gravitational potential well. For example, radiative cooling can cool the dense gas to form stars, thereby removing low entropy gas and raising the average temperature of the hot phase. This occurs with greater efficiency in lower mass systems, resulting in a tilt in the total mass--temperature relation.
Furthermore, AGN feedback can raise the average entropy of the ICM, particularly in the cluster core, leading to a higher average temperature. Indeed, comparing with our simulation that repeats the $(40 \, h^{-1} \, \mathrm{Mpc})^3$ periodic volume but without AGN we find on average a small increase ($\sim 0.05$ dex) in the spectroscopic and mass-weighted temperature when including AGN feedback. We lack enough massive haloes in this volume to constrain the slope of the total mass--temperature relation, however previous numerical studies have shown that AGN feedback plays a role in driving the slope away from the self-similar expectation (e.g. \citealt{Fabjan2011, Pike2014, LeBrun2017}).

The majority of observational studies also measure a steeper than self-similar slope, such as \cite{Arnaud2007} ($1.71 \pm 0.09$), \cite{Reichert2011} ($1.76 \pm 0.08$) and \cite{Lieu2016} ($1.67 \pm 0.12$). Conversely, some studies yield results consistent with self-similarity, such as \cite{Kettula2015} ($1.52^{+0.17}_{-0.16}$) and \cite{Mantz2016a} ($1.52 \pm 0.11$).
Observed cluster samples are typically limited to objects at or below $z \sim 1$ where the slope predicted by \MACSIS, T18 and the mass-weighted \fable\ relation agree on a roughly redshift-independent value of $\beta_{\mathrm{M-T}} \sim 1.7$, in good agreement with the majority of observational constraints.

The slope of the total mass--spectroscopic temperature relation (solid line) is significantly shallower than the mass-weighted temperature relation (dashed line) at all redshifts and is consistent with self-similarity. We have investigated this discrepancy and find that the spectroscopic temperature is biased slightly low compared with the mass-weighted temperature in galaxy groups ($T \lesssim 2$ keV) and biased somewhat high in massive clusters ($T \gtrsim 5$ keV).

At $z=0$ the spectroscopic temperature is approximately $0.1$ dex lower than the mass-weighted temperature at $\lesssim 2$ keV, as shown in fig. B4 of Paper~I and confirmed for our extended sample.
At these temperatures we find that the mock X-ray spectra are poorly fit by a single-temperature model, which is biased towards the lower temperature component(s) of the spectrum and tends to underestimate the X-ray continuum emission at high energies.
\cite{Mazzotta2004} find the same qualitative result in two-temperature thermal spectra for which the lower-temperature component is $\lesssim 2$ keV.
Indeed, we find that a two-temperature model is a significantly better fit to our mock spectra, as found by \cite{DePlaa2017} for a large sample of clusters, groups and elliptical galaxies observed with \textit{XMM-Newton}.
We have used a single-temperature fit for consistency with the majority of observational constraints but caution that this can underestimate the mass-weighted temperature in the galaxy group regime. This causes a significant tilt in our best-fitting total mass--spectroscopic temperature relation due to the high proportion of galaxy groups in our sample.
We find a similar level of bias for simulated \textit{Chandra} observations with realistic exposure times ($10$~ks), which suggests that current observational constraints may also be affected. Indeed, studies of the total mass--temperature relation in the galaxy group regime measure a slope slightly higher than, but statistically consistent with, our $z=0$ result ($1.56^{+0.06}_{-0.06}$), such as \cite{Sun2009} ($1.65 \pm 0.04$), \cite{Eckmiller2011} ($1.68 \pm 0.20$), \cite{Kettula2013a} ($1.48^{+0.13}_{-0.09}$) and \cite{Lovisari2015} ($1.65 \pm 0.07$), all of which derive the spectroscopic temperature from a single-temperature fit.
It is possible that the bias is caused by an excess of cool, X-ray emitting gas in our simulated galaxy groups, however this is difficult to constrain with observations. Our tests have shown that, if such gas is present in our simulations, it is neither gravitationally bound in substructures nor does it belong to the separated cooling phase of gas identified in Appendix B of Paper~I.

The bias at low temperatures can be avoided by limiting our sample to higher masses ($M_{500} \gtrsim 10^{14} M_{\odot}$). However, the mass--temperature slope remains low (see Appendix~\ref{A:mass_dependence}) due to an opposite spectroscopic temperature bias at the high mass end.
In Paper~I we found that the temperature and entropy profiles of \fable\ clusters show signs of over-heating in the central regions due to our relatively simple model for radio-mode AGN feedback. Because the density, and thus the X-ray emissivity, of the ICM increases towards the cluster centre, this causes the spectroscopic temperature to be biased high relative to the mass-weighted temperature.
Indeed, we find that excising the cluster core ($r < 0.15 \, r_{500}$) from the temperature computation largely removes the spectroscopic temperature bias at the high mass end.
If we also avoid the bias at the low mass end by restricting our sample to higher temperatures ($\gtrsim 2$ keV), we find that the slope of the total mass--spectroscopic temperature relation is in good agreement with the mass-weighted one.
We note that a more sophisticated model for AGN feedback may address the spectroscopic temperature bias at the high mass end by bringing the thermodynamic profiles of our simulated clusters into better agreement with observations in the central regions, as we discuss in more detail in Paper~I.

\fable\ and \MACSIS\ predict a roughly redshift-independent total mass--temperature slope within the uncertainties.
In contrast, T18 predict a mild decrease in the slope with redshift at $z \gtrsim 1$. They attribute this evolution to relatively cool gas at high redshift that has yet to thermalise in low-mass objects, whereas the most massive systems have been heated by strong shocks driven by minor and major mergers.
This could indicate more intense AGN activity at high redshift in \fable\ and \MACSIS\ compared with T18, which would raise the temperature of the gas preferentially in the lowest mass systems and cause a steepening of the total mass--temperature relation with increasing redshift. Indeed, T18 show this to be the case in their simulations with and without AGN feedback.

\paragraph{Total mass--temperature normalisation}
The middle panel of Fig.~\ref{fig:M-T} shows the normalisation of the total mass--temperature relation as a function of redshift. In \fable\ we find a mild positive evolution in the normalisation when using the mass-weighted temperature (dashed line). This implies that objects of a given mass at high redshift are cooler than expected from the self-similar model.
In contrast, the evolution of the spectroscopic temperature-based relation is consistent with the self-similar prediction. The difference between this and the mass-weighted temperature relation is due to redshifting of low-energy X-ray emission below the X-ray bandpass ($0.2$--$10$ keV), which causes the spectroscopic temperature to be biased high. We have confirmed this effect by fitting our mock X-ray spectra in the rest frame of the source, in which case the evolution of the normalisation is almost identical to the mass-weighted case.

All three simulations predict a positive evolution in the normalisation, although it is somewhat stronger in \MACSIS\ (e.g. an increase of $\sim 35$ per cent from $z=0$ to $z = 1.5$ compared with an increase of $\sim 15$ per cent for T18 and the mass-weighted temperature relation of \fable). The difference may be driven by the slight redshift dependence of the spectroscopic temperature bias in \MACSIS\ clusters, as demonstrated in fig.~7 of B17. They attribute this temperature bias to relatively cool X-ray emitting gas in the outskirts of massive clusters, although it is unclear whether this gas has a physical origin or is an unphysical artefact of the hydrodynamics scheme \citep{Henson2017}.
Interestingly, \cite{LeBrun2017} also find a positive evolution in the normalisation even in a simulation without radiative cooling, star formation or feedback. This implies that the redshift evolution is driven by the merger history of clusters rather than non-gravitational physics. Indeed, \cite{LeBrun2017} find that the ratio of the total kinetic energy of the ICM to the total thermal energy strongly increases with increasing redshift for a given halo mass due to the increasing importance of mergers and associated lack of thermalisation. We have confirmed the same redshift trend of the kinetic-to-thermal energy ratio in \fable\ (not shown), as do T18. This implies that clusters of a given mass possess greater non-thermal pressure support from bulk motions and turbulence at higher redshift, resulting in a lower temperature required for virial equilibrium.

The fact that simulation studies find a positive evolution in the total mass--temperature normalisation regardless of the precise physical modelling implies that this prediction is fairly robust. Yet this appears to be in mild tension with the results of \cite{Reichert2011} who use observational data from various literature sources to show that the evolution of the total mass--temperature relation is consistent with the self-similar prediction out to $z \sim 1.4$.
This is similar to our spectroscopic temperature relation, which suggests that part of the discrepancy may be the result of a redshift-dependent spectroscopic temperature bias similar to that found in our mock X-ray analysis.
Further observational constraints on the redshift evolution of the total mass--temperature normalisation are required with which to compare the simulation predictions however, particularly as the \cite{Reichert2011} result may be adversely affected by sample selection biases, which are of particular concern in imhomogeneous datasets drawn from multiple sources.

The \textit{Athena} X-ray observatory will provide an excellent opportunity to constrain the biases in previous analyses and to test the simulation predictions \citep{Nandra2013, Barcons2017}.
Indeed, our mock \textit{Athena} X-IFU observations suggest that \textit{Athena}, with careful consideration of a possibly redshift-dependent spectroscopic temperature bias, should observe a significant positive evolution in the total mass--temperature relation out to $z \sim 2$. The size of this evolution (or potentially its sign) will place constraints on the non-gravitational physics important in galaxy cluster formation and provide an opportunity to distinguish between different physical models.

\paragraph{Total mass--temperature intrinsic scatter}
The intrinsic scatter about the total mass--temperature relation at $z=0$ is $0.10^{+0.02}_{-0.01}$ and $0.07^{+0.01}_{-0.01}$ dex for the spectroscopic and mass-weighted temperature relations, respectively. These values are consistent with the observational studies of \cite{Eckmiller2011} ($0.117$), \cite{Kettula2013a} ($0.12 \pm 0.03$) and \cite{Kettula2015} ($0.07 \pm 0.04$) but smaller than \cite{Lieu2016} ($0.18 \pm 0.03$) and larger than the relaxed cluster samples of \cite{Arnaud2007} ($0.064$) and \cite{Mantz2016a} ($0.058 \pm 0.008$).
The T18 results are consistent with ours within the uncertainties. The intrinsic scatter at fixed mass is also in good agreement with \MACSIS\ (not shown).
Interestingly, we find that excising the core from the temperature computation and/or using a spherical rather than a projected aperture has a negligible effect on the intrinsic scatter.

\subsubsection{$Y_{\mathrm{X}}$--total mass scaling relation}\label{subsubsec:YX-M}

\begin{figure*}
\begin{center}
{\includegraphics[width=\factorthree\textwidth]{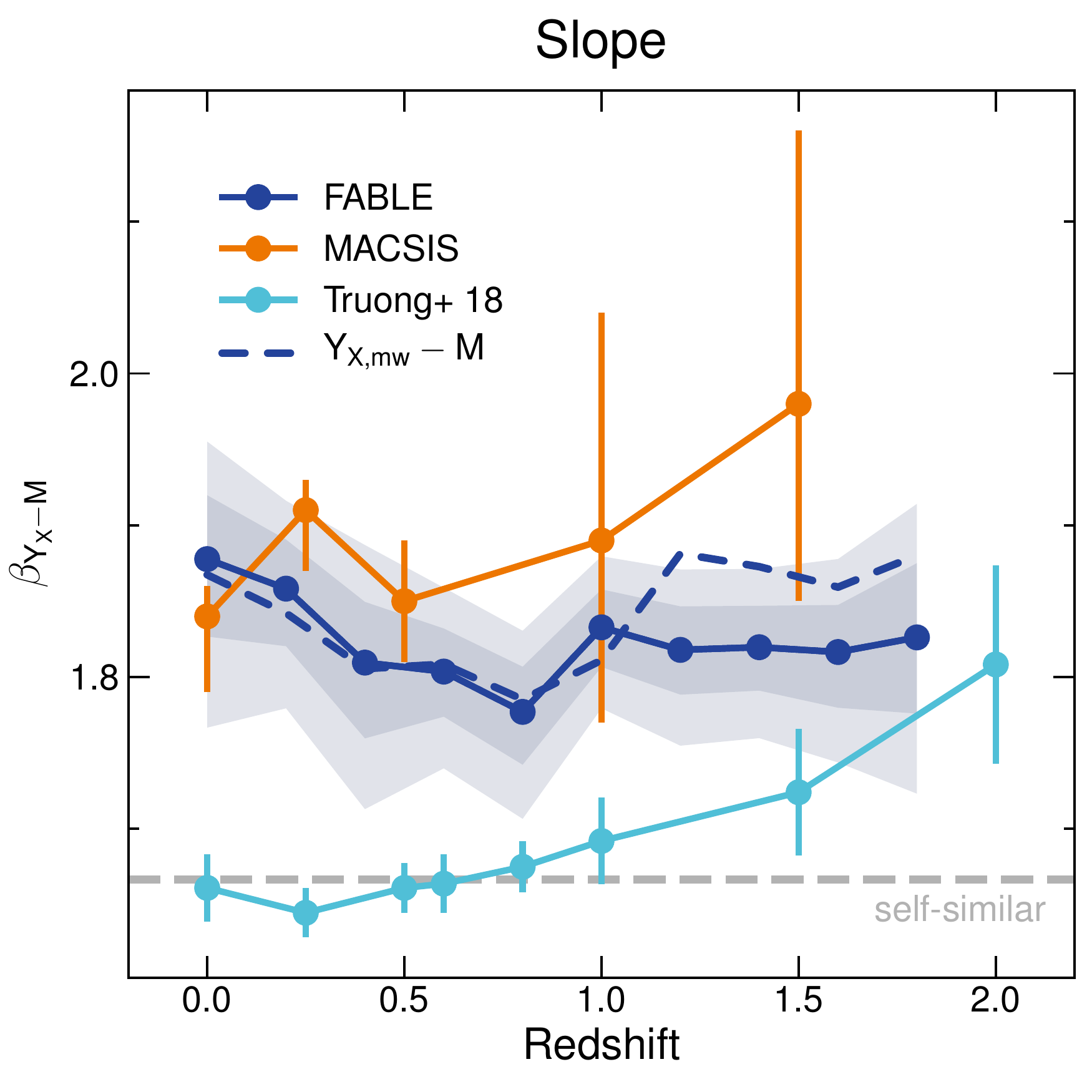}}
{\includegraphics[width=\factorthree\textwidth]{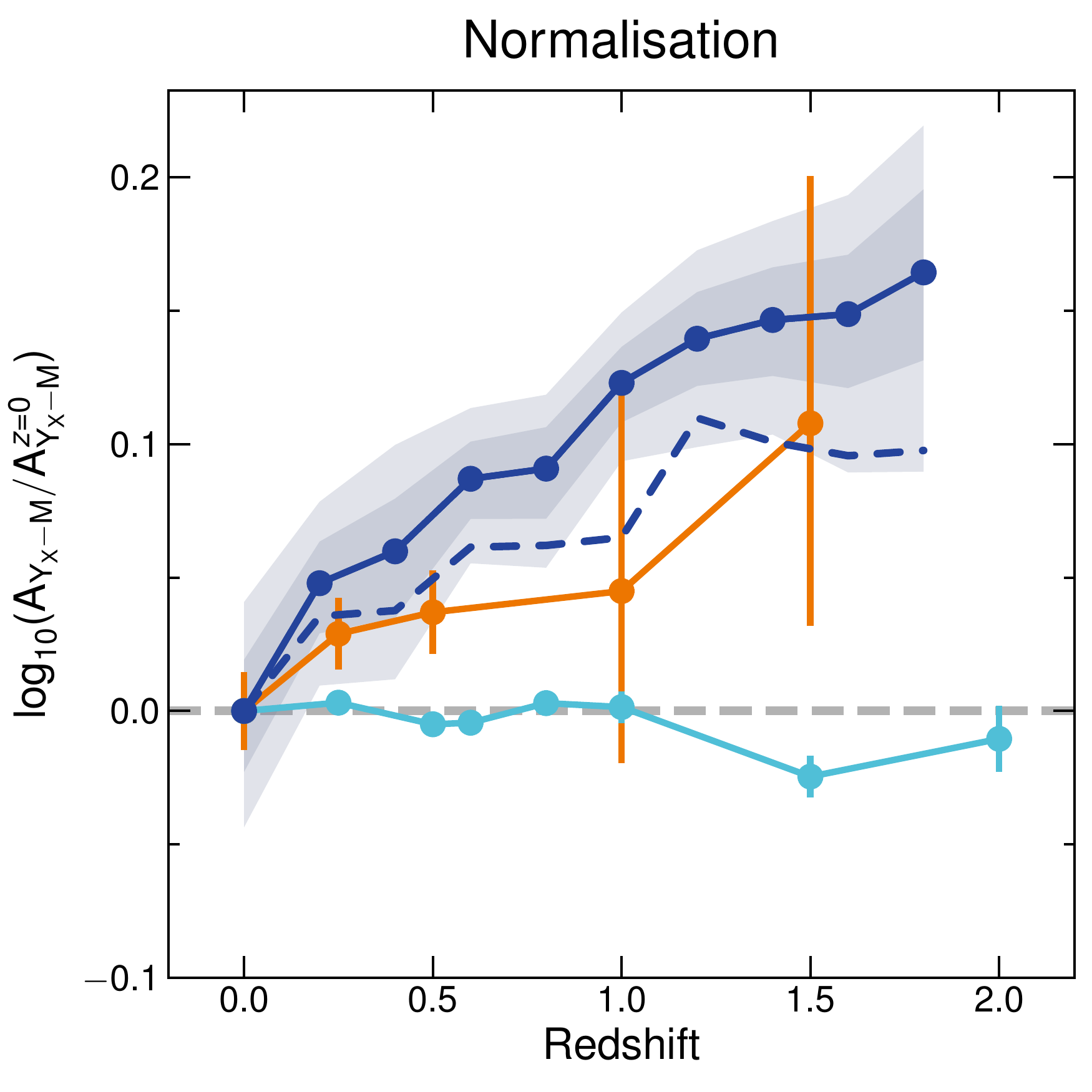}}
{\includegraphics[width=\factorthree\textwidth]{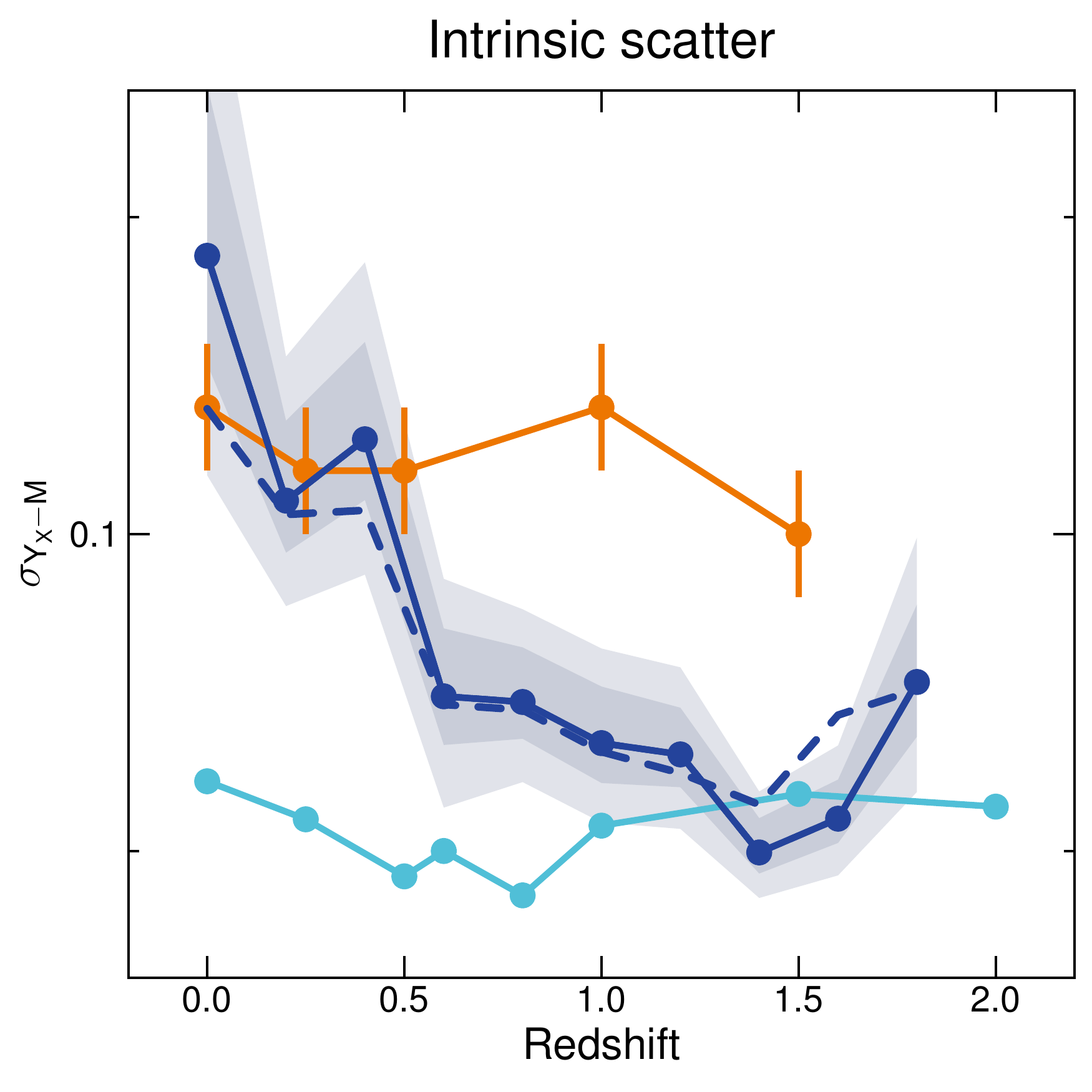}} \\
{\includegraphics[width=\factorthree\textwidth]{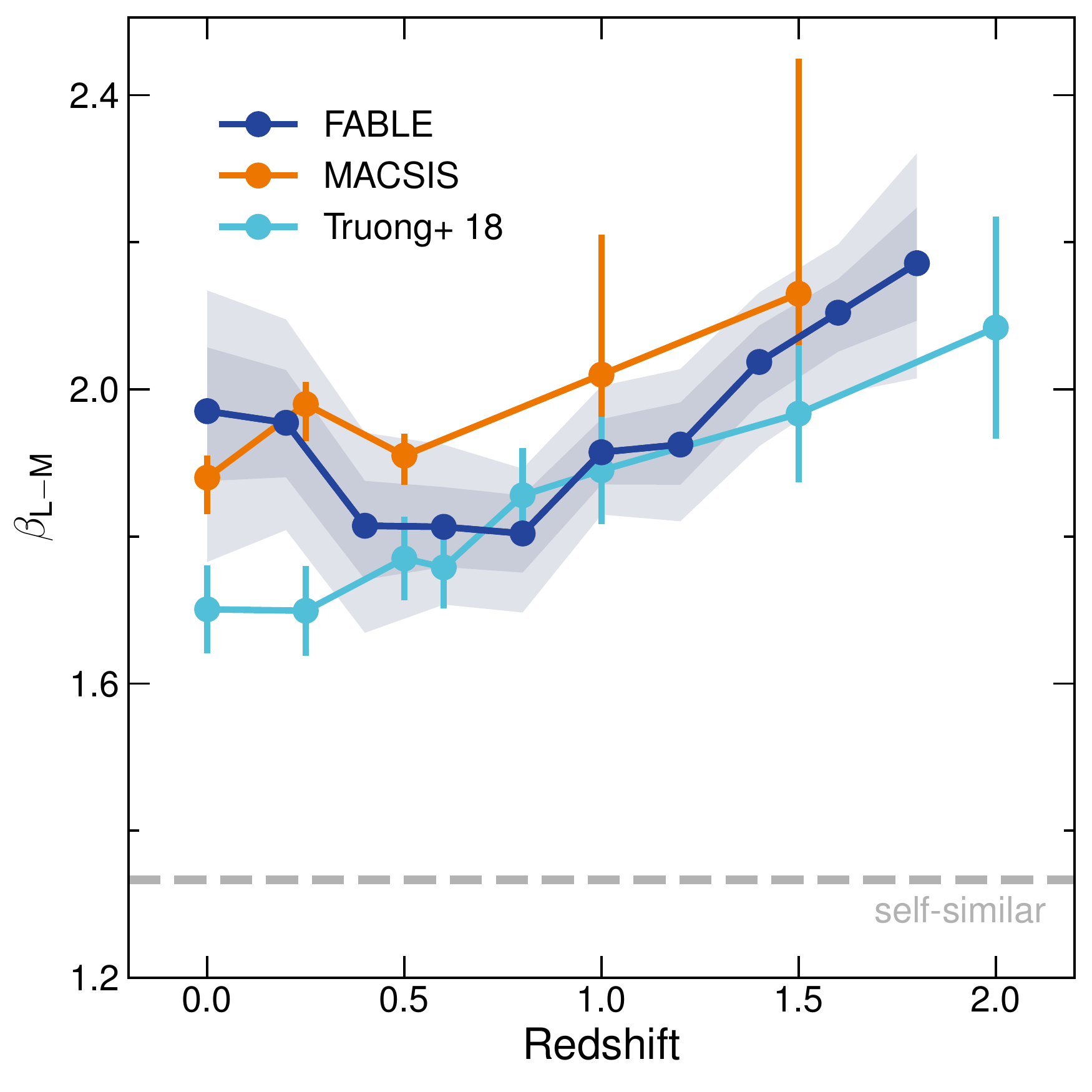}}
{\includegraphics[width=\factorthree\textwidth]{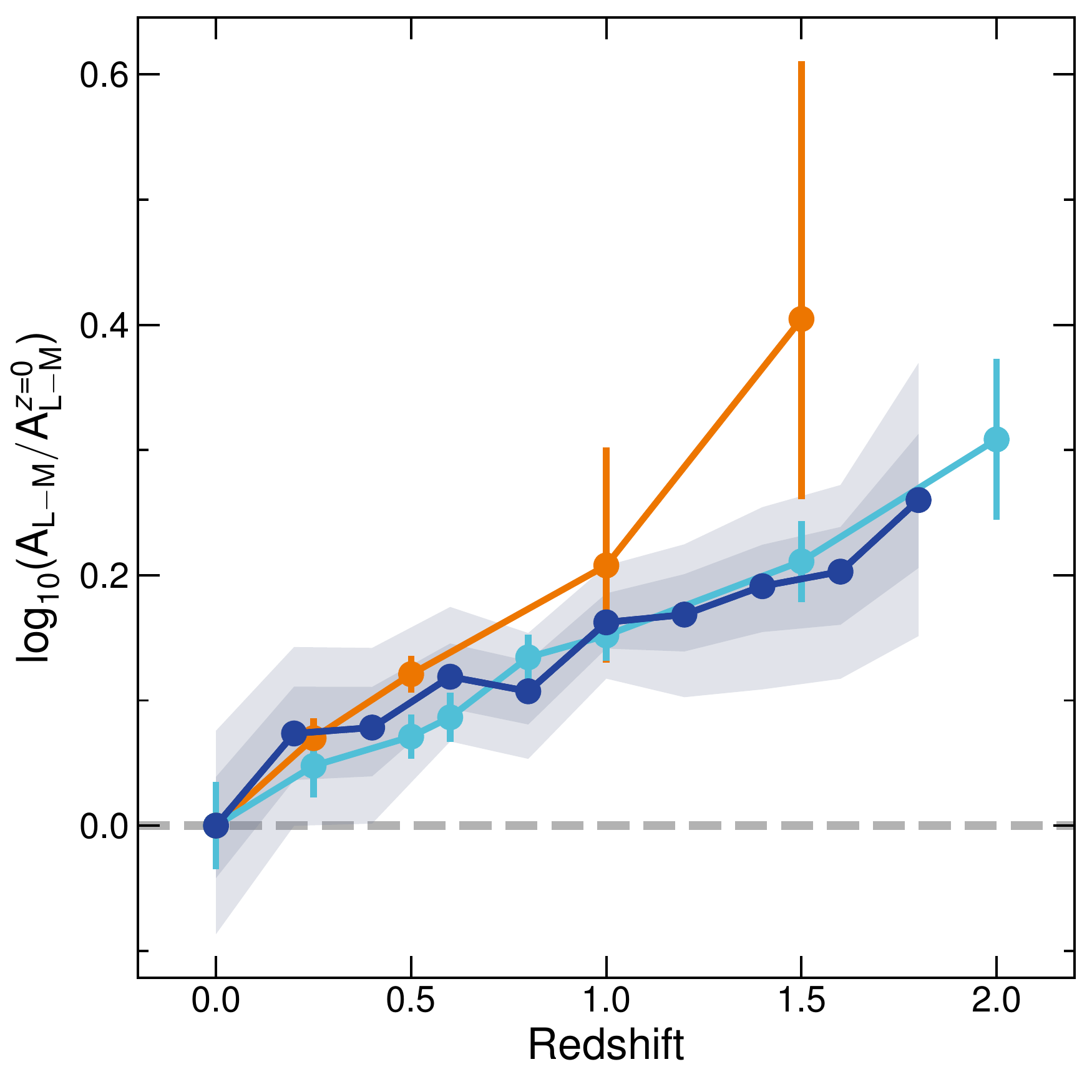}}
{\includegraphics[width=\factorthree\textwidth]{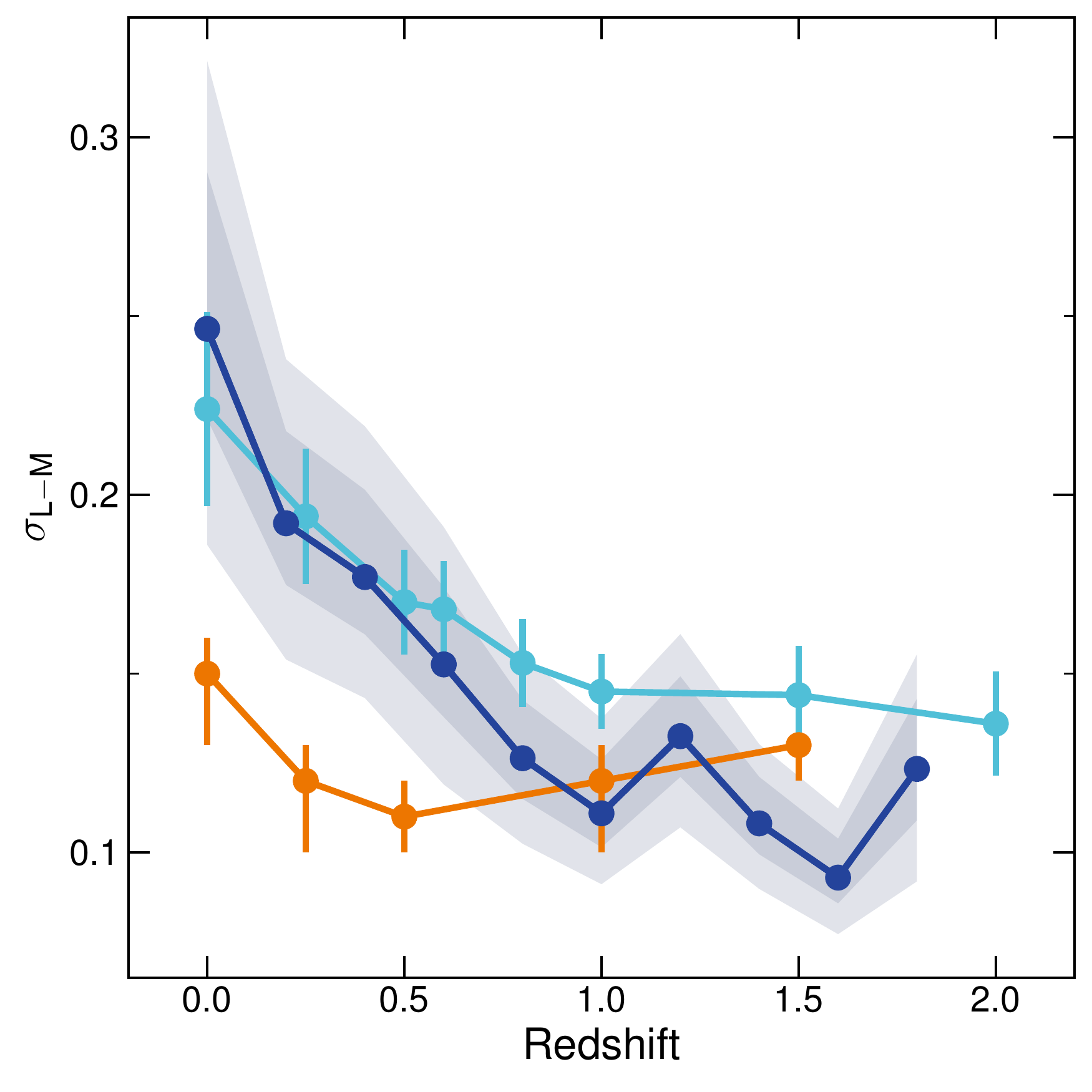}} \\
{\includegraphics[width=\factorthree\textwidth]{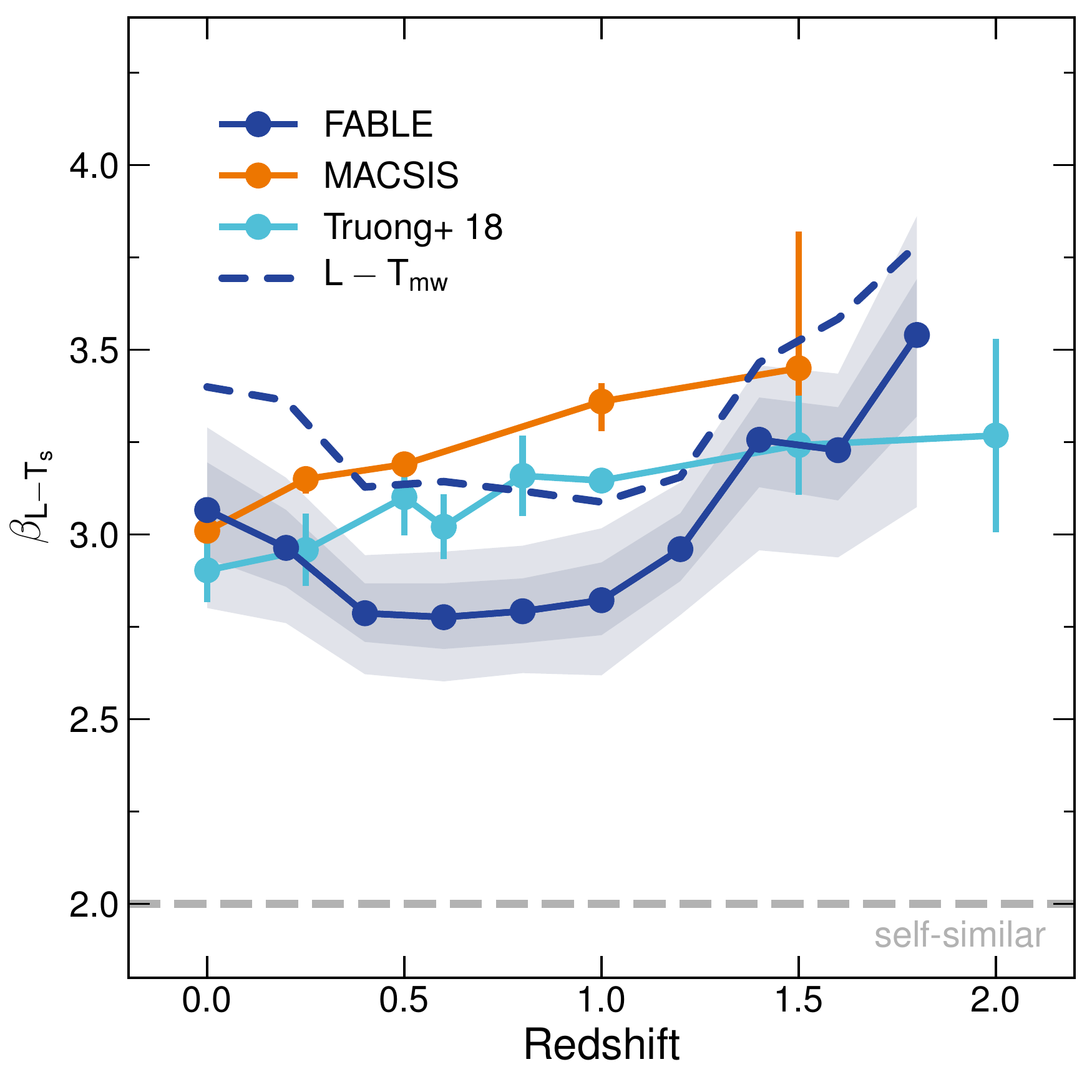}}
{\includegraphics[width=\factorthree\textwidth]{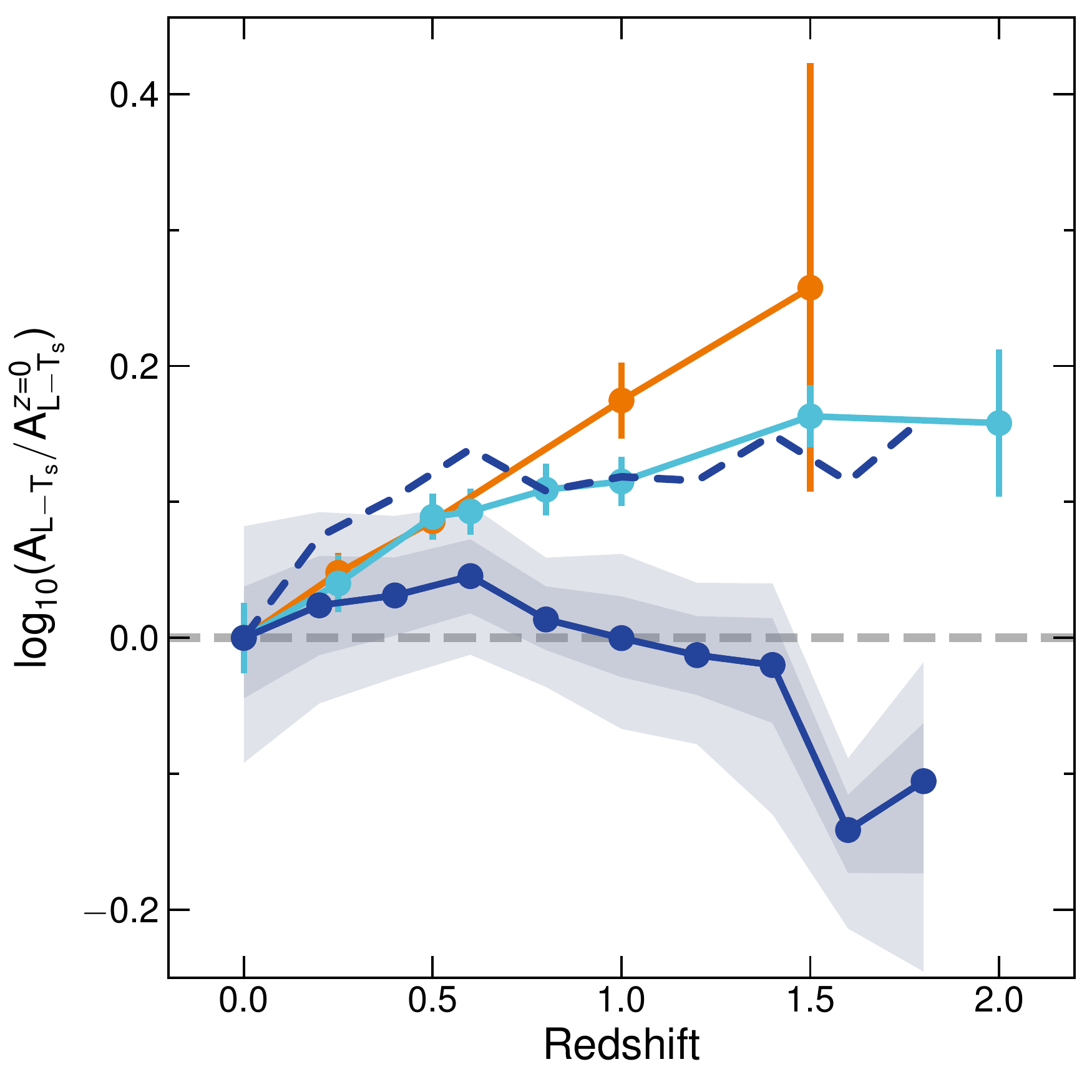}}
{\includegraphics[width=\factorthree\textwidth]{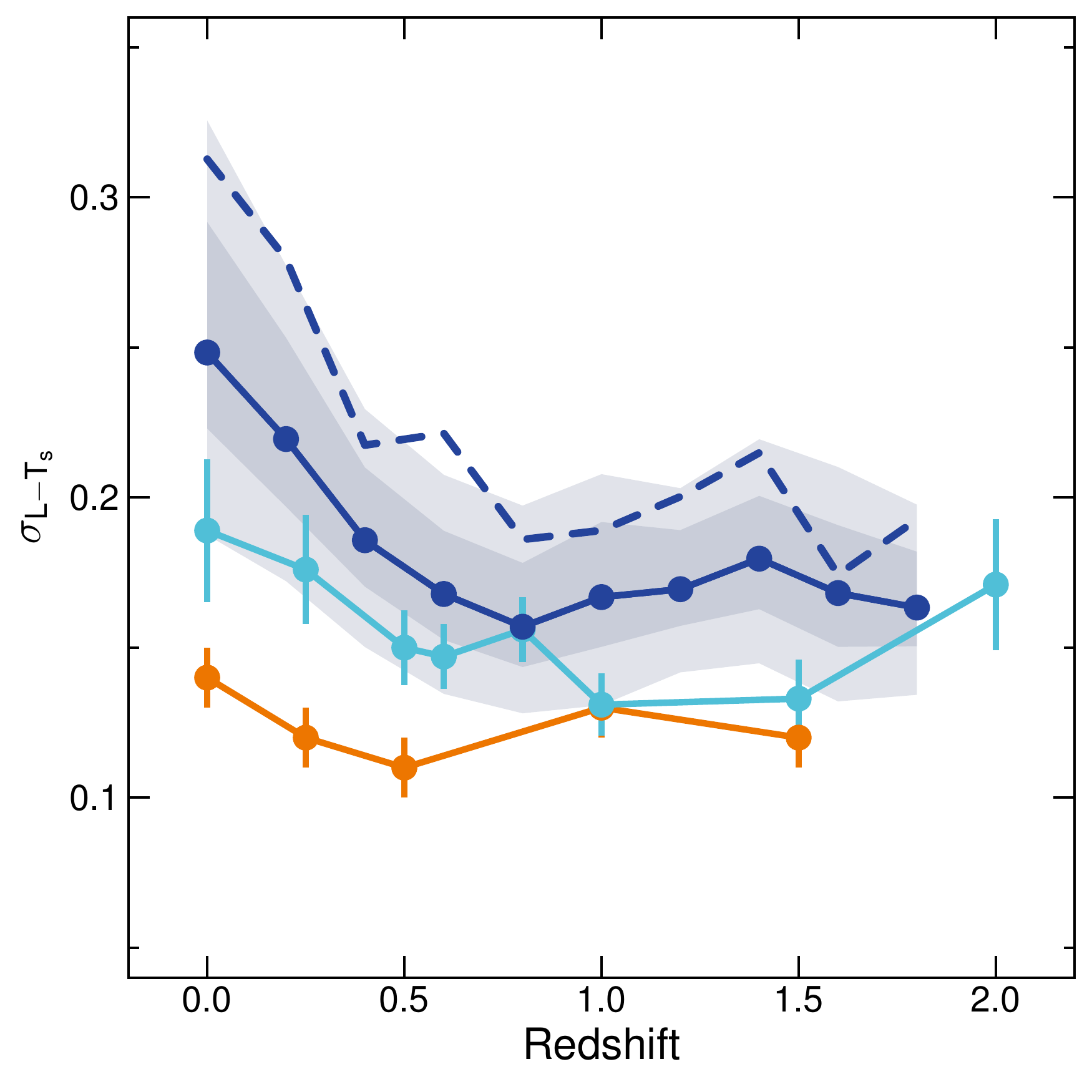}} \\
\end{center}
\caption{As Fig.~\ref{fig:Mgas-M} and \ref{fig:M-T} for the redshift evolution of the $Y_{\mathrm{X}}$--total mass relation (top row), the X-ray luminosity--total mass relation (middle row) and the X-ray luminosity--temperature relation (bottom row). Blue solid and dashed lines correspond to the \fable\ relations based on the spectroscopic or mass-weighted temperature, respectively.
  The uncertainties on the intrinsic scatter of the $Y_{\mathrm{X}}$--total mass relation are unknown for T18.
  }
    \label{fig:evol}
\end{figure*}

The top row of Fig.~\ref{fig:evol} shows the redshift evolution of the best-fitting $Y_{\mathrm{X}}$--total mass relation, where $Y_{\mathrm{X}}$ is defined as the product of the gas mass within $r_{500}$ and the average temperature within the core-excised aperture $(0.15$--$1) \, r_{500}$.
We calculate $Y_{\mathrm{X}}$ using either the spectroscopic temperature (solid line) or the mass-weighted temperature (dashed line).

\paragraph{$Y_{\mathrm{X}}$--total mass slope}
At low redshift the $Y_{\mathrm{X}}$--total mass relation is significantly steeper than the self-similar expectation ($\beta_{Y_{\mathrm{X}}-M} = 5/3$), as has been found in several previous simulation studies (e.g. \citealt{Short2010, Stanek2010, Planelles2014, LeBrun2017}). For example, at $z=0$ we find a slope of $1.88^{+0.04}_{-0.05}$ for the spectroscopic temperature-based relation, which is consistent with the \MACSIS\ result ($1.84^{+0.02}_{-0.05}$) as well as the observational constraints of \cite{Arnaud2007} ($1.82 \pm 0.09$), \cite{Eckmiller2011} ($1.82 \pm 0.07$) and \cite{Mahdavi2013} ($1.79 \pm 0.22$).
T18 on the other hand find a self-similar slope ($1.66 \pm 0.02$ at $z=0$), consistent with the simulations of \cite{Fabjan2011} and \cite{Biffi2014} and the observational findings of \cite{Lovisari2015} ($1.67 \pm 0.08$) and \cite{Mantz2016} ($1.63 \pm 0.04$).

The slope of the $Y_{\mathrm{X}}$--total mass relation is approximately equal to the product of the gas mass--total mass slope and the inverse of the total mass--temperature slope. Both components deviate from their self-similar values but in opposite directions. This is one of the motivating reasons for using the $Y_{\mathrm{X}}$ variable as a mass proxy \citep{Kravtsov2006}. For example, in T18 the steeper than self-similar gas mass--total mass and total mass--temperature slopes cancel, yielding an approximately self-similar $Y_{\mathrm{X}}$--total mass relation at $z \lesssim 1$.
In contrast, \fable\ and \MACSIS\ predict a steeper than self-similar $Y_{\mathrm{X}}$--total mass relation at all redshifts because the gas mass--total mass relation deviates from self-similarity to a greater degree than the total mass--temperature relation. Physically, non-gravitational processes such as star formation and AGN feedback remove gas from the ICM and raise the average temperature of the hot phase (by removing cold gas via star formation and raising the entropy of the gas via thermal energy injection by AGN).
The \fable\ and \MACSIS\ results, as well as observations that measure a steeper than self-similar slope, imply that the removal of gas steepens the $Y_{\mathrm{X}}$--total mass relation to a greater degree than the corresponding increase in the temperature of the remaining gas.

The slope is approximately independent of redshift within the uncertainties, in agreement with \MACSIS. Similarly, T18 find a slope that is constant up to $z \sim 1$ although there is an increase in slope toward higher redshifts that reflects the increase in slope of their gas mass--total mass relation (due to the shrinking mass range of their sample) and the decrease in slope of their total mass--temperature relation (due to incomplete thermalisation of gas in low-mass objects), as discussed in the previous two sections.

\paragraph{$Y_{\mathrm{X}}$--total mass normalisation}
A number of previous simulation studies have found that the normalisation of the $Y_{\mathrm{X}}-M_{500}$ relation evolves in a self-similar manner (e.g. \citealt{Kravtsov2006, Nagai2007c, Short2010, Fabjan2011}), in agreement with the T18 result.
In contrast, \fable\ and \MACSIS\ predict a positive, albeit mild, evolution with respect to self-similarity. This reflects the positive evolution of the gas mass--total mass normalisation that is somewhat, but not completely, offset by the mild positive evolution in the total mass--temperature relation.
\cite{LeBrun2017} also find a mild positive evolution in their low-mass sample, although for high-mass haloes ($M_{500} > 10^{14} M_{\odot}$) the situation is reversed.
We point out that beyond self-similar evolution in the $Y_{\mathrm{X}}-M_{500}$ relation would have important implications for the use of $Y_{\mathrm{X}}$ as a mass proxy, as many observational studies assume self-similar evolution when estimating total masses from this relation (e.g. \citealt{Maughan2008, Maughan2012, Bartalucci2017c}).

\paragraph{$Y_{\mathrm{X}}$--total mass intrinsic scatter}
Most previous simulation works predict a low level of intrinsic scatter in the $Y_{\mathrm{X}}$--total mass relation at $z=0$, such as \cite{Short2010} ($0.05$), \cite{Stanek2010} ($0.05$), \cite{Fabjan2011} ($0.05$) and \cite{Planelles2014} ($0.08$).
In fact, many observational studies find that it is dominated by the statistical scatter (e.g. \citealt{Sun2009, Vikhlinin2009a, Lovisari2015}).
Observations that are able to constrain the intrinsic scatter vary significantly in their measurements, for example, \cite{Arnaud2007} ($0.04$), \cite{Mahdavi2013} ($0.10 \pm 0.02$), \cite{Mantz2016} ($0.080 \pm 0.007$) and \cite{Eckmiller2011} ($0.264$).

The intrinsic scatter in \fable\ at $z=0$ is $0.14^{+0.02}_{-0.02}$ and $0.12^{+0.03}_{-0.02}$ for the spectroscopic and mass-weighted temperature-based relations respectively. These values are consistent with the \MACSIS\ value ($0.12 \pm 0.01$) but larger than most other theoretical and observational constraints, with the exception of \cite{Eckmiller2011} who measure an intrinsic scatter of $0.350$ and $0.218$ dex for their group and cluster samples, respectively, and $0.264$ dex for their combined sample.
Like \cite{Eckmiller2011} we also find a slight mass dependence in the scatter, although it is stronger for the mass-weighted temperature than the spectroscopic one. For example, when including only cluster-scale haloes with $M_{500} > 10^{14} M_{\odot}$ in the $z=0$ sample we find a smaller intrinsic scatter of $0.08^{+0.03}_{-0.01}$ for the mass-weighted temperature relation but a similar scatter for the spectroscopic temperature relation ($0.13^{+0.04}_{-0.02}$). This is because the latter has increased intrinsic scatter at the high mass end due to scatter in the spectroscopic temperature bias in massive clusters.

Unlike \MACSIS\ and T18, which predict a roughly constant scatter with redshift, the intrinsic scatter decreases from $z=0$ to $z \approx 1$. This largely reflects the redshift trend of the intrinsic scatter in the gas mass--total mass relation (Fig.~\ref{fig:Mgas-M}) as discussed in Section~\ref{subsubsec:Mgas-M}.

\subsubsection{X-ray luminosity--total mass scaling relation}\label{subsubsec:L-M}

The best-fitting parameters of the bolometric X-ray luminosity--total mass relation are shown in the middle row of Fig.~\ref{fig:evol}.

\paragraph{X-ray luminosity--total mass slope}
The bolometric X-ray luminosity of the ICM largely depends on the total mass of gas and its distribution, with line emission becoming dominant only in low-temperature ($T \lesssim 2$ keV) systems. As such, the evolution of the X-ray luminosity--total mass relation shares many similarities with that of the gas mass--total mass relation.
Indeed, the slopes of both relations are significantly steeper than self-similar at all redshifts in \fable, \MACSIS\ and T18.
At $z=0$ we find a slope of $1.97^{+0.10}_{-0.08}$, consistent with the \MACSIS\ prediction ($1.88^{+0.03}_{-0.05}$) but slightly steeper than T18 ($1.70 \pm 0.06$).
Observational studies tend to agree on a steeper than self-similar slope, although the exact value can vary. Most observational constraints are in good agreement with the \fable\ and \MACSIS\ predictions, for example, \cite{Maughan2007} ($1.96 \pm 0.10$), \cite{Pratt2009} ($1.96 \pm 0.11$ or $2.08 \pm 0.13$ after correcting for Malmquist bias) and \cite{Giles2017} ($2.22 \pm 0.24$).

We point out that our use of a projected aperture can cause the X-ray luminosity to be boosted by hot gas along the line of sight, largely from overlapping haloes.
In the general field, significant overlap between haloes is fairly rare. However, for the secondary haloes in our zoom-in simulations this occurs more frequently and has a non-negligible effect on the X-ray luminosity--total mass slope.
Indeed, the increase in slope found by switching from a projected to a spherical aperture of the same radius is small compared with the uncertainties (e.g. from $1.97^{+0.10}_{-0.08}$ to $2.02^{+0.11}_{-0.09}$ at $z=0$) but is systematic across all redshift bins.
In addition, excluding the core ($< 0.15 \, r_{500}$) from the spherically integrated X-ray luminosity leads to overall slightly shallower slopes, as was also found in T18. Coincidentally, this change is approximately equal and opposite to the effect of projection so that the slope and its evolution are unchanged by switching from a projected aperture to a spherical, core-excised aperture as used for \MACSIS\ and T18.

The slope of the X-ray luminosity--total mass relation shows some redshift dependence. At $z \lesssim 1$ this reflects the change in slope of the gas mass--total mass relation (Section~\ref{subsubsec:Mgas-M}).
At $z \gtrsim 1$ the slope of the X-ray luminosity--total mass relation increases with increasing redshift similarly to \MACSIS. Like B17 we attribute this redshift evolution to the combined redshift trends of the gas mass--total mass and total mass--temperature slopes, which are negligible on their own but combine into a mild redshift evolution in the X-ray luminosity--total mass slope. The \fable\ and \MACSIS\ total mass--temperature slopes show opposing redshift trends at $z \gtrsim 1$, however, the \fable\ sample is significantly less massive than \MACSIS, particularly at high redshift. This modifies the temperature dependence of the X-ray luminosity due to the increasing contribution of line emission at lower temperatures (e.g. \citealt{Maughan2013}). Indeed, with a more massive sample of haloes with $M_{500} > 10^{14} E(z)^{0.5} M_{\odot}$ we find no significant redshift evolution in the X-ray luminosity--total mass slope.

\paragraph{X-ray luminosity--total mass normalisation}
The normalisation of the X-ray luminosity--total mass relation has a positive redshift evolution, being $\sim 60$ per cent higher at $z=1.5$ than at $z=0$ relative to self-similarity. \MACSIS\ and T18 predict a very similar increase in normalisation within the uncertainties. This largely reflects the evolution in the gas mass at fixed total mass (see Section~\ref{subsubsec:Mgas-M}).

When measuring the X-ray luminosity within a spherical, core-excised aperture, the normalisation evolves slightly faster (e.g. $\sim 75$ per cent between $z=0$ and $z=1.5$). The difference is almost entirely due to projection effects rather than excision of the cluster core. At low redshift, the X-ray luminosity at fixed total mass is biased high by hot gas along the line of sight that lies predominantly in hotter, more massive objects. At higher redshift, the projection bias at fixed mass is smaller because there are fewer objects above that mass. Hence, the normalisation evolves less rapidly when the luminosity is projected rather than spherically integrated. We note, however, that we are likely overestimating the effect of projection since we do not perform background subtraction in our X-ray analysis as would be carried out in real observations.

We caution that, because the slope of the X-ray luminosity--total mass relation varies substantially with redshift, the evolution of the normalisation is sensitive to the choice of pivot point. In particular, choosing a larger (smaller) mass for the pivot point makes the apparent evolution in the normalisation more (less) positive. For example, at pivot points of $10^{14} M_{\odot}$ and $5 \times 10^{14} M_{\odot}$ keV the increase in normalisation between $z=0$ and $z=1.5$ relative to self-similarity is $\sim 45$ and $\sim 70$ per cent, respectively. A similar change is found for the other simulation relations as they evolve similarly in slope.

\paragraph{X-ray luminosity--total mass intrinsic scatter}
The intrinsic scatter of the X-ray luminosity--total mass relation is $0.25^{+0.04}_{-0.03}$ at $z=0$, which agrees with the T18 value ($0.22 \pm 0.03$) but is somewhat higher than \MACSIS\ ($0.15^{+0.01}_{-0.02}$).
Observational constraints on the intrinsic scatter span a similar range of values, for example \cite{Sun2012} and \cite{Giles2017} measure $0.25 \pm 0.05$ and $0.30 \pm 0.05$, respectively, while \cite{Maughan2007}, \cite{Pratt2009} and \cite{Mantz2010a} derive smaller values of $0.17 \pm 0.02$, $0.166 \pm 0.026$ and $0.185 \pm 0.019$, respectively.

Much of the variation between studies can be attributed to sample selection. For example, lower mass haloes exhibit somewhat larger intrinsic scatter so that restricting our sample to haloes with $M_{500} > 6 \times 10^{14} E(z)^{-0.5} M_{\odot}$ slightly reduces the scatter at all redshifts (e.g. from $0.25^{+0.04}_{-0.03}$ to $0.21^{+0.04}_{-0.02}$ at $z=0$).
Furthermore, several studies have found that the scatter in X-ray luminosity at fixed mass or temperature is dominated by the cluster core regions, particularly for mixed samples of relaxed, cool core clusters and unrelaxed, morphologically disturbed ones (e.g. \citealt{Markevitch1998, Maughan2007, Pratt2009, Maughan2012}). Indeed, when excising the core from the projected aperture we find that the scatter at $z=0$ decreases slightly from $0.25^{+0.04}_{-0.03}$ to $0.22^{+0.04}_{-0.03}$, with a similar decrease at higher redshifts. This is a smaller effect than is typically found in observations however, which may be due to a lack of strong cool core clusters in our sample.

At $z \lesssim 0.5$ all three simulations predict a drop in the intrinsic scatter with increasing redshift. This is in qualitative agreement with \cite{Mantz2016} who find tentative evidence for evolution in the intrinsic scatter of X-ray luminosity at fixed mass for a large sample of clusters at $z \lesssim 0.5$.
At $0.5 \lesssim z \lesssim 1$ the intrinsic scatter in \fable\ and T18 continues to fall with increasing redshift. This is consistent with \cite{Maughan2007} who measure a significantly larger intrinsic scatter for a subset of clusters at $0.1 < z < 0.5$ compared with a subset at $0.5 < z < 1.3$ (approximately a factor of two increase).
At $z \gtrsim 1$ the simulations converge on a roughly redshift-independent scatter of $0.10$--$0.15$ dex.

\subsubsection{X-ray luminosity--temperature scaling relation}\label{subsubsec:L-T}

The bottom row of Fig.~\ref{fig:evol} shows the best-fitting parameters of the X-ray luminosity--temperature relation. Solid and dashed lines correspond to the relation based on the spectroscopic and mass-weighted temperature, respectively.

\paragraph{X-ray luminosity--temperature slope}
The slope of the X-ray luminosity--temperature relation is steeper than the self-similar expectation ($\beta_{L-T} = 2$) at all redshifts in the \fable, \MACSIS\ and T18 simulations. At $z=0$ the slope of the \fable\ relation is $3.07^{+0.11}_{-0.15}$ and $3.40^{+0.19}_{-0.20}$ for the spectroscopic and mass-weighted temperature-based relations, respectively.
These values are slightly higher than T18 ($2.903 \pm 0.086$) but statistically consistent with \MACSIS\ ($3.07 \pm 0.04$) and the majority of observational constraints, for example, \cite{Pratt2009} ($3.35 \pm 0.32$), \cite{Hilton2012} ($3.04 \pm 0.16$), \cite{Sun2012} ($3.03 \pm 0.01$), \cite{Giles2016} ($3.08 \pm 0.15$) and \cite{Zou2016} ($3.28 \pm 0.33$).

The slope increases slightly when measuring the luminosity and temperature within a spherical rather than a projected aperture (from $3.07^{+0.11}_{-0.15}$ to $3.23^{+0.15}_{-0.15}$ at $z=0$ and similarly at higher redshift). This is because gas that overlaps with the cluster in projection causes the luminosity and spectroscopic temperature to be biased high and low, respectively, particularly in low-mass objects.
Excising the cluster core has a negligible effect on the X-ray luminosity--temperature slope as the X-ray luminosity--total mass and total mass--temperature slopes show opposing behaviour.

\paragraph{X-ray luminosity--temperature normalisation}
The X-ray luminosity at fixed mass-weighted temperature evolves positively with respect to self-similarity, with clusters at the pivot temperature ($3$ keV) having $\sim 35$ per cent higher luminosity at $z=1.5$ compared with $z=0$ relative to the self-similar expectation. In contrast, the spectroscopic temperature-based relation shows negligible evolution due to redshifting of the low-energy X-ray emission beyond the X-ray bandpass (see Section~\ref{subsubsec:M-T}).
As for the X-ray luminosity--total mass relation, the evolution in the normalisation is somewhat dependent on the pivot point. In particular, using a higher (lower) pivot point causes the normalisation to evolve more (less) positively in each case. For example, at a pivot point of $5$ keV the increase in X-ray luminosity between $z=0$ and $z=1.5$ rises to $\sim 45$ per cent.

The positive evolution in normalisation predicted by \MACSIS, T18 and the mass-weighted temperature relation of \fable\ agrees with the results of \cite{Giles2016} for the XXL-100-GC sample, which spans a redshift range of $0.05 < z < 1.05$. Assuming a redshift-independent slope, \cite{Giles2016} find that the normalisation of the X-ray luminosity--temperature relation evolves as $E(z)^{1.64 \pm 0.77}$. This corresponds to an increase in normalisation of $\approx 0.15$ dex between $z=0$ and $z=1$ relative to self-similarity, which is in good agreement with the simulation predictions, albeit with large uncertainties.
In contrast, the observational studies by \cite{Reichert2011}, \cite{Hilton2012} and \cite{Clerc2014} measure a negative evolution in the normalisation of the X-ray luminosity--temperature relation up to $z \sim 1.5$.
Potential explanations for these discrepancies are discussed in \cite{Giles2016} and relate mainly to selection bias and the choice of local baseline relation.

It is worth pointing out that the positive evolution indicated in Fig.~\ref{fig:evol} represents an even greater departure from self-similarity than it appears. This is because the factor $E(z)$ that we have incorporated into the normalisation of the X-ray luminosity--temperature relation in order to factor out the expected self-similar evolution (see Section~\ref{subsec:fitting} and Appendix~\ref{A:SS}) is no longer the appropriate expected scaling given that the slope of the X-ray luminosity--total mass relation has been shown to depart from self-similarity.
If we assume that the X-ray luminosity--total mass and total mass--temperature relations take the form of equation~\ref{eq:powerlaw} with slopes $\beta_{LM}$ and $\beta_{MT}$ and $E(z)$ exponents $\gamma_{LM}$ and $\gamma_{MT}$, respectively, then these equations can be combined to show that the luminosity, $L$, scales with temperature, $T$, as
\begin{equation}
  L \propto E(z)^{(\gamma_{LM} \, + \, \gamma_{MT} \beta_{LM})} \; T^{\beta_{LM} \beta_{MT}}.
\end{equation}
From this equation it is clear that, even if the X-ray luminosity--total mass and total mass--temperature relations evolve self-similarly (i.e. $\gamma_{LM}$ and $\gamma_{MT}$ equal their self-similar values), any departure of the slope $\beta_{LM}$ from the self-similar value will alter the expected evolution of the X-ray luminosity--temperature relation. As $\beta_{LM}$ is steeper than the self-similar value (Section~\ref{subsubsec:L-M}) and $\gamma_{MT} < 0$ (both in the self-similar scenario and in our simulations), self-similar evolution would imply that the normalisation of the X-ray luminosity--temperature relation should evolve less rapidly than $E(z)$. For example, using the $z=0$ value for $\beta_{LM}$ and the self-similar values $\gamma_{LM} = 7/3$ and $\gamma_{MT} = -1$ yields a normalisation that evolves approximately as $E(z)^{0.4}$. As a result, the positive evolution of the normalisation shown in Fig.~\ref{fig:evol} somewhat underestimates the departure from the expected evolution. Similar reasoning applies to the \MACSIS\ and T18 relations, which also predict a steeper than self-similar X-ray luminosity--total mass slope.

\paragraph{X-ray luminosity--temperature intrinsic scatter}
The intrinsic scatter about the X-ray luminosity--temperature relation is somewhat higher than the \MACSIS\ and T18 predictions due to the increased scatter of lower mass haloes. When restricting our $z=0$ sample to haloes with $M_{500} > 10^{14} M_{\odot}$ (the same as \MACSIS\ and T18 at $z=0$) the intrinsic scatter drops from $0.25^{+0.04}_{-0.03}$ to $0.18^{+0.05}_{-0.02}$ for the spectroscopic temperature relation, which lies in between the $z=0$ values for \MACSIS\ ($0.14 \pm 0.01$) and T18 ($0.19 \pm 0.02$).
These values are somewhat smaller than the observational constraints of \cite{Pratt2009} ($0.32 \pm 0.058$), \cite{Hilton2012} ($0.27 \pm 0.03$), \cite{Maughan2012} ($0.29 \pm 0.02$) and \cite{Sun2012} ($0.24 \pm 0.01$) but consistent with recent observational constraints from \cite{Giles2016} ($0.20 \pm 0.03$) and \cite{Zou2016} ($0.20 \pm 0.05$).

We find that the intrinsic scatter drops slightly with increasing redshift, falling from $0.25^{+0.04}_{-0.03}$ at $z=0$ to $0.17^{+0.02}_{-0.02}$ at $z=1$ and remaining roughly constant at higher redshift. This is in good agreement with the results of \cite{Hilton2012} who measure the evolution of the X-ray luminosity--temperature relation with 211 clusters up to $z \sim 1.5$ and find a similar drop in the intrinsic scatter with redshift, falling from $0.33 \pm 0.04$ dex at $z<0.25$ to $0.24 \pm 0.05$ at $0.5 < z < 1.5$.

\section{SZ--total mass relation}\label{sec:SZ}
A complementary means of measuring the thermodynamic properties of gas in groups and clusters is via the thermal SZ effect, which arises from the inverse Compton scattering of CMB photons from energetic electrons in the hot ICM. By integrating the Comptonisation parameter over the volume of a system we obtain an SZ ``flux'', $Y_{\mathrm{SZ}}$, that is proportional to the total thermal energy of the hot gas.
Specifically,
\begin{equation}\label{eq:SZ}
  D_A^2(z) \, Y_{\mathrm{SZ}} \equiv \frac{\sigma_T k_B}{m_e c^2} \int n_e T_e \, dV \, ,
\end{equation}
where $D_A(z)$ is the angular diameter distance at redshift $z$,  $n_e$ is the electron number density, $T_e$ is the electron temperature and the constants $\sigma_T$, $k_B$, $m_e$ and $c$ are the Thomson cross-section, Boltzmann constant, electron mass and the speed of light, respectively.

In this section we explore the scaling between SZ flux, $Y_{\mathrm{SZ}}$, and halo mass, $M_{500}$.
We fit the $Y_{\mathrm{SZ}}-M_{500}$ relations with a power law as described in Section~\ref{subsec:fitting} to the sample defined in Section~\ref{subsec:sample}.
In Section~\ref{subsec:SZ_comp} we compare the \fable\ relation to observed clusters from the \textit{Planck} and SPT-SZ catalogues out to $z \sim 0.8$ and $z \sim 1.2$, respectively, measuring $Y_{\mathrm{SZ}}$ within an aperture appropriate for the comparison. In Section~\ref{subsec:Y500} we study the redshift evolution of the $Y_{500}-M_{500}$ relation ($Y_{\mathrm{SZ}}$ measured within a radius $r_{500}$) and in Section~\ref{subsec:counts} we investigate how different predictions for the $Y_{500}-M_{500}$ relation and its redshift evolution affect the expected number of clusters in an SZ-selected survey such as SPT-3G.

\subsection{Comparison with \textit{Planck} and SPT}\label{subsec:SZ_comp}
The second \textit{Planck} SZ source catalogue \citep{PlanckXXVII2015} is the largest SZ-selected sample of galaxy clusters to-date, containing over one thousand confirmed clusters out to $z \sim 1$. The catalogue contains estimates for the integrated flux of each cluster measured within a circular aperture of radius $5 \, r_{500}$, which we denote $\mathrm{Y_{5r_{500}}}$.
Total cluster masses, $M_{500}$, are estimated assuming a scaling relation between the SZ flux and $M_{500}$ calibrated on X-ray hydrostatic masses \citep{PlanckXX2014} given the \textit{Planck} posterior information on the size-flux correlation. These mass estimates will depend on the assumed scaling relation, although \cite{PlanckXXVII2015} show that they agree with external X-ray and optical data with low scatter.
Of the 1653 clusters in the full \textit{Planck} catalogue we exclude 586 clusters that do not have an $M_{500}$ estimate, that are heavily contaminated by infra-red emission ($\mathrm{IR\_FLAG}=1$) or that correspond to low-reliability detections ($\mathrm{Q\_NEURAL} < 0.4$).
For the simulated objects we calculate fluxes within a spherical aperture of radius $5 \, r_{500}$. Using a spherical rather than a circular aperture has a negligible effect on $\mathrm{Y_{5r_{500}}}$ over the range of halo masses in common with the \textit{Planck} clusters ($\gtrsim 10^{14} M_{\odot}$) and thus does not affect our comparison. However, for lower mass systems the SZ flux measured within a circular aperture can be boosted by hot gas that overlaps in projection and for this reason we use a spherical aperture to avoid biasing the best-fitting $\mathrm{Y_{5r_{500}}-M_{500}}$ relation, which is fit to $M_{500} \geq 3 \times 10^{13} \, E(z)^{-0.5} \: M_{\odot}$.

\begin{figure*}
  \includegraphics[width=0.83\textwidth]{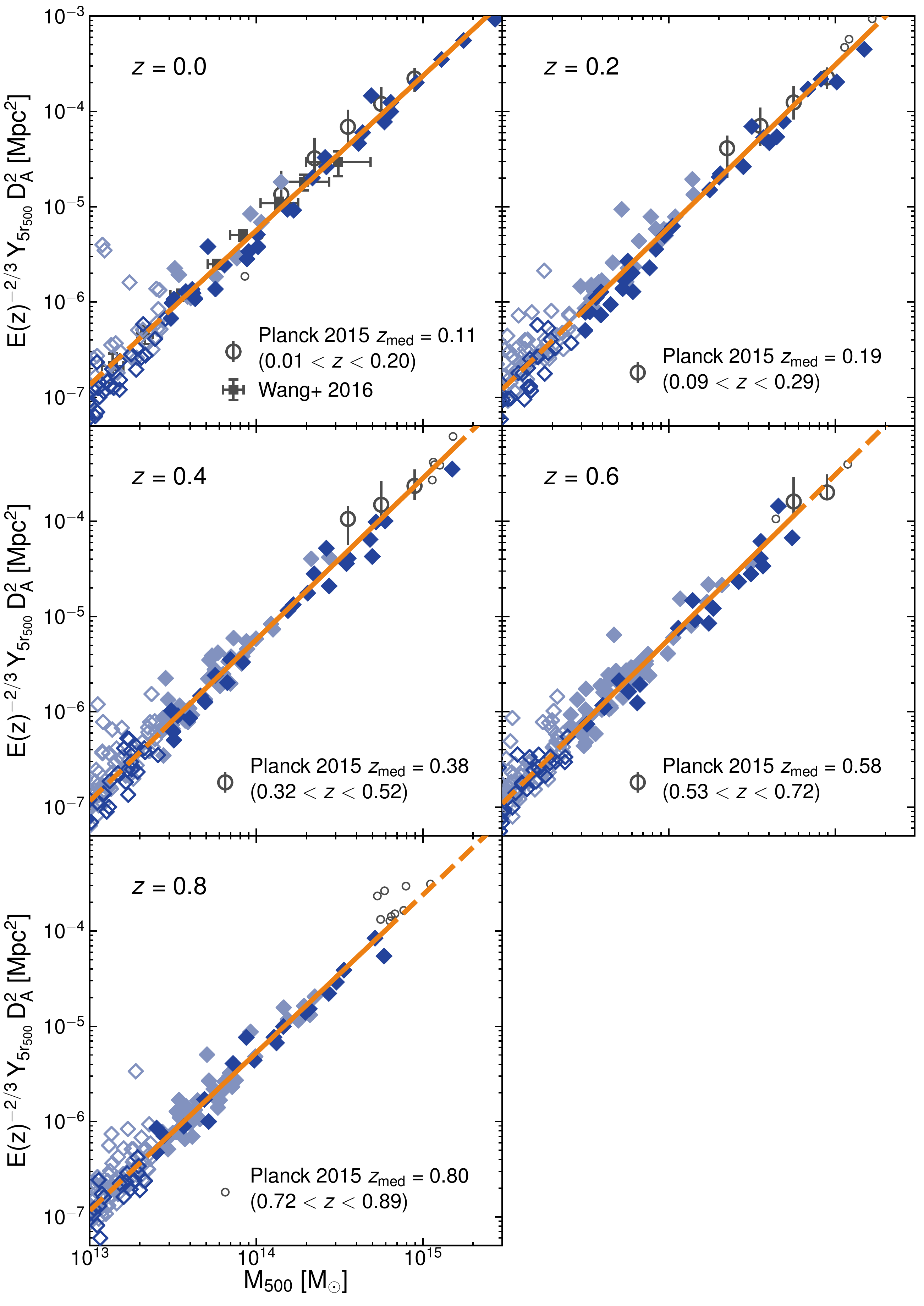}
  \caption{The SZ signal, $\mathrm{Y_{5r_{500}}}$, as a function of total mass for \fable\ haloes (diamonds; styles as in Fig.~\ref{fig:xray_z04}) at various redshifts compared to \textit{Planck} data (grey circles). $\mathrm{Y_{5r_{500}}}$ is the Comptonisation parameter integrated within a spherical aperture of radius $5 \, r_{500}$. We scale $\mathrm{Y_{5r_{500}}}$ by the square of the angular diameter distance and multiply by $E(z)^{-2/3}$ to factor out self-similar redshift evolution.
	The solid line shows the best-fitting power-law relation to haloes of mass $M_{500} > 3 \times 10^{13} \, E(z)^{-0.5} M_{\odot}$ (filled diamonds), which becomes dashed in the region of extrapolation.
    Open circles with error bars show the median relation and 1-sigma intrinsic scatter in total mass bins of width 0.2 dex for \textit{Planck} clusters in the redshift range indicated in the panel legends. For bins containing fewer than ten objects we plot individual clusters as small open circles.
        }
    \label{fig:Y-M_Planck}
\end{figure*}

In Figure~\ref{fig:Y-M_Planck} we plot $\mathrm{Y_{5r_{500}}}$ as a function of $M_{500}$ at five redshifts between $z=0$ and $z=0.8$.
We scale $\mathrm{Y_{5r_{500}}}$ by the square of the angular diameter distance of each cluster and multiply by the self-similar scaling factor, $E(z)^{-2/3}$.
At each redshift we select a comparison sample of \textit{Planck} clusters with a similar median redshift. The redshift range that defines each sample is given in the legend. Due to the large number of sources, we bin the \textit{Planck} clusters in halo mass bins of width 0.2 dex and plot the median and 1-sigma intrinsic scatter. For bins containing fewer than ten objects we plot the individual clusters.
In addition, at $z=0$ we compare with the SZ flux measurements of \cite{Planck2013XI} for a large sample of locally brightest galaxies (LBGs). These galaxies were selected to be predominantly central galaxies and therefore form a natural extension of the cluster relation to lower masses. \cite{Planck2013XI} bin the LBGs by stellar mass and then estimate an ``effective'' halo mass for each bin using the semi-analytic galaxy formation simulation of \cite{Guo2011}. Since this procedure has a non-negligible dependence on the semi-analytic model \citep{Wang2016}, we compare to the \cite{Planck2013XI} relation recalibrated by \cite{Wang2016} using weak lensing masses. We convert the $Y_{500}$ values reported in \cite{Planck2013XI} back into the measured flux, $\mathrm{Y_{5r_{500}}}$, by multiplying by the factor 1.796, which is the conversion factor corresponding to the spatial template used in their matched filter, the universal pressure profile \citep{Arnaud2010}.

At $z=0$ the \fable\ relation is in excellent agreement with the \cite{Wang2016} LBG relation over their full mass range ($3 \times 10^{13} M_{\odot} \lesssim M_{500} \lesssim 3 \times 10^{14} M_{\odot}$). In comparison, the \textit{Planck} clusters have slightly higher SZ flux at fixed halo mass. The most likely explanation for this offset is an X-ray hydrostatic mass bias of $\sim 30$ per cent, which would bias the \textit{Planck} cluster relation relative to the weak lensing-calibrated LBG relation and the simulations. This is consistent with the $\sim 30$ per cent X-ray mass bias required to explain the offset in the total mass--spectroscopic temperature relation between \fable\ and observational constraints based on X-ray hydrostatic masses (see e.g. Fig.~\ref{fig:xray_z04} and \ref{fig:xray_z1}).
A similar offset between the \textit{Planck} clusters and the \fable\ relation persists to higher redshift, although at $z \geq 0.6$ we rely on the extrapolation of the best-fitting relation as there are few, if any, \fable\ systems as massive as the \textit{Planck} clusters. This suggests that the X-ray mass bias remains at a similar level with increasing redshift, as found for example in \cite{Nagai2007a}, \cite{Henson2017} and \cite{Hurier2017}.

The best-fitting power law relation is a good description of the $\mathrm{Y_{5r_{500}}} - M_{500}$ scaling over a wide range of masses, from massive clusters ($\sim 10^{15} M_{\odot}$) to low mass galaxy groups ($\sim 10^{13} M_{\odot}$). The slope of the relation is consistent with the self-similar prediction at all redshifts. We also find no evidence for a systematic redshift trend in the normalisation or intrinsic scatter.
This is understandable as the SZ flux is integrated out to several times the virial radius and is thus fairly insensitive to non-gravitational processes such as AGN feedback.

\begin{figure*}
  \includegraphics[width=0.83\textwidth]{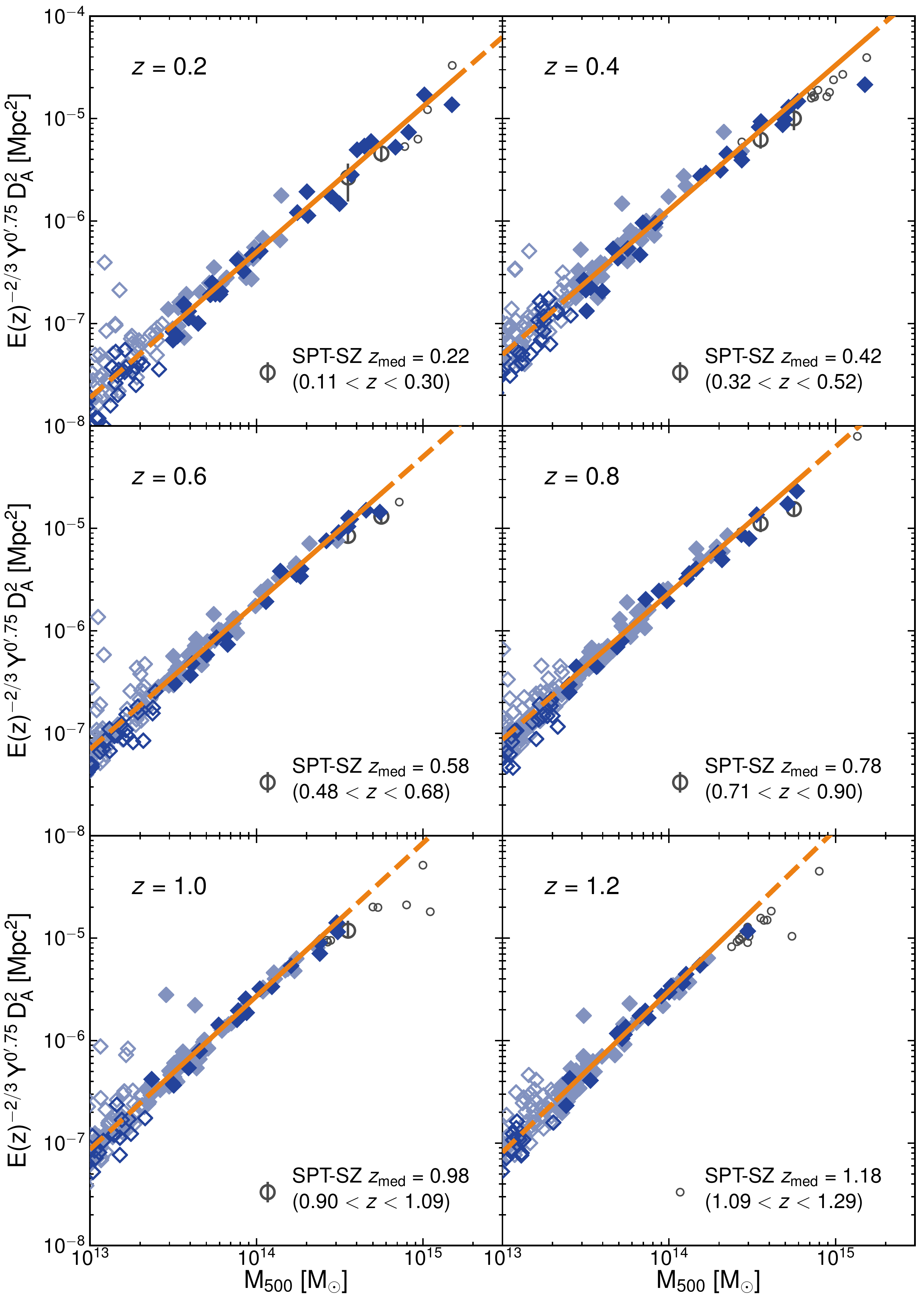}
        \caption{The SZ signal integrated within a circular aperture of radius 0.75 arcminutes as a function of total mass at different redshifts compared with clusters detected in the 2500~deg$^2$ SPT-SZ survey \citep{Bleem2015, Bocquet2018}. Symbol styles are similar to those in Fig.~\ref{fig:Y-M_Planck}.
        }
    \label{fig:Y-M_SPT}
\end{figure*}

In Figure~\ref{fig:Y-M_SPT} we perform a similar comparison with galaxy clusters discovered in the 2500 deg$^2$ SPT-SZ survey \citep{Bleem2015} at $0.2 \leq z \leq 1.2$. We use the cluster catalogue described in \cite{Bocquet2018}, which uses an updated calibration of the photometric redshifts from \cite{Bleem2015} plus additional spectroscopic redshifts.
\cite{Bocquet2018} also perform a weak-lensing calibrated cosmological analysis of the SPT data and derive cluster mass estimates marginalized over the uncertainties in the scaling relation and the cosmological parameters. We use these mass estimates in our comparison but note that our conclusions are unchanged if instead we use the fiducial mass estimates presented in \cite{Bleem2015}.
The SPT SZ flux measurements are integrated within a circular aperture of radius $0.75$ arcminutes. We measure the SZ flux within the same radius, integrating the entire simulation volume along the line-of-sight but ignoring low-resolution elements in the zoom-in simulations.

The \fable\ clusters are a good match to the SPT data at all redshifts. The best-fitting relation tends to overestimate the SZ flux of the most massive SPT clusters, however this is because the fit is biased towards the lowest mass systems, which prefer a steeper slope. Although we do not have a large number of haloes in a common mass range with the SPT clusters, those which do overlap are in excellent agreement.
The handful of outliers with high SZ flux at the low mass end of the relation are objects that overlap with more massive haloes in projection, which boosts their apparent flux. Note, however, that these are mostly secondary haloes in the zoom-in simulations for which the probability of such an alignment is increased.

Whereas the relatively large \textit{Planck} aperture integrates the SZ flux to the cluster outskirts, SPT is most sensitive to the cluster core regions. For example, at $z=0.2$ the 0.75 arcminute aperture radius corresponds to a physical radius of only $153$ kpc. For a typical mass of $M_{500} \sim 5 \times 10^{14} M_{\odot}$ with $r_{500} \sim 1$ Mpc this corresponds to roughly $0.15 \, r_{500}$, which characterises the cluster core. Therefore, our agreement with the SPT and \textit{Planck} clusters suggests that the SZ flux is realistically distributed from the core to the cluster outskirts. Indeed, we showed in Paper~I that the pressure profiles of \fable\ clusters are in good agreement with observations of local clusters from the REXCESS sample \citep{Bohringer2007, Pratt2010}.

\subsection{The \texorpdfstring{$Y_{500}-M_{500}$}{Y500-M500} relation}\label{subsec:Y500}
In this section we investigate the $Y_{500}-M_{500}$ relation, where $Y_{500}$ is the SZ flux measured within a spherical aperture of radius $r_{500}$.
In Fig.~\ref{fig:Y500-M500_z0} we show the $Y_{500}-M_{500}$ relation at $z=0$ and in Fig.~\ref{fig:Y500-M500_vs_z} we plot the parameters of the best-fitting relation as a function of redshift.
We compare to the best-fitting relations from the \MACSIS\ and \cite{Planelles2017} simulations as well as observationally derived relations from \cite{Andersson2011}, \cite{PlanckXX2014} and \cite{Nagarajan2018}.

The simulations reported in \cite{Planelles2017} were analysed in terms of their X-ray properties in T18, which we have compared with the \fable\ and \MACSIS\ results in Section~\ref{sec:xray}. However we note that \cite{Planelles2017} use a lower mass sample than T18, which consists of $\sim 100$ clusters and groups with $M_{500} \gtrsim 4.1 \times 10^{13} M_{\odot}$ at $z=0$.
\cite{Andersson2011} derive the $Y_{500}-M_{500}$ relation for a sample of 15 SPT clusters observed with \textit{Chandra} and \textit{XMM-Newton}.
Cluster candidates with the highest signal-to-noise ratio were chosen from 178 deg$^2$ of the sky surveyed by the SPT in 2008 \citep{Vanderlinde2010}. Total masses are estimated from the $M_{500}-Y_{\mathrm{X}}$ relation of \cite{Vikhlinin2009a}, which is calibrated on X-ray hydrostatic mass estimates.
We note that $Y_{\mathrm{X}}$ is related but not equal to $Y_{500}$ because the two quantities rely on differently weighted gas temperatures (X-ray spectroscopic and mass-weighted, respectively).
Indeed, the $Y_{500}-Y_{\mathrm{X}}$ relation derived in \cite{Andersson2011} has an intrinsic scatter of $0.09 \pm 0.04$ dex and its normalisation implies an average $Y_{500}/Y_{\mathrm{X}}$ ratio of $0.82 \pm 0.07$.
\cite{PlanckXX2014} derive a baseline $Y_{500}-M_{500}$ relation for use in their cosmological analysis of SZ cluster counts.
The relation is derived from 71 clusters in the \textit{Planck} cosmological sample with \textit{XMM-Newton} observations and mass estimates from the relation between $Y_X$ and the X-ray hydrostatic mass $M_{500}$ established for 20 local relaxed clusters by \cite{Arnaud2010}.
The $Y_{500}-Y_{\mathrm{X}}$ relation for this sample has an intrinsic scatter of $0.07 \pm 0.01$ dex and an average $Y_{500}/Y_{\mathrm{X}}$ ratio of $0.94 \pm 0.02$ \citep{PlanckXXIX2014}. The baseline relation includes a mass bias parameter, $b$, which we take to be zero.
Finally, \cite{Nagarajan2018} calibrate the $Y_{500}-M_{500}$ relation with SZ measurements from the APEX-SZ experiment and total cluster mass estimates derived from weak lensing. The sample consists of 27 clusters selected from the ROSAT All-Sky Survey.

\subsubsection{Comparison at $z \approx 0$}

\begin{figure}
  \includegraphics[width=\columnwidth]{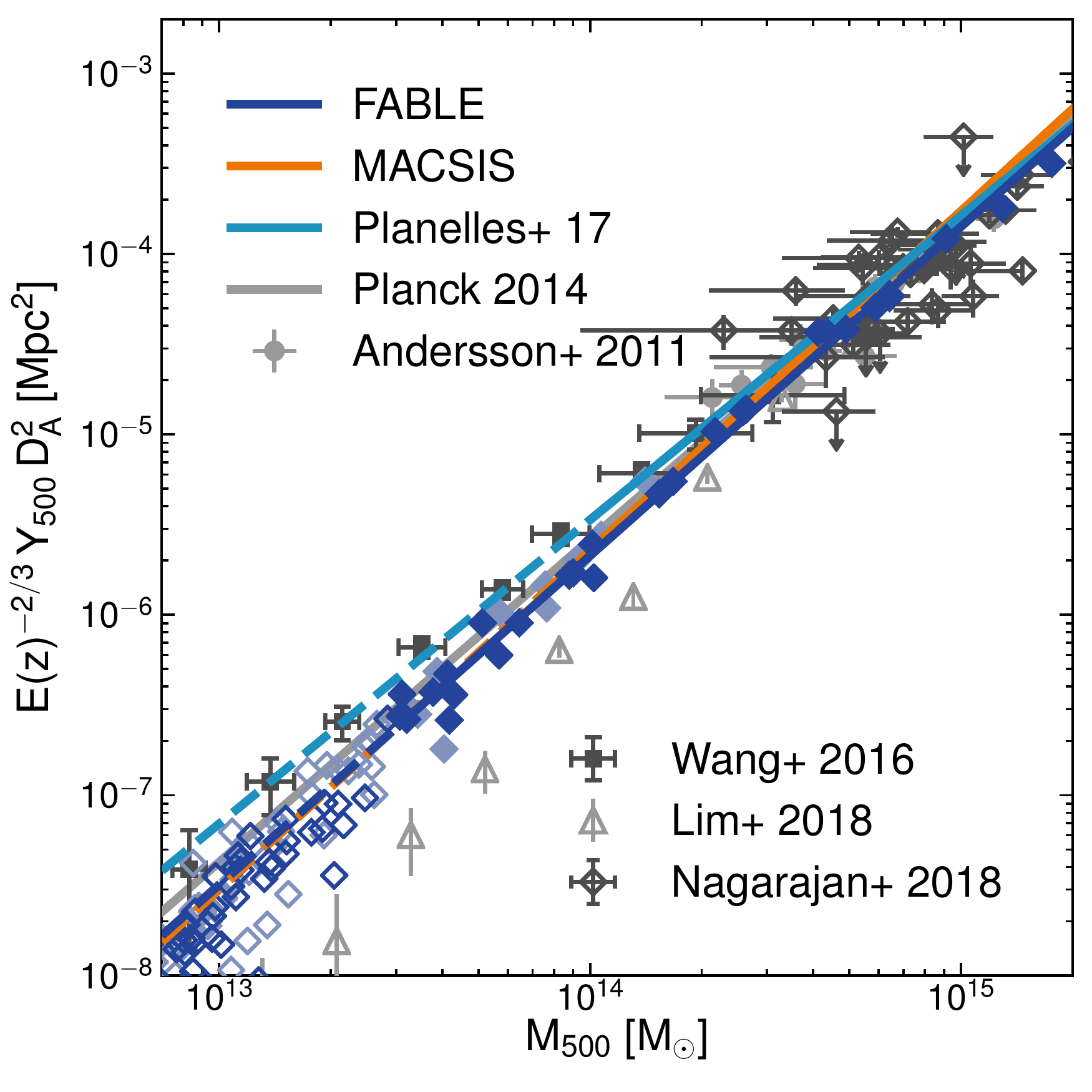}
\caption{The SZ signal integrated within a spherical aperture of radius $r_{500}$ as a function of total mass at $z=0$ (diamonds; styles as in Fig.~\ref{fig:xray_z04}). Solid lines indicate the best-fitting relations for \fable\ (dark blue), \MACSIS\ (orange) and the \protect\cite{Planelles2017} simulations (light blue), which become dashed in the region of extrapolation. The grey solid line is the baseline relation of \protect\cite{PlanckXX2014}. Grey error bars show data from \protect\cite{Andersson2011}, \protect\cite{Wang2016}, \protect\cite{Lim2018} and \protect\cite{Nagarajan2018}.
  }
    \label{fig:Y500-M500_z0}
\end{figure}

To study the reliability of our model at the low mass end of the $Y_{500}-M_{500}$ scaling relation, we compare to data from \cite{Wang2016} and \cite{Lim2018}. We compare to the \cite{Wang2016} weak lensing-calibrated relation of the LBG sample using the $Y_{500}$ values given in \cite{Planck2013XI}. \cite{Lim2018} use the matched filter approach applied to the \textit{Planck} all-sky maps to measure the SZ effect produced by local galaxy groups. The sample constitutes galaxy groups identified with a halo-based group finder from several large surveys with halo masses assigned based on the stellar mass of member galaxies. The groups are stacked via their halo mass and the mean SZ flux calculated for each bin.

At $z=0$ (Fig.~\ref{fig:Y500-M500_z0}) the predicted and observed best-fitting relations are in excellent agreement in the cluster regime, overlapping to within $\sim 0.1$ dex at $\sim 5 \times 10^{14} M_{\odot}$. This supports previous results that show the $Y_{500}-M_{500}$ relation is relatively insensitive to the precise implementation of non-gravitational physics in simulations (e.g. \citealt{Battaglia2012, LeBrun2014, Pike2014}).

On the other hand, there is a significant dispersion between relations in the regime of low-mass clusters and galaxy groups ($M_{500} \lesssim 3 \times 10^{14} M_{\odot}$).
On the one hand, \cite{Andersson2011}, \cite{PlanckXX2014} and \cite{Nagarajan2018} fit the SZ flux--mass relation only on cluster scales and so it is perhaps unsurprising that the best-fitting relations of these studies do not fully agree at relatively low masses. On the other hand, observational constraints at the low mass end by \cite{Wang2016} and \cite{Lim2018} present an equally large discrepancy (compare squares and diamonds in Fig.~\ref{fig:Y500-M500_z0}).

The \cite{Wang2016} data lie slightly above the \fable\ $Y_{500}-M_{500}$ relation despite good agreement in $Y_{5r500}-M_{500}$ (Figure~\ref{fig:Y-M_Planck}). This suggests that there is a bias in the method used in \cite{Planck2013XI} to infer $Y_{500}$ from the measured flux, $Y_{5r500}$. Indeed, \cite{LeBrun2015} generate mock synthetic thermal SZ maps from the cosmo-OWLS suite of simulations to show that $Y_{500}$ estimated using the \cite{Planck2013XI} spatial template is biased high, for instance by a factor of two at $M_{500} \sim 3 \times 10^{13} M_{\odot}$. In fact the bias found in \cite{LeBrun2015} accounts for almost all of the offset between the \cite{Wang2016} and \fable\ relations, and similarly for the extrapolated \MACSIS\ relation.

\cite{Lim2018} obtain a significantly lower amplitude for galaxy groups compared with the \cite{Wang2016} weak lensing calibrated LBG relation and the original \cite{Planck2013XI} LBG relation.
\cite{Lim2018} discuss two possible reasons for this difference. First is that they match all groups in the sample during the matched filtering process simultaneously, which accounts for projection by larger haloes along the line-of-sight, while \cite{Planck2013XI} matches individual filters separately. Second, \cite{Planck2013XI} bin haloes by the stellar mass of the central galaxy, which may mix haloes of different masses.
On the other hand, \fable, \MACSIS\ and \cite{Planelles2017} prefer a significantly higher normalisation than \cite{Lim2018} that is much closer to the \cite{Wang2016} relation.
Furthermore, the \cite{Lim2018} relation is significantly steeper than the constraints at cluster scales, which implies a break in the $Y_{500}-M_{500}$ relation at $M_{500} \sim 3 \times 10^{14} M_{\odot}$. Such a change in slope is not apparent in the simulations shown here, although there is some evidence of a break in the cosmo-OWLS simulations (see fig. 2 of \citealt{Lim2018}).

\subsubsection{Evolution of the $Y_{500}-M_{500}$ relation}\label{subsubsec:Y500-M500_evol}
\begin{figure*}
\begin{center}
{\includegraphics[width=\factorthree\textwidth]{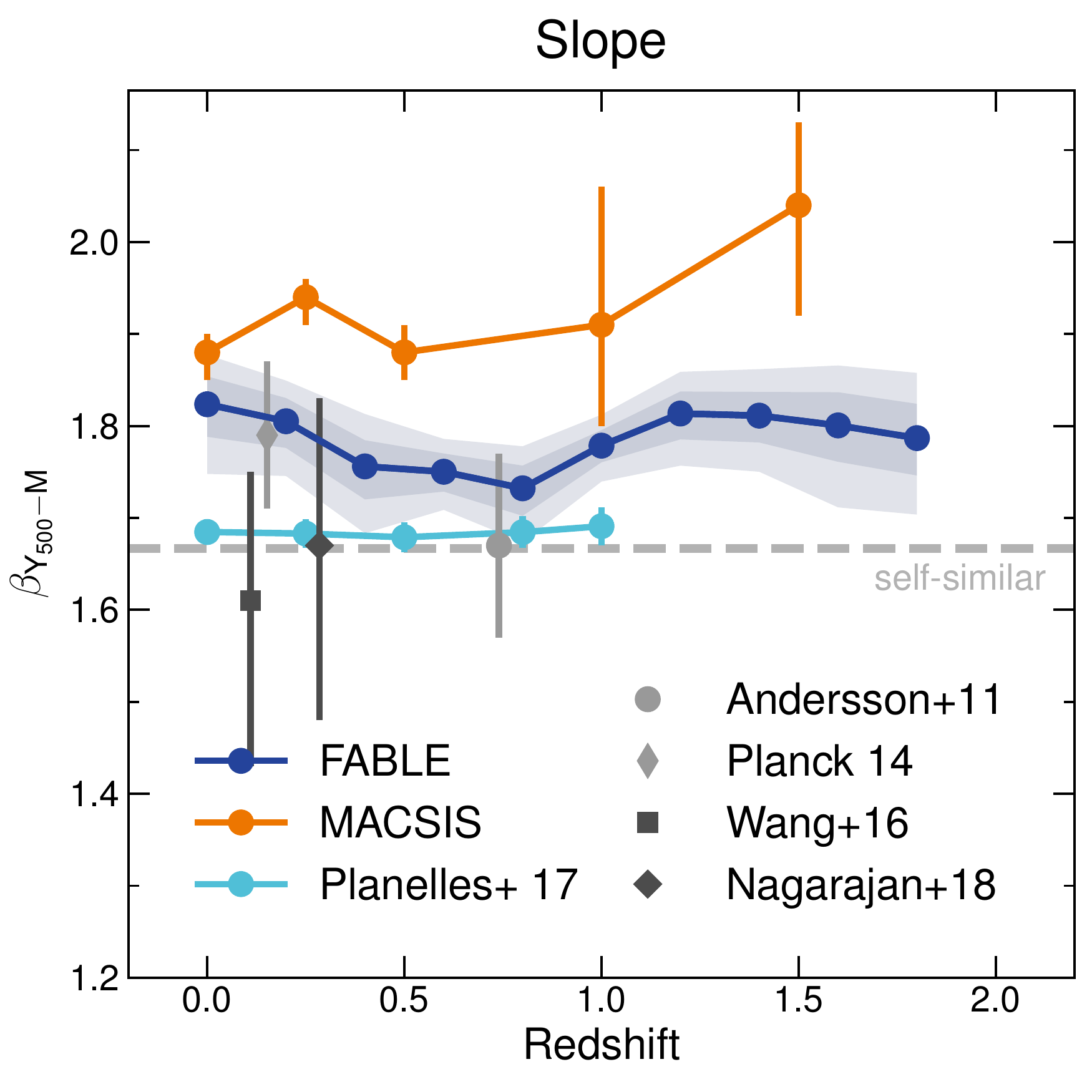}}
{\includegraphics[width=\factorthree\textwidth]{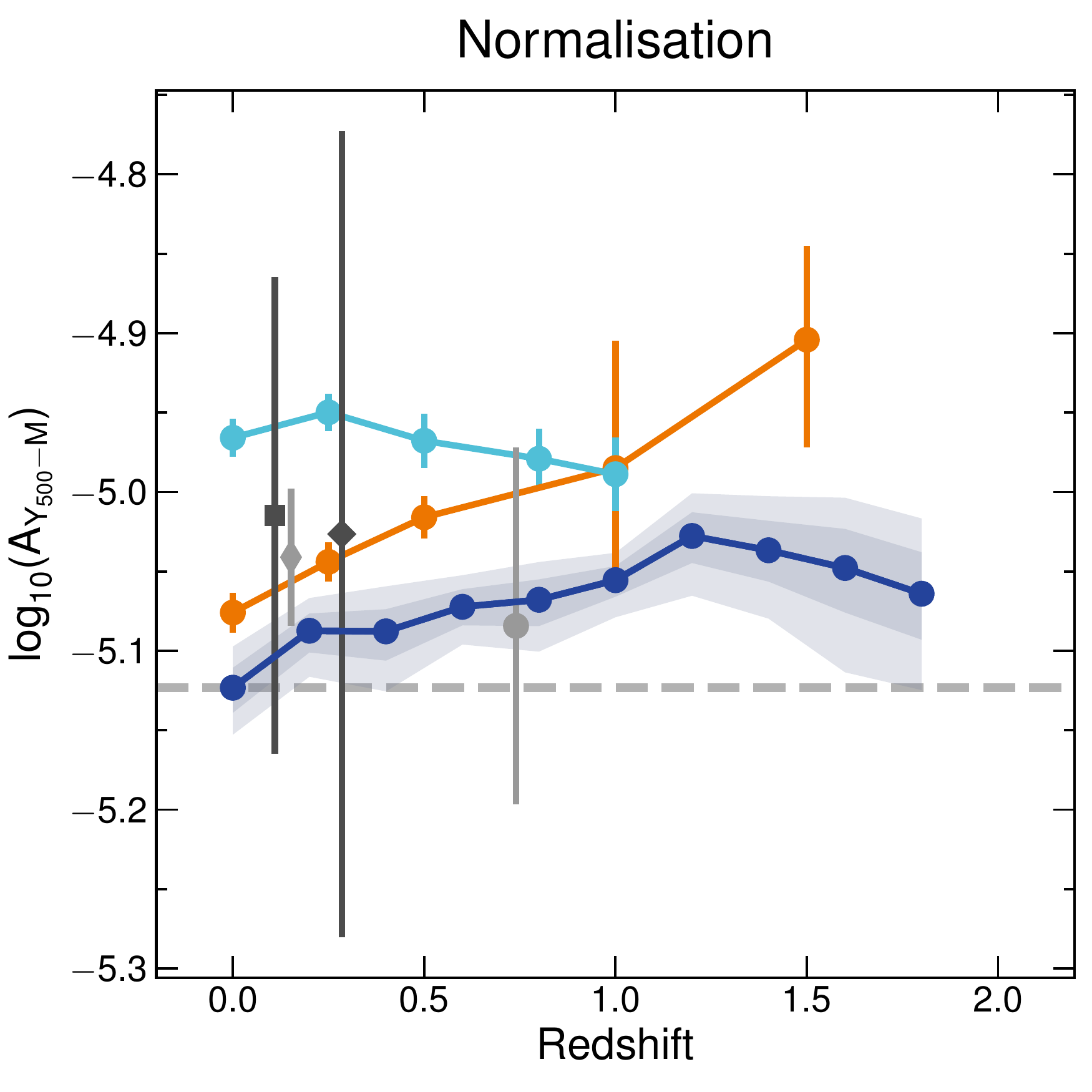}}
{\includegraphics[width=\factorthree\textwidth]{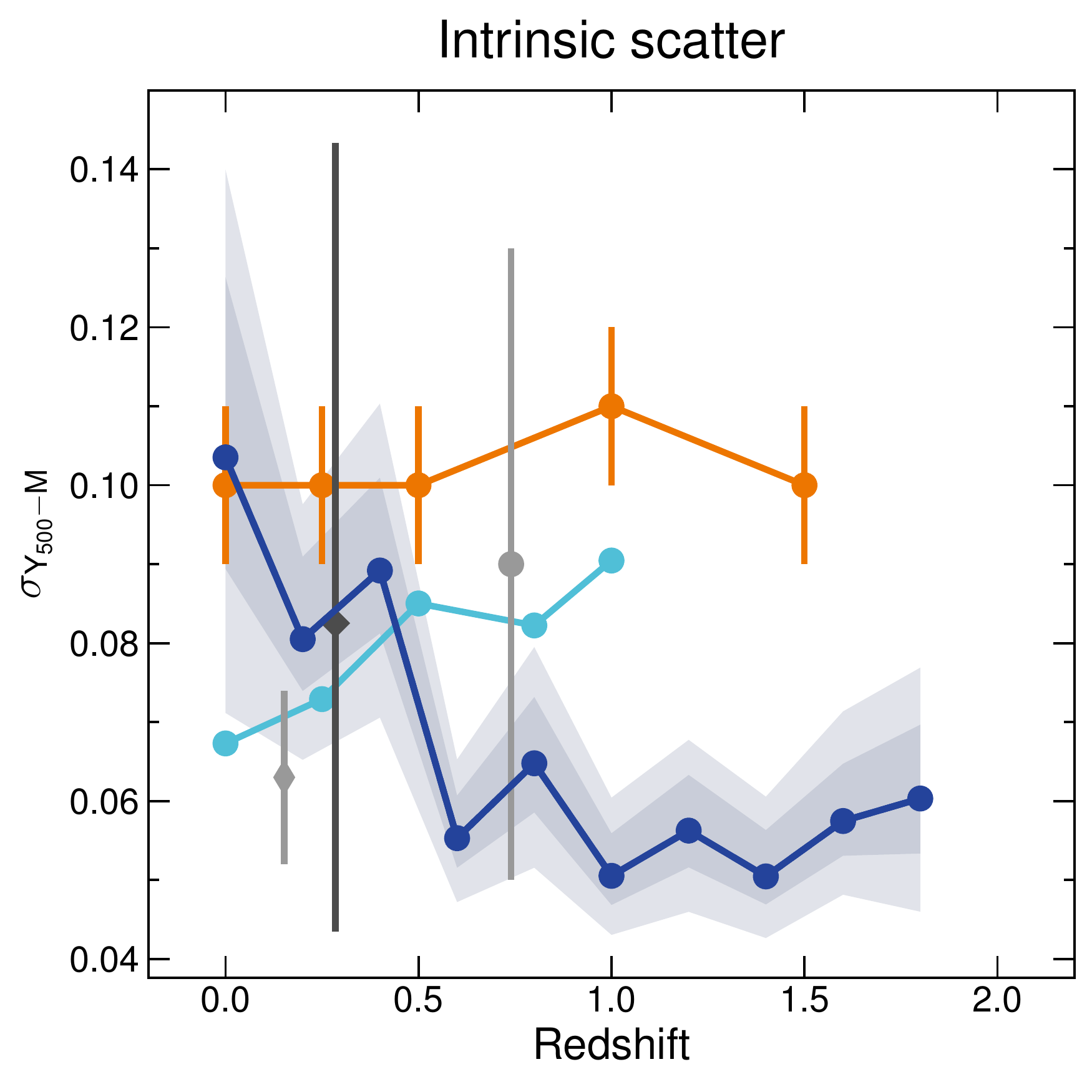}}
\end{center}
\caption{The redshift evolution of the $Y_{500}-M_{500}$ relation, where $Y_{500}$ is the SZ signal integrated within a spherical aperture of radius $r_{500}$. Panels show the slope, normalisation and intrinsic scatter of the best-fitting power-law relation from left to right, respectively. Symbol styles are the same as Fig.~\ref{fig:Mgas-M}. The normalisation is calculated at $M_{500} = 2 \times 10^{14} M_{\odot}$. The light blue curves show the results of \protect\cite{Planelles2017}, who analyse the same set of simulations as T18 but using a different sample selection as described in the text.
  Observational constraints on the parameters are plotted as grey error bars at the median redshift of the sample.
  Note that the \protect\cite{PlanckXX2014} normalisation assumes zero mass bias and the error bar indicates only the statistical uncertainty not including the uncertainty in their X-ray mass bias parameter.
  }
    \label{fig:Y500-M500_vs_z}
\end{figure*}

Fig.~\ref{fig:Y500-M500_vs_z} shows the redshift evolution of the slope, normalisation and intrinsic scatter of the best-fitting $Y_{500}-M_{500}$ relation. For comparison we plot observational constraints on these parameters from the aforementioned studies at the median redshift of their respective samples.

\paragraph{$Y_{500}-M_{500}$ slope}
The slope of the \fable\ relation is approximately constant with redshift at $\beta_{Y_{500}-M} \sim 1.8$, which is slightly steeper than self-similar ($\beta_{Y_{500}-M} = 5/3$). This value is in good agreement with \cite{PlanckXX2014} but slightly high compared with \cite{Andersson2011}, \cite{Wang2016} and \cite{Nagarajan2018}. On the other hand, limiting our sample to massive objects ($M_{500} > 10^{14} E(z)^{-0.5} M_{\odot}$) yields to a shallower slope that is consistent with most observational constraints and the self-similar prediction within the uncertainties (see discussion in Appendix~\ref{A:mass_dependence}).
\cite{Planelles2017} predict a close to self-similar slope out to $z=1$, whereas the \MACSIS\ relation is significantly steeper than self-similar out to $z=1.5$. Note, however, that the \MACSIS\ relation is based on X-ray hydrostatic masses, which leads to a somewhat steeper slope. The lack of redshift evolution in the slope is consistent with a number of previous numerical studies, including \cite{Battaglia2012}, \cite{Pike2014} and \cite{Sembolini2014}.

\paragraph{$Y_{500}-M_{500}$ normalisation}
The middle panel of Fig.~\ref{fig:Y500-M500_vs_z} plots the (absolute) normalisation at the pivot point ($M_{500} = 2 \times 10^{14} M_{\odot}$) as a function of redshift.
The observational constraints are consistent, albeit with large uncertainties, and do not seem to prefer one simulation prediction over another.
We caution that the \cite{Andersson2011} and \cite{PlanckXX2014} constraints, which are based on X-ray hydrostatic mass estimates, possess additional uncertainty associated with a possible X-ray mass bias. For example, \cite{PlanckXX2014} assume a flat prior on the mass bias that allows X-ray hydrostatic masses to be biased low by as much as $30$ per cent, which they derive from comparisons with simulations. This corresponds to lowering the normalisation by as much as $0.28$ dex compared with no bias (as assumed in Fig.~\ref{fig:Y500-M500_vs_z}). The mass bias thus clearly dominates the uncertainty in the \cite{PlanckXX2014} normalisation.

Both the \fable\ and \MACSIS\ relations evolve positively in normalisation to at least $z \sim 1.2$. In both cases this reflects the evolution in the gas mass--total mass relation (see Section~\ref{subsubsec:Mgas-M}).
Significant evolution in the normalisation of the $Y_{500}-M_{500}$ relation could have important consequences for constraining cosmological parameters from SZ cluster counts. Indeed, \cite{PlanckXX2014} find that the limiting factor in their cosmological analysis is the modelling of the $Y_{500}-M_{500}$ relation, in particular its normalisation.
Fortunately, the change in normalisation predicted by \fable\ and \MACSIS\ is small compared with current observational uncertainties, especially considering the uncertainty associated with a potential X-ray mass bias.
Yet as ongoing and future SZ surveys beat down the statistical uncertainties and weak lensing analyses provide more, unbiased mass estimates, our results suggest that future cosmological studies may have to account for beyond self-similar evolution in the mass calibration.

\paragraph{$Y_{500}-M_{500}$ intrinsic scatter}
As for the gas mass--total mass relation (Section~\ref{subsubsec:Mgas-M}), the intrinsic scatter at $z=0$ predicted by \fable\ and \MACSIS\ ($0.10^{+0.02}_{-0.01}$ and $0.10^{+0.01}_{-0.01}$ dex respectively) is slightly higher than \cite{Planelles2017} ($0.067$ dex).
Other simulation works predict even less scatter, such as \cite{Pike2014} ($0.034$) and \cite{LeBrun2017} ($\approx 0.04$). Current observational constraints span a range of values -- e.g., \cite{Arnaud2007} ($0.039$), \cite{PlanckXX2014} ($0.063 \pm 0.011$), \cite{Nagarajan2018} ($0.08^{+0.06}_{-0.04}$) and \cite{Sereno2015} ($0.07$--$0.15$) -- and cannot yet distinguish between models.
We find that the intrinsic scatter is smaller for higher mass haloes, as also found in \cite{Planelles2017}. For example, when limiting our sample to masses $M_{500} > 10^{14} M_{\odot}$ at $z=0$ we find that the intrinsic scatter drops from $0.10^{+0.02}_{-0.01}$ to $0.07^{+0.01}_{-0.03}$, which is consistent with the majority of observational constraints.

The simulations make quite different predictions for the redshift evolution of the intrinsic scatter. \MACSIS\ predicts a roughly redshift-independent scatter while \fable\ and \cite{Planelles2017} find a mild increase or decrease with increasing redshift, respectively. The difference is most likely related to the sample selection, in particular the mass distribution. Indeed, we find that limiting our sample to higher mass haloes lowers the intrinsic scatter preferentially at low redshifts, which reduces the apparent redshift evolution. Similarly, \cite{Planelles2017} find constant scatter with redshift when focusing on massive clusters.

\subsection{Predicted SZ cluster counts}\label{subsec:counts}
In this section we investigate how the choice of $Y_{500}-M_{500}$ relation and its redshift evolution affects predictions for the number of clusters detected in an SZ-selected survey.
We are motivated by several ongoing and future SZ surveys such as SPT-3G \citep{Benson2014, Bender2018}, Advanced ACTpol \citep{Henderson2016} and CMB-S4 \citep{Abazajian2016}, which are expected to vastly expand the number of known clusters and groups. For example, \cite{Benson2014} predict that the ongoing SPT-3G survey will detect $\sim 5000$ clusters over 2500 deg$^2$ compared with $\sim 500$ detected in SPT-SZ, while the proposed CMB-S4 experiment could identify approximately 45,000 to 140,000 clusters depending on the instrument configuration \citep{Abazajian2016}).

For an SZ-selected survey the predicted cluster counts depend sensitively on the assumed relationship between the SZ signal and halo mass.
This relation can be constrained from current surveys, however the uncertainties are large and results can vary significantly between studies.
In addition, the increased sensitivity of future SZ surveys will enable them to detect lower mass clusters than are present in current observational samples. For example, \cite{Benson2014} predict that SPT-3G will detect clusters to a lower mass limit of $M_{500} \sim 10^{14} M_{\odot}$ compared to $\sim 3 \times 10^{14} M_{\odot}$ for SPT-SZ. As a result, the predicted cluster counts are dependent on the extrapolation of the SZ signal--total mass relation to lower masses than existing surveys are able to probe. Extrapolation from higher masses is particularly uncertain because the effects of feedback are expected to be greater in the low mass regime.
Furthermore, many of these new clusters will be at higher redshift than existing samples so that the predicted cluster counts depend also on the assumed redshift evolution of the relation.
Here we use some of the $Y_{500}-M_{500}$ relations described in the previous section to study how such aspects can affect the predicted number of clusters as a function of redshift.

\subsubsection{Calculating cluster counts}\label{subsubsec:calc_counts}
For each $Y_{500}-M_{500}$ relation we calculate the expected number of clusters with $Y_{500}$ greater than a chosen threshold, $Y_{\mathrm{500, lim}}$, in redshift bins of width $\delta z = 0.1$.
For each redshift bin we calculate a halo mass function with mass bins of width $0.01$ dex at the central redshift. We generate the mass function using the \texttt{hmf} code \citep{Murray2013a} with a \cite{Tinker2008} fitting function and a transfer function computed using the Code for Anisotropies in the Microwave Background (CAMB; \citealt{Lewis2000}) for which we use the default \texttt{hmf} parameters.
Our choice of fitting function is based on the cosmological analysis of SZ cluster counts from \textit{Planck} \citep{PlanckXX2014, PlanckXXIV2015} and SPT-SZ \citep{Haan2016, Bocquet2018}, which adopt the \cite{Tinker2008} mass function. There are a number of other fits in the literature and these can vary depending on, for example, the size and resolution of the simulations or the choice of halo-finder (e.g. \citealt{Jenkins2001, Courtin2010, Crocce2010, Angulo2012, Ishiyama2015}). Furthermore, the halo mass function can be affected by baryonic effects, which tends to lower the number density of haloes of a given mass (e.g. \citealt{Velliscig2014, Bocquet2016}). Although a different choice for the fitting function will lead to different predictions for the absolute number counts, we do not expect this to significantly affect the relative difference between predicted counts, which is the main focus of this analysis.

We assume that haloes are normally distributed in $\mathrm{log}_{10}(Y_{500})$ at fixed mass with mean and standard deviation equal to the best-fitting $Y_{500}-M_{500}$ relation and its intrinsic scatter.\footnote{We find that, in our simulations, the intrinsic scatter near the threshold is close to the global intrinsic scatter for the thresholds under consideration.} Consequently, the number density of haloes in each halo mass bin is scaled by the fraction of clusters that are expected to lie above $Y_{\mathrm{500, lim}}$ at the given halo mass.
The total cluster count is obtained by integrating over all mass bins and multiplying by the volume of the redshift bin as seen in a 2500 deg$^2$ survey.
For the observational constraints we assume redshift-independent intrinsic scatter and extrapolate the normalisation of the $Y_{500}-M_{500}$ relation in redshift assuming self-similarity. For the simulation predictions we linearly interpolate on the best-fitting parameters, which are reported at the redshifts shown in Fig.~\ref{fig:Y500-M500_vs_z}.
Note that we take into account the different values of $H_0$ assumed by different studies but do not otherwise account for their differing cosmologies. We ignore measurement errors for simplicity.

In practice, clusters are not selected on the integrated SZ flux, such as $Y_{500}$, but on a related quantity such as the signal-to-noise ratio. In the case of the SPT-SZ sample \citep{Bleem2015} and the SPT-3G predictions \citep{Benson2014}, clusters are selected above a threshold value for the SZ detection significance, $\zeta$, which is a measure of the signal-to-noise across all filter scales \citep{Vanderlinde2010}.
Unfortunately, a fixed threshold in $\zeta$ does not necessarily correspond to an integrated SZ flux threshold, $Y_{\mathrm{500, lim}}$, that is constant with redshift. For example, confusion with primary CMB fluctuations and atmospheric noise suppresses the detection significance of low-redshift clusters (e.g. \citealt{Vanderlinde2010}) while the detectability of high-redshift or low-mass clusters depends sensitively on the instrument resolution.
\cite{Andersson2011} derive a relationship between $\zeta$ and $Y_{500}$ using simulated SPT observations and a subsample of SPT-SZ clusters. Their relation implies that, for a fixed $\zeta$ threshold, the corresponding $Y_{\mathrm{500, lim}}$ varies with redshift as $E(z)^{-0.25}$.
This scaling is consistent with the redshift evolution of the halo mass threshold of the SPT-SZ 2500 deg$^2$ sample\footnote{If we assume that $Y_{500}$ is related to $M_{500}$ via the self-similar scaling relation, then the threshold mass corresponding to $Y_{\mathrm{500, lim}} \sim E(z)^{-0.25}$ varies as $E(z)^{\frac{3}{5}(-\frac{2}{3} - 0.25)} \approx E(z)^{-0.55}$, consistent with the $\sim E(z)^{-0.5}$ scaling of the minimum mass of the SPT-SZ 2500 deg$^2$ sample, as shown in Appendix~\ref{A:mass_dependence}.}.
We therefore assume that $Y_{\mathrm{500, lim}}$ scales with redshift as $E(z)^{-0.25}$ and use the SPT-SZ 2500 deg$^2$ results as a baseline for comparison.

We investigate two values for $Y_{\mathrm{500, lim}}$, which we refer to as `high' and `low' detection thresholds in the following. In Section~\ref{subsubsec:counts_HighThresh} we test a high threshold flux of $Y_{\mathrm{500, lim}} = \, 3.75 \times 10^{-5} \; E(z)^{-0.25}$~Mpc$^2$. This value was chosen such that the total number of clusters at $z \leq 1.8$ is approximately equal to the total number of detected clusters in the SPT-SZ 2500 deg$^2$ survey, assuming the $Y_{500}-M_{500}$ relation of \cite{Andersson2011} and our fiducial cosmology.
To our knowledge, the \cite{Andersson2011} relation provides the most reliable estimate of the $Y_{500}-M_{500}$ relation underlying the full SPT-SZ cluster sample.\footnote{\cite{Saliwanchik2015} use a novel method to derive $Y_{500}$ for a similar sample of clusters to \cite{Andersson2011}. They do not derive a best-fitting $Y_{500}-M_{500}$ relation, however they compare their measurements with overlapping data from \cite{Andersson2011} and find consistent results.}
In Section~\ref{subsubsec:counts_LowThresh} we investigate a low detection threshold of $Y_{\mathrm{500, lim}} = 1.1 \times 10^{-5} \; E(z)^{-0.25}$~Mpc$^2$. Assuming the \cite{Andersson2011} $Y_{500}-M_{500}$ relation this corresponds to $\approx 5000$ clusters at $z < 2$ for a 2500 deg$^2$ survey, equivalent to the total number of predicted clusters for SPT-3G over the same survey area \citep{Benson2014}.

The cluster counts as a function of redshift corresponding to the high and low detection thresholds are shown in the left- and right-hand panels of Fig.~\ref{fig:counts}, respectively.

\subsubsection{X-ray mass bias in the \MACSIS\ relation}\label{subsubsec:counts_MACSIS}
\begin{figure*}
  \begin{center}
    {\includegraphics[width=0.497\textwidth]{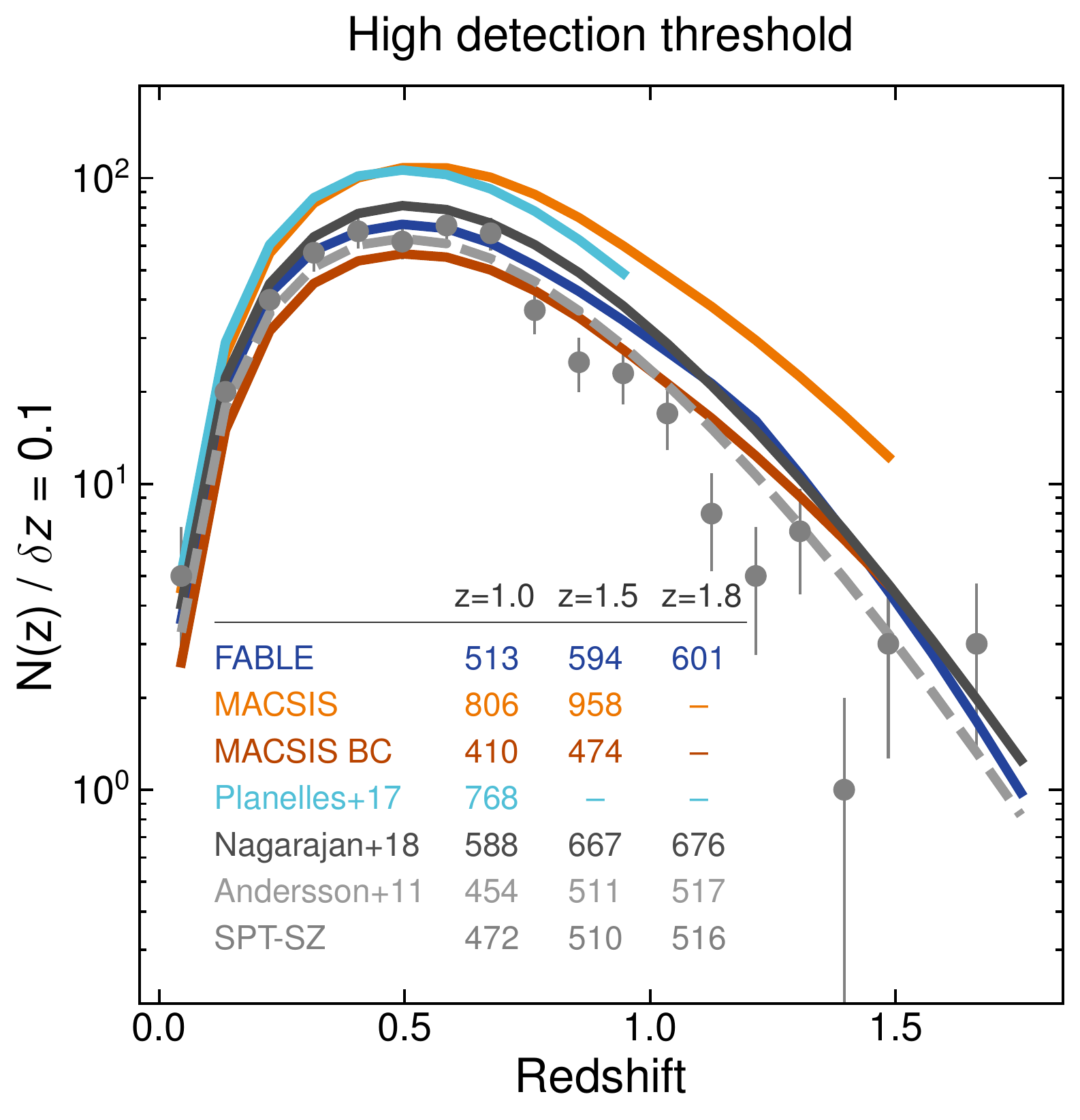}}
    {\includegraphics[width=0.497\textwidth]{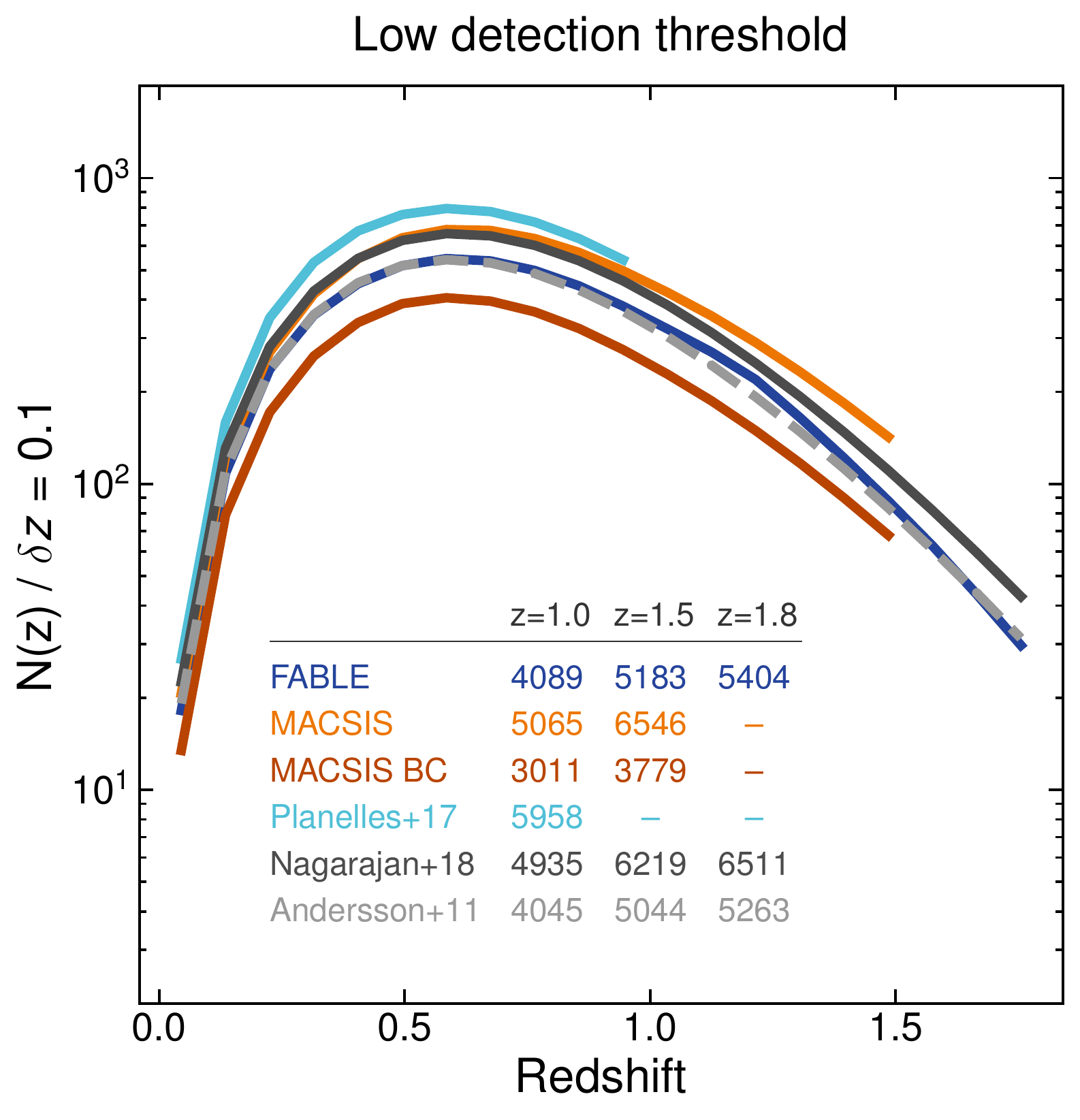}}
  \end{center}
\caption{The number of SZ-detectable haloes per redshift bin of width $\delta z = 0.1$ over 2500 deg$^2$ of the sky corresponding to different assumptions for the underlying $Y_{500}-M_{500}$ relation (solid and dashed curves) for two choices of the threshold flux: a high detection threshold, $Y_{\mathrm{500, lim}} \, {\mathrm{D_A^2}} = 3.75 \times 10^{-5} \; E(z)^{-0.25} \: {\mathrm{Mpc^2}}$ (left), and a low detection threshold, $Y_{\mathrm{500, lim}} \, {\mathrm{D_A^2}} = 1.1 \times 10^{-5} \; E(z)^{-0.25} \: {\mathrm{Mpc^2}}$ (right).
Curves correspond to the best-fitting $Y_{500}-M_{500}$ relations of \fable\ (dark blue), \MACSIS\ (orange), \protect\cite{Planelles2017} (light blue), \protect\cite{Nagarajan2018} (solid grey) and \protect\cite{Andersson2011} (dashed grey). The red curve shows the \MACSIS\ counts after accounting for an X-ray mass bias of $20$ per cent in their best-fitting $Y_{500}-M_{500}$ relation via the method described in Section~\ref{subsubsec:counts_MACSIS}.
Circles with error bars show the cluster counts of the 2500 deg$^2$ SPT-SZ survey using the updated redshifts from \protect\cite{Bocquet2018} in redshift bins of width $\delta z = 0.1$ (error bars equal the square-root of the number of clusters).
The inset table lists the total number of haloes between $z=0$ and $z = 1.0, 1.5$ and $1.8$ for each relation.
  }
    \label{fig:counts}
\end{figure*}

The predicted cluster counts corresponding to the \MACSIS\ $Y_{500}-M_{500}$ relation derived in B17 (orange lines in Fig.~\ref{fig:Y500-M500_vs_z}) are shown in orange in Fig.~\ref{fig:counts}. This relation is, however, based on X-ray hydrostatic masses, which \cite{Henson2017} show underestimate the true mass by $\sim 20$ per cent at $M_{500} \approx 10^{14} M_{\odot}$ and $\sim 35$ per cent at $M_{500} \approx 10^{15} M_{\odot}$. Equating the true mass from the halo mass function with the X-ray hydrostatic mass used in the \MACSIS\ relation therefore biases the predicted cluster counts high.
We estimate the impact of such a bias on the predicted counts by first converting the true mass from the halo mass function to an X-ray hydrostatic mass assuming a constant mass bias. We then retrieve the corresponding $Y_{500}$ from the \MACSIS\ relation and rescale the aperture to the true $r_{500}$ using the universal pressure profile \citep{Arnaud2010}. The relevant mass bias is that near the detection threshold, which we take to be $20$ per cent.
The counts calculated in this way are shown in red in Fig.~\ref{fig:counts} and are referred to as `bias-corrected' in the following.

\subsubsection{High SZ detection threshold}\label{subsubsec:counts_HighThresh}
For $Y_{\mathrm{500, lim}} = \, 3.75 \times 10^{-5} \; E(z)^{-0.25}$~Mpc$^2$ (left-hand panel of Fig.~\ref{fig:counts}) we find a wide spread in the predicted cluster counts as a function of redshift for the different $Y_{500}-M_{500}$ relations. Indeed, the predicted number of clusters below $z = 1$ varies from $\approx 400$ for \MACSIS\ (bias-corrected) and $\approx 500$ for \fable\ to $\approx 800$ for the \cite{Planelles2017} relation.
This is driven by small differences in the normalisation and slope of the $Y_{500}-M_{500}$ relations (see Fig.~\ref{fig:Y500-M500_vs_z}).
For example, \cite{Planelles2017} predict a relatively high normalisation for the $Y_{500}-M_{500}$ relation, which results in a relatively low mass threshold and a corresponding increase in the number of clusters. In other words, clusters in the \cite{Planelles2017} simulations possess a comparatively high SZ flux at fixed mass, which translates to a higher detection rate in an SZ-selected survey.
In the opposite sense, \fable\ predicts a comparatively low normalisation and thus relatively fewer clusters.
The difference in intrinsic scatter between relations is small enough that it has negligible effect on the relative cluster counts.

Fig.~\ref{fig:counts} clearly demonstrates the sensitivity of the predicted cluster counts to mass bias in the underlying mass-observable relation. Comparing the \MACSIS\ cluster counts with and without the mass bias correction described in Section~\ref{subsubsec:counts_MACSIS} implies that an X-ray mass bias of $20$ per cent can lead to a factor of two difference in the expected number of clusters in an SZ-selected survey. This reinforces the need for accurate mass calibration in future cluster surveys that attempt to constrain cosmology from cluster counts (see e.g. \citealt{Pratt2019} and references therein).
Yet we also point out that there remains a significant spread in the predicted cluster counts even after mass bias is taken into account, as demonstrated by the difference in cluster counts between \MACSIS, the \fable\ and \cite{Planelles2017} relations based on true masses and the \cite{Nagarajan2018} relation based on (relatively unbiased) weak lensing masses.

The shape of the cluster counts distribution with redshift is similar between the relations. This reflects the general lack of redshift evolution in the relations beyond that expected in the self-similar model, as evidenced by the flatness of the curves in Fig.~\ref{fig:Y500-M500_vs_z}.
\fable\ and \MACSIS\ predict a mild positive evolution in the normalisation relative to self-similarity, which lowers the mass threshold corresponding to $Y_{\mathrm{500, lim}}$ with increasing redshift thereby slightly boosting the number of clusters at high redshift.

All of the $Y_{500}-M_{500}$ relations predict a slightly larger number of clusters than the \cite{Andersson2011} relation. If the \cite{Andersson2011} relation accurately describes the $Y_{500}-M_{500}$ relation of the SPT-SZ sample then this suggests that cluster counts predicted from SPT-SZ data alone, such as the SPT-3G predictions, may be underestimated.
We also point out that the \cite{Andersson2011} relation is based on X-ray hydrostatic masses, which likely biases the predicted counts somewhat high.
On the other hand, the \cite{Andersson2011} relation may be biased compared with the full SPT-SZ sample since the \cite{Andersson2011} clusters are on average more massive. As such the best-fitting relation may not accurately describe the relationship between $Y_{500}$ and $M_{500}$ at masses close to the lower mass threshold of the SPT-SZ survey, which could have a significant impact on the predicted cluster counts given the steepness of the cluster mass function.

In the left-hand panel of Fig.~\ref{fig:counts} we also plot cluster counts from the 2500 deg$^2$ SPT-SZ survey binned by redshift \citep{Bocquet2018}.
We caution that comparison to these data is dependent on the value of $Y_{\mathrm{500, lim}}$, which we have chosen such that the \cite{Andersson2011} $Y_{500}-M_{500}$ relation yields approximately the same number of clusters as SPT-SZ at $z \leq 1.8$.
Therefore, comparing the predictions (curves) to the SPT-SZ counts (error bars) only describes the level of agreement with the \cite{Andersson2011} prediction (dashed line) and not the actual SPT-SZ data, at least in terms of their normalisation.

The shape of the \cite{Andersson2011} curve is a reasonable match to the SPT-SZ counts, however there is a slight underestimate (overestimate) at low (high) redshifts. This is also the case when using the $\Lambda\mathrm{CDM}$ cosmological constraints derived from the SPT-SZ cluster data in \cite{Bocquet2018}. There are a number of possible explanations for this discrepancy.
Firstly, the prediction implicitly assumes that all haloes with $Y_{500} \geq Y_{\mathrm{500, lim}}$ in a given redshift bin are detected. However, for a real SZ survey the completeness will depend on aspects of the observations such as the noise level and instrument resolution as well as the intrinsic scatter in the $\zeta-Y_{500}$ relation, which we do not take into account here.
Secondly, the discrepancy could be due to the choice of fitting function, which can vary due to properties of the underlying simulations and/or baryonic effects (see discussion in Section~\ref{subsubsec:calc_counts}).
Thirdly, the redshift trend of cluster counts is also fairly sensitive to the redshift evolution of the halo mass threshold.
For example, assuming that $Y_{\mathrm{500, lim}}$ is constant with redshift brings the \cite{Andersson2011} curve into better agreement with the SPT-SZ counts compared to the $E(z)^{-0.25}$ scaling, which suggests that the redshift dependence of the $\zeta-Y_{500}$ relation may be even weaker than the \cite{Andersson2011} estimate.
On the other hand, the halo mass threshold depends not only on the redshift evolution of $Y_{\mathrm{500, lim}}$ but also the redshift evolution of the normalisation of the $Y_{500}-M_{500}$ relation. In our analysis these choices are degenerate with one another as long as the slope of the $Y_{500}-M_{500}$ relation is independent of redshift, as it is in the self-similar expectation. Hence the discrepancy between the SPT-SZ counts and the \cite{Andersson2011}--based prediction could also be resolved if the normalisation of the $Y_{500}-M_{500}$ relation evolves less rapidly with redshift than the self-similar expectation of $E(z)^{-2/3}$. This is not supported by the simulation predictions however, which yield either a redshift-independent normalisation consistent with self-similarity as in \cite{Planelles2017} or, in the case of \fable\ and \MACSIS, a normalisation that evolves with redshift faster than self-similar.

\subsubsection{Low SZ detection threshold}\label{subsubsec:counts_LowThresh}
The right-hand panel of Fig.~\ref{fig:counts} corresponds to the low detection threshold ($Y_{\mathrm{500, lim}} = \, 1.1 \times 10^{-5} E(z)^{-0.25}$~Mpc$^2$).
There are some subtle changes in the relative counts distributions as compared with the high detection threshold due to the difference in the slopes of the relations (see left-hand panel of Fig.~\ref{fig:Y500-M500_vs_z}). For example, when switching from the high to the low detection threshold, the relatively steep slope of the \MACSIS\ relation leads to a smaller increase in the predicted cluster counts compared with the other relations, which have shallower slopes. Hence, the fractional offset between \MACSIS\ and the other relations is larger than for the high detection threshold.
A similar, although smaller, effect is found for \fable, which predicts a somewhat steeper slope than \cite{Planelles2017}, \cite{Nagarajan2018} and \cite{Andersson2011}, who find close to self-similar slopes.
These results demonstrate how relatively small changes in the slope of the best-fitting $Y_{500}-M_{500}$ relation can have a significant impact on the expected cluster counts, especially as we push towards lower masses with future surveys.

The mild positive evolution in the normalisation of the \fable\ and \MACSIS\ relations has a small but appreciable effect on the shape of the cluster counts distribution.
For example, the \MACSIS\ and \cite{Nagarajan2018} counts are similar in shape at $z \lesssim 0.5$, but the increase in normalisation with increasing redshift in the \MACSIS\ relation yields relatively more clusters at high redshift.
In the case of \fable, the slight increase in normalisation with increasing redshift relative to self-similarity is offset by the drop in intrinsic scatter with increasing redshift (right-hand panel of Fig.~\ref{fig:Y500-M500_vs_z}), yielding a cluster counts distribution very similar to that of the \cite{Andersson2011} relation, which we assume evolves self-similarly.
This example highlights the potential impact of a redshift-dependent intrinsic scatter in the mass--observable scaling relations as found in the \fable\ $Y_{500}-M_{500}$ relation and several of the X-ray scaling relations (see Section~\ref{subsec:evol}).

The total number of clusters varies considerably between the relations, ranging from $\sim 3000$ for \MACSIS\ (bias-corrected) and $\sim 4000$ for \fable\ and \cite{Andersson2011} to $\sim 6000$ for \cite{Planelles2017} at $z \leq 1.0$. Similarly large differences are also found at higher redshift.
This highlights the need for improvements in the measurement of the $Y_{500}-M_{500}$ relation, not only for predicting the outcomes of future SZ surveys but also for taking advantage of such surveys for probing cosmology. Indeed, cosmological constraints from current surveys are primarily limited by uncertainties in the calibration of the SZ signal--total mass relation (e.g. \citealt{Sehgal2011, VonderLinden2014, Bocquet2015, PlanckXXIV2015}).
Furthermore, simulations such as \fable\ and \MACSIS\ hint towards a mild redshift evolution in the relations beyond the self-similar expectation that may prove important to the next generation of SZ experiments, such as CMB-S4, which will find clusters out to the highest redshifts where they exist.

\section{Conclusions}\label{sec:conclusions}

We have studied the redshift evolution of the X-ray and SZ scaling relations in the \fable\ suite of cosmological hydrodynamical simulations.
The simulations are performed using the {\sc arepo} moving-mesh code with a set of physical models based on those of Illustris \citep{Genel2014, Vogelsberger2014, Sijacki2015} but with improved modelling of supernovae and AGN feedback (see Paper~I).
For the present study we have greatly expanded our sample of simulated galaxy groups and clusters by performing an additional 21 zoom-in simulations in addition to the original 6 presented in Paper~I. Our extended suite of 27 high-resolution zoom-in simulations spans a wide halo mass range, from low-mass groups ($\sim 3 \times 10^{13} M_{\odot}$) to massive clusters ($\sim 3 \times 10^{15} M_{\odot}$).

Using our expanded sample we have investigated six scaling relations: gas mass--total mass, total mass--temperature, $Y_{\mathrm{X}}$--total mass, X-ray luminosity--total mass, X-ray luminosity--temperature and the SZ flux--total mass relation. First we examined the reliability of our model by comparing the predicted scaling relations to observations at intermediate to high redshift ($z \lesssim 1$; Sections~\ref{subsec:comp} and \ref{subsec:SZ_comp}), in extension to the $z=0$ comparison presented in Paper~I. Subsequently we investigated the redshift evolution of the slope, normalisation and intrinsic scatter of the scaling relations out to $z \approx 2$ (Sections~\ref{subsec:evol} and \ref{subsubsec:Y500-M500_evol}). With comparison to other recent simulation predictions we investigated how simulation predictions for the evolution of the X-ray scaling relations can vary given different choices for the physical modelling.
For all of the relations studied we find significant deviations from the simple self-similar expectation (Fig.~\ref{fig:Mgas-M}-\ref{fig:evol} and Fig.~\ref{fig:Y500-M500_vs_z}).
These predictions relate directly to the outcomes of future experiments such as CMB-S4 \citep{Abazajian2016} and the \textit{Athena} X-ray observatory \citep{Nandra2013}. In particular, our mock X-ray analysis mimics observations with the planned \textit{Athena} X-IFU instrument, which we expect to find significant deviations from self-similarity in the evolution of the X-ray scaling relations out to $z \sim 2$.
This could have important implications for cluster cosmology, particularly as ongoing and future cluster surveys push to higher redshift. We illustrate this point in Section~\ref{subsec:counts} by comparing predicted cluster counts for an SZ-selected survey for different assumptions about the underlying SZ flux--total mass relation.
We provide a summary of our main results below.

\begin{itemize}
\item At $z=0.4$ and $z=1$, the gas mass--total mass, $Y_{\mathrm{X}}$--total mass and X-ray luminosity--total mass relations are in good agreement with observational constraints based on X-ray hydrostatic mass estimates (Fig.~\ref{fig:xray_z04} and \ref{fig:xray_z1}).
On the other hand, the X-ray luminosity--spectroscopic temperature relation lies on the upper end of the observed scatter.
Comparison to observations based on weak lensing masses as opposed to X-ray hydrostatic masses suggests that \fable\ clusters possess slightly overestimated gas masses and X-ray luminosities at fixed total mass, although we find no evidence for a significant redshift dependence in the offset.

\item We find that the slopes of the relations are in good agreement with the majority of observations at low redshift. Furthermore, the slopes deviate significantly from the self-similar predictions in all cases and at all redshifts ($0 < z < 1.8$; left-hand panels of Fig.~\ref{fig:Mgas-M}-\ref{fig:evol} and Fig.~\ref{fig:Y500-M500_vs_z}). The same qualitative result is found in the recent studies of B17, for the \MACSIS\ simulations, and T18, for a set of simulations with AGN feedback. The predicted slopes are consistent with the \MACSIS\ and T18 predictions at $z \lesssim 2$, with the exception of the gas mass--total mass and $Y_{\mathrm{X}}$--total mass relations for which T18 predict somewhat shallower slopes.

\item The scaling relations of gas mass, $Y_{\mathrm{X}}$, SZ flux and X-ray luminosity with total mass are all steeper than the self-similar expectation. This is largely due to the effects of non-gravitational physics, such as star formation and feedback. In particular, AGN feedback can expel gas with greater efficiency from lower mass haloes due to their shallower potentials wells. The fact that T18 predict a shallower gas mass--total mass slope than \fable\ and \MACSIS\ (Fig.~\ref{fig:Mgas-M}) suggests that gas expulsion via AGN feedback is less efficient in their simulations, at least at $z \lesssim 2$. This may stem from differences in the frequency of thermal energy injection by AGN feedback, which is continuous in the T18 model but operates on a duty cycle in \fable\ and \MACSIS.

\item The total mass--temperature slope based on the spectroscopic temperature is only marginally steeper than self-similar due to spectroscopic temperature biases at the low- and high-mass end of the relation. On the other hand, the mass-weighted temperature relation is significantly steeper than self-similar, in good agreement with \MACSIS\ and T18 (Fig.~\ref{fig:M-T}). This is due to the combined action of radiative cooling, star formation and AGN feedback, which raises the temperature of the ICM and acts with greater efficiency in lower mass haloes.

\item We find no strong evidence for a redshift evolution in the slopes of the relations, with the exception of the X-ray luminosity-based relations for which we see a mild steepening with increasing redshift (Fig.~\ref{fig:evol}).
We also find a mild increase in slope with decreasing redshift at $z \lesssim 0.6$ for the gas mass, $Y_{\mathrm{X}}$ and X-ray luminosity-based relations, which may be driven by the increasing prevalence of radio-mode AGN feedback in our simulations.

\item The normalisation of the scaling relations evolves positively with respect to self-similarity in all cases (middle panels of Fig.~\ref{fig:Mgas-M}-\ref{fig:evol} and Fig.~\ref{fig:Y500-M500_vs_z}).
The positive evolution of the gas mass, X-ray luminosity and SZ flux-based relations is because haloes of a given mass are denser at higher redshift, which raises the energy required to expel gas beyond $r_{500}$.
  The evolution of the total mass--temperature relation is due to increasing non-thermal pressure support from kinetic motions in the ICM with increasing redshift, as found in \cite{LeBrun2017} for the cosmo-OWLS simulations. Contrary to the results of T18, this is not sufficient to offset the evolution in the gas mass at fixed mass, and so the $Y_{\mathrm{X}}$--total mass relation evolves positively with respect to self-similarity.
The spectroscopic temperature bias increases with increasing redshift due to redshifting of the low-energy X-ray emission below the X-ray bandpass. This largely offsets the otherwise positive evolution of the total mass--temperature and X-ray luminosity--temperature relations so that they appear to evolve self-similarly when based on spectroscopic temperatures as opposed to mass-weighted temperatures.

\item The intrinsic scatter is in good agreement with the majority of observational and theoretical constraints at low redshift, although in some cases it is biased high by the scattering of low-mass objects in our comparatively low-mass sample. \fable\ and \MACSIS\ predict a larger scatter in the gas mass at fixed total mass compared with T18, which may be related to differences in the duty cycle of AGN feedback. The scatter tends to decrease with increasing redshift (right-hand panels of Fig.~\ref{fig:Mgas-M}-\ref{fig:evol} and Fig.~\ref{fig:Y500-M500_vs_z}), consistent with \cite{LeBrun2017} for a large, volume-limited sample. This is likely a combination of the decreasing efficiency of AGN feedback with redshift and the increased influence of radio-mode AGN feedback at $z \lesssim 1$ in our model.

\item The scaling relation between SZ flux and total mass is in good agreement with \textit{Planck} and SPT-SZ clusters out to $z \sim 0.8$ and $z \sim 1.2$, respectively (Fig.~\ref{fig:Y-M_Planck} and \ref{fig:Y-M_SPT}).
The $Y_{500}-M_{500}$ relation at $z=0$ also agrees well with local cluster observations (Fig.~\ref{fig:Y500-M500_z0}).
At group scales, the simulation predictions favour the relation of \cite{Wang2016} for locally brightest galaxies (recalibrated from \citealt{Planck2013XI}) over that of \cite{Lim2018} for local galaxy groups.

\item We show that the number and redshift distribution of clusters expected from an SZ-selected survey depends sensitively on the assumed slope, normalisation and redshift evolution of the $Y_{500}-M_{500}$ relation (Fig.~\ref{fig:counts}).
  Relatively small differences in the normalisation ($\sim 0.1$ dex) can result in predicted cluster counts that differ by more than a thousand for an SZ-selected survey similar to SPT-3G (with an expected total of $\sim 5000$ clusters).
  Furthermore, the steeper than self-similar $Y_{500}-M_{500}$ relations of \fable\ and \MACSIS\ yield relatively fewer clusters at low redshift compared with the close to self-similar prediction of \cite{Planelles2017}, whereas the mild positive evolution in their normalisations leads to a relatively higher number of high-redshift clusters.
  This has important consequences for SZ cluster cosmology, which depends on our ability to link observed SZ counts to the underlying cluster mass function.

\end{itemize}

As ongoing and future cluster surveys detect lower mass systems at higher redshifts, our ability to leverage this new data to probe cosmology will depend on our understanding of the mass and redshift dependence of the observable--mass relations.
With \fable\ we predict significant deviations from the self-similar expectation in terms of the slope of the relations and the redshift evolution of their normalisation.
Fortunately for observational studies, the assumption of a redshift-independent slope seems to be robust for the relations examined here.
On the other hand, none of the relations we have examined evolves self-similarly in normalisation, including commonly-used mass proxies such as the $Y_{\mathrm{X}}$ parameter.
This could have a significant impact on the expected number of high-redshift clusters from future surveys (see e.g. Section~\ref{subsec:counts}) and, relatedly, our ability to constrain cosmology using cluster abundances.
Furthermore, the intrinsic scatter of the scaling relations tends to decrease with increasing redshift (and increasing mass; see Appendix~\ref{A:mass_dependence}), which implies the need for a more complex parametrization of the intrinsic scatter than the mass- and redshift-independent scatter that is assumed in most current observational studies.

Other avenues for future research with \fable\ include the covariance of cluster observables and the impact of ICM structure on observable properties.
In addition, we plan to utilise the relatively high resolution of the \fable\ simulations to study the galactic component of clusters, including the growth of brightest cluster galaxies and the origin of stellar mass build-up in clusters over cosmic time.

\section*{Acknowledgements}
We are grateful to Volker Springel for making the {\sc arepo} moving-mesh code available to us and to the Illustris collaboration for their development of the Illustris galaxy formation model, which provided an excellent starting point for the development of the \fable\ project.
NAH is supported by the Science and Technology Facilities Council (STFC).
EP acknowledges support by the Kavli Foundation.
DS acknowledges support by the STFC and the ERC Starting Grant 638707 ``Black holes and their host galaxies: co-evolution across cosmic time''.
This work made use of the following DiRAC facilities (\href{www.dirac.ac.uk}{www.dirac.ac.uk}):
the Data Analytic system at the University of Cambridge [funded by a BIS National E-infrastructure capital grant (ST/K001590/1), STFC capital grants ST/H008861/1 and ST/H00887X/1, and STFC DiRAC Operations grant ST/K00333X/1], the Cambridge Service for Data Driven Discovery (CSD3) [the DiRAC component funded by BEIS capital funding via STFC capital grants ST/P002307/1 and ST/R002452/1 and STFC operations grant ST/R00689X/1] and the COSMA Data Centric system at Durham University [funded by BEIS capital funding via STFC capital grants ST/P002293/1, ST/R002371/1 and ST/S002502/1, Durham University and STFC operations grant ST/R000832/1]. DiRAC is part of the National E-Infrastructure.

\bibliographystyle{mnras}
\bibliography{Paper2}

\clearpage
\appendix
\section{Self-similar scaling}\label{A:SS}
In the self-similar scenario, galaxy clusters are simply scaled versions of each other and their properties depend only on their total mass \citep{White1978, Kaiser1986}.
Despite its simplicity, the self-similar model provides predictions for the functional form of correlations between cluster properties that are remarkably similar to observational results for massive clusters (e.g. \citealt{Mantz2010a, Maughan2012, Mantz2016a}).
Self-similarity therefore provides a useful baseline for studying cluster scaling relations, in particular their evolution.

If we define the total mass of a cluster, $M_{\Delta}$, to be the mass enclosing a spherical region of radius $r_{\Delta}$ with an average density $\Delta$ times the critical density of the Universe, $\rho_{\mathrm{crit}}$, such that
\begin{equation}\label{eq:SO_mass}
  M_{\Delta} = \frac{4 \pi}{3} \Delta \, \rho_{\mathrm{crit}}(z) \,  r_{\Delta}^3 ,
\end{equation}
then, under the assumption of self-similarity, the redshift evolution of the scaling relations is due only to the evolving critical density of the Universe.
The critical density at redshift $z$ is defined as
\begin{equation}\label{eq:rho_crit}
\rho_{\mathrm{crit}} \equiv \frac{3H(z)^2}{8 \pi G} = E(z)^2 \frac{3H_0^2}{8 \pi G}
\end{equation}
where $H(z)$ is the Hubble parameter and
\begin{equation}
E(z) \equiv \frac{H(z)}{H_0} = \sqrt{\Omega_m (1+z)^3 + \Omega_{\Lambda}}
\end{equation}
is the Hubble parameter normalised to its present-day value, $H_0$, in a $\mathrm{\Lambda CDM}$ Universe.
Equations \ref{eq:SO_mass} and \ref{eq:rho_crit} lead to the useful scaling between cluster size and mass
\begin{equation}\label{eq:SO_rad}
  r_{\Delta} \propto M_{\Delta}^{1/3} E(z)^{-2/3}.
\end{equation}

Under the assumption that the intracluster medium is in hydrostatic equilibrium within the cluster gravitational potential, the total cluster mass within $r_{\Delta}$ can be expressed in terms of the density and temperature profiles, $\rho_g(r)$ and $T(r)$, as
\begin{equation}\label{eq:hydro_eq}
  M_{\Delta} = \frac{k}{G \mu m_p} r_{\Delta} T(r_{\Delta})  \left. \left[ \frac{d\,\mathrm{ln}\,\rho_g(r)}{d\,\mathrm{ln}\,r} + \frac{d\,\mathrm{ln}\,T(r)}{d\,\mathrm{ln}\,r} \right] \right\rvert_{r = r_{\Delta}} ,
\end{equation}
where $k$ is the Boltzmann constant, $\mu$ is the mean molecular weight and $m_p$ is the proton mass. (e.g. \citealt{Sarazin1986}).
If we assume that the density and temperature profiles themselves are self-similar, then the logarithmic slopes of the profiles are independent of mass and $T(r_{\Delta})$ is proportional to $T_{\Delta}$, the average temperature within $r_{\Delta}$ \citep{Kravtsov2012}. The average temperature is therefore related to the total mass via
\begin{equation}\label{eq:SS_temp}
  T_{\Delta} \propto \frac{M_{\Delta}}{r_{\Delta}} \propto E(z)^{2/3} M_{\Delta}^{2/3} ,
\end{equation}
where $r_{\Delta}$ has been substituted from equation~\ref{eq:SO_rad}.

Assuming that the bolometric X-ray emission of a cluster is dominated by thermal bremsstrahlung, the bolometric X-ray luminosity can be written as
\begin{equation}
  L_{\Delta}^{\mathrm{bol}} \propto \rho_g^2 \, r_{\Delta}^3 \, \Lambda(T) \propto \frac{M^2_{\mathrm{gas, \Delta}}}{r_{\Delta}^3} \, T^{1/2}
\end{equation}
where $M_{\mathrm{gas, \Delta}} \equiv \rho_g r_{\Delta}^3$ is the total gas mass and $\Lambda(T) \propto T^{1/2}$ is the cooling function for bolometric emission (e.g. \citealt{Sarazin1986}).
Given that the gas mass is the integral of the gas density profile, which we have assumed to scale self-similarly, the gas mass can be expressed as
\begin{equation}\label{eq:SS_gasmass}
  M_{\mathrm{gas, \Delta}} \propto M_{\Delta} .
\end{equation}
The bolometric X-ray luminosity is thus related to mass and temperature via
\begin{equation}
  L_{\Delta}^{\mathrm{bol}} \propto \frac{M_{\Delta}^2}{r_{\Delta}^3} \, T_{\Delta}^{1/2} \propto E(z)^2 \, M_{\Delta} \, T_{\Delta}^{1/2}
\end{equation}
where $r_{\Delta}$ has been substituted from equation~\ref{eq:SO_rad}.
Substituting for temperature or mass from equation~\ref{eq:SS_temp} gives the self-similar scaling relations
\begin{equation}
  L_{\Delta}^{\mathrm{bol}} \propto E(z)^{7/3} M_{\Delta}^{4/3}
\end{equation}
and
\begin{equation}
  L_{\Delta}^{\mathrm{bol}} \propto E(z) \, T_{\Delta}^2.
\end{equation}

Lastly, we can relate the integrated SZ signal, $Y_{\mathrm{SZ, \Delta}}$, and its X-ray analogue, $Y_{\mathrm{X}}$, to the gas mass and temperature as $Y_{\mathrm{SZ, \Delta}} \propto Y_{\mathrm{X, \Delta}} \equiv M_{\mathrm{gas, \Delta}} \, T_{\Delta}$. Substituting for temperature from equation~\ref{eq:SS_temp} and relating the gas mass to the total mass with equation~\ref{eq:SS_gasmass} gives the self-similar relations
\begin{equation}
 Y_{\mathrm{SZ, \Delta}} \propto Y_{\mathrm{X, \Delta}} \propto E(z)^{2/3} \, M_{\Delta}^{5/3}.
\end{equation}

\section{Choice of lower mass threshold}
The sample used to derive our best-fitting X-ray and SZ scaling relations consists of haloes above an evolving mass threshold of $M_{500} > 3 \times 10^{13} E(z)^{-0.5} M_{\odot}$ (Section~\ref{subsec:sample}). In this appendix we investigate the sensitivity of the best-fitting parameters to the choice of mass threshold. We discuss the effect of varying the size of the mass threshold in \ref{A:mass_dependence} while in \ref{A:mass_evol_dependence} we vary its redshift dependence.

\begin{figure*}
  \begin{center}
  {\includegraphics[width=\factorthree\textwidth]{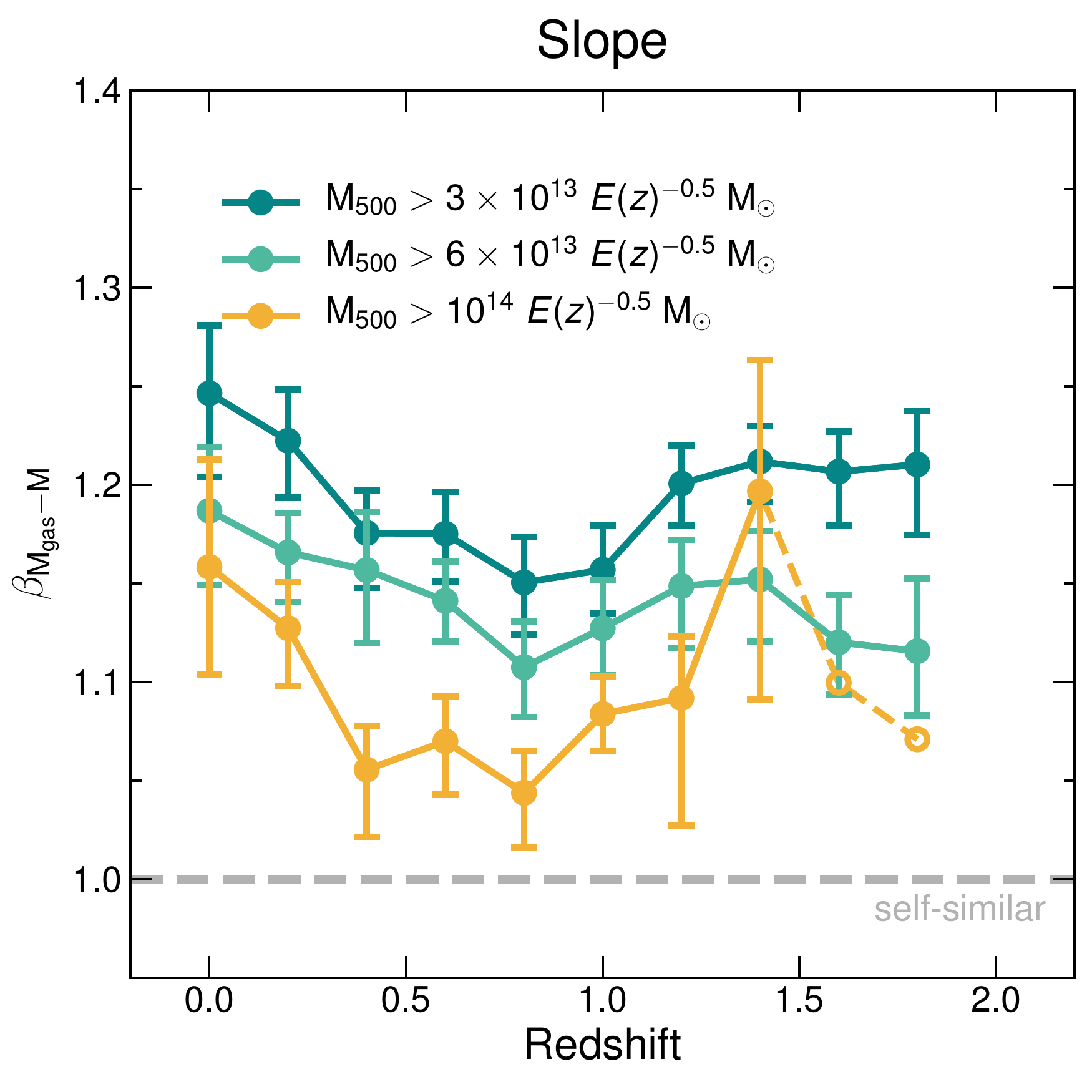}}
  {\includegraphics[width=\factorthree\textwidth]{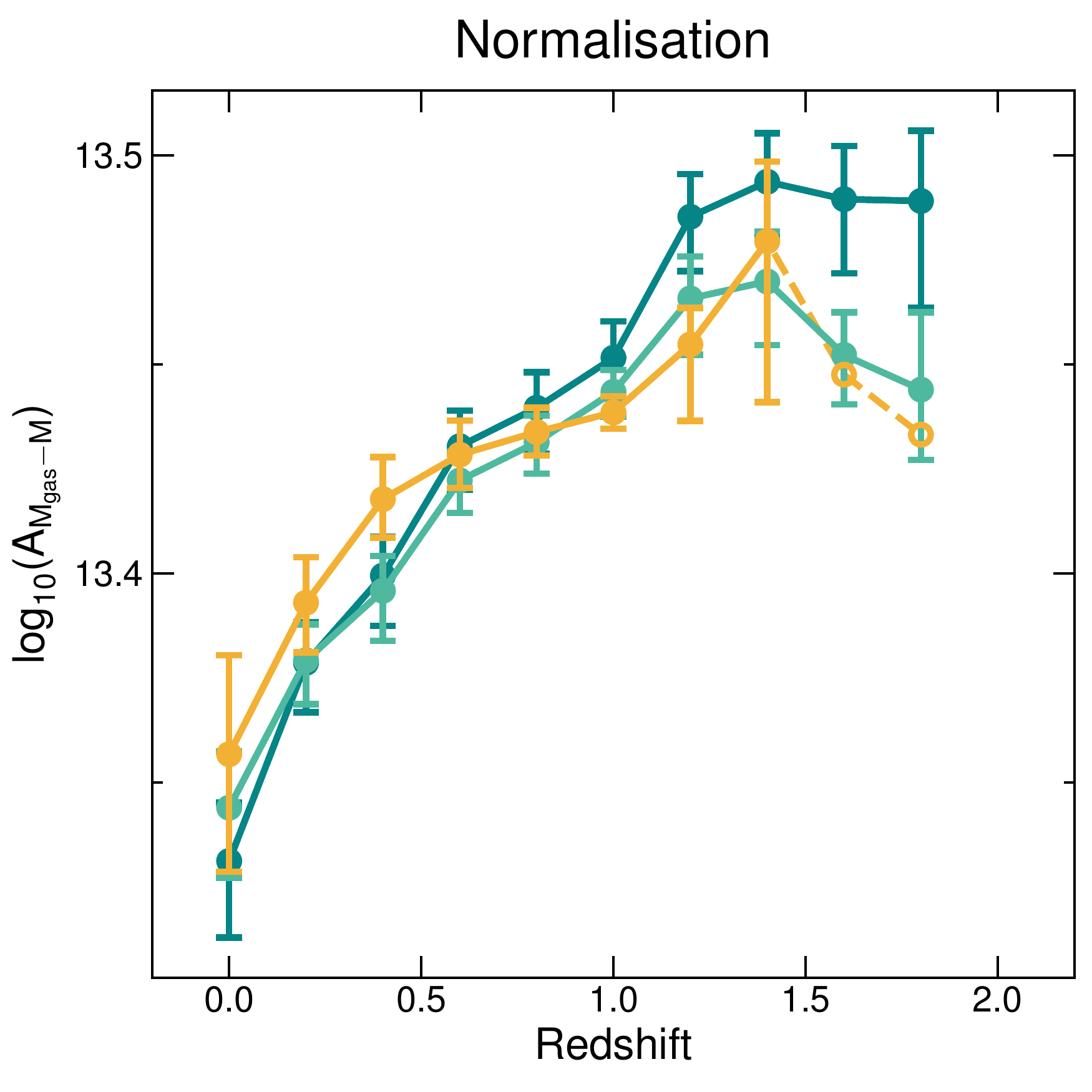}}
  {\includegraphics[width=\factorthree\textwidth]{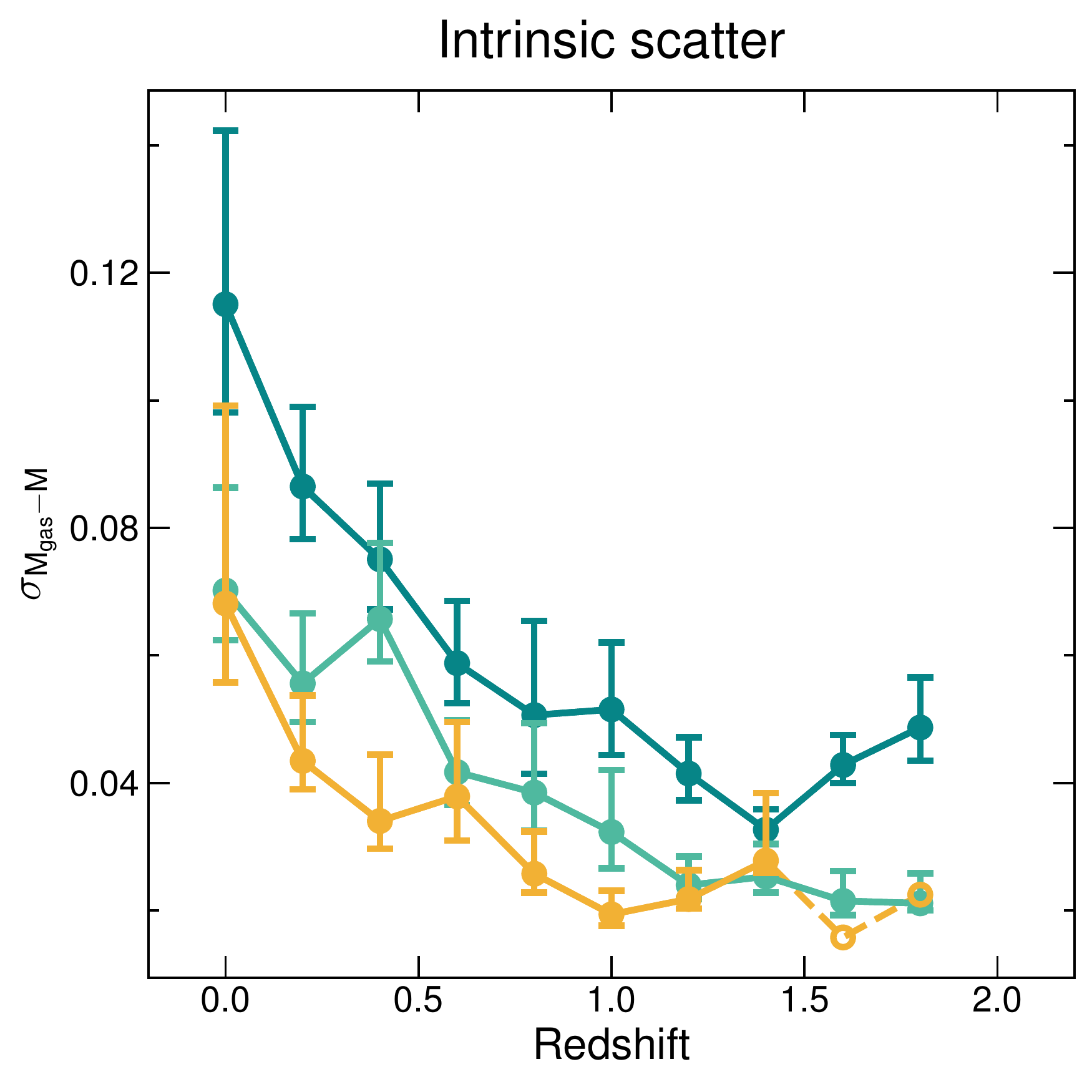}} \\
  {\includegraphics[width=\factorthree\textwidth]{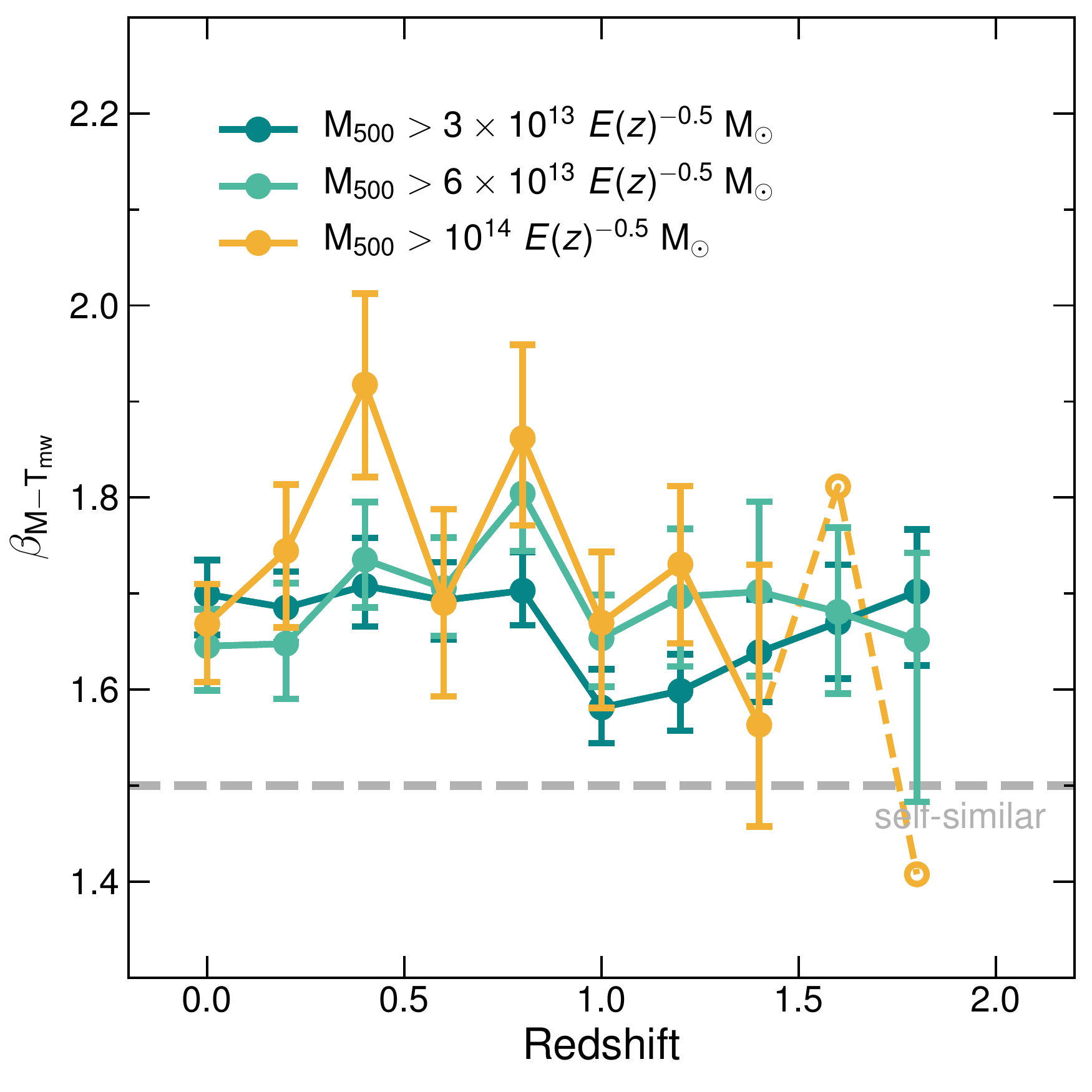}}
  {\includegraphics[width=\factorthree\textwidth]{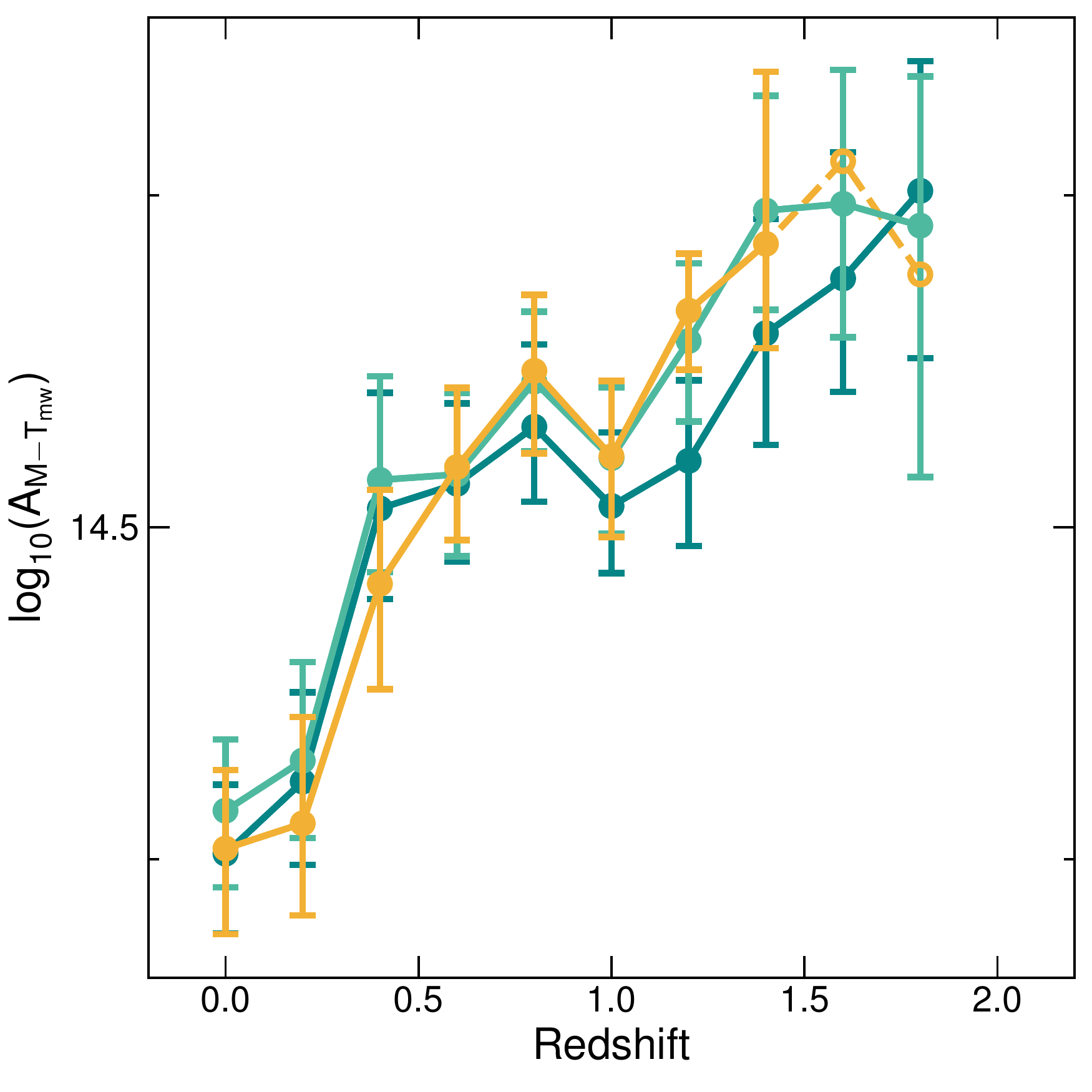}}
  {\includegraphics[width=\factorthree\textwidth]{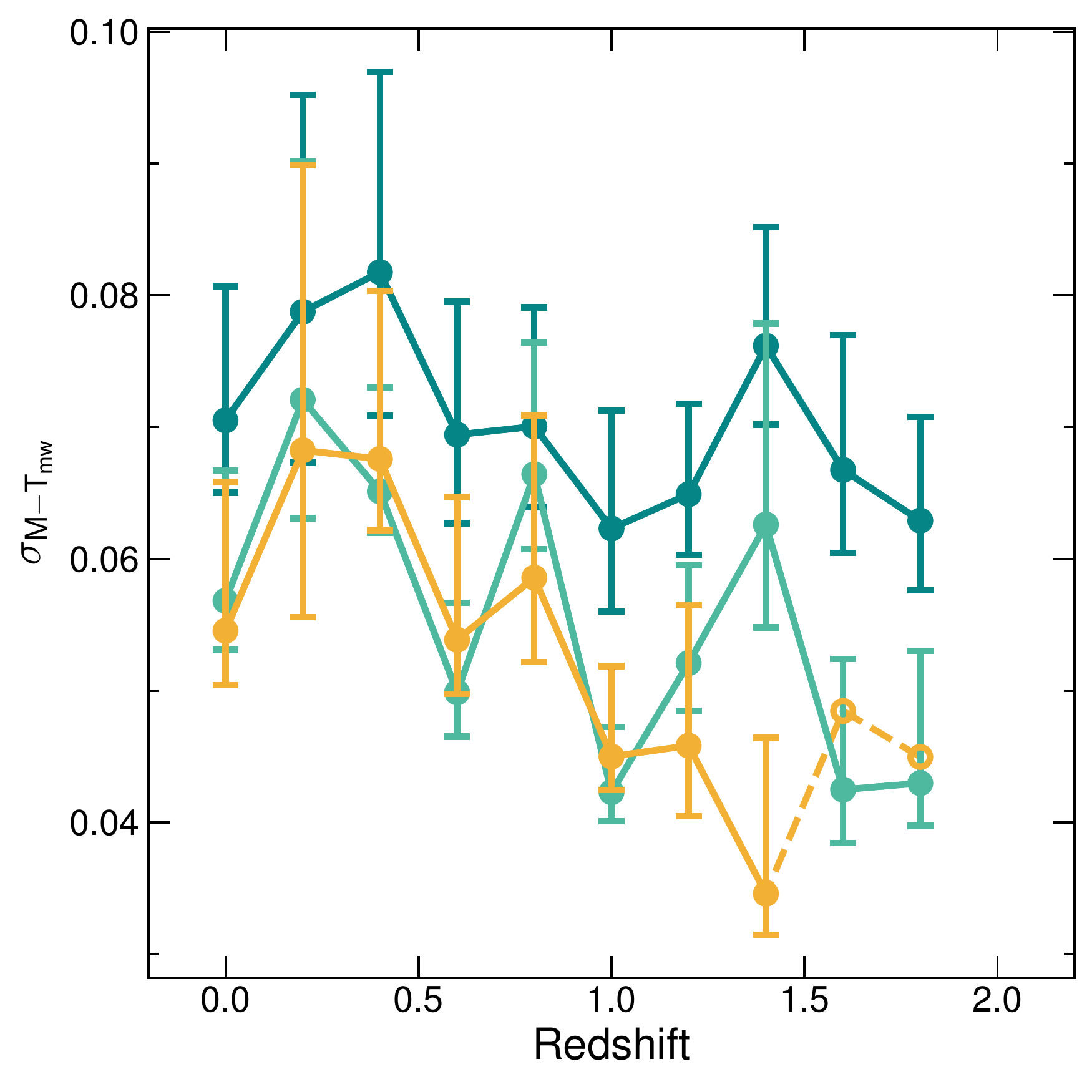}} \\
  \end{center}
  \caption{The redshift evolution of the gas mass--total mass (top row) and total mass--mass-weighted temperature (bottom row) relations and their dependence on the lower mass threshold of the sample. The panels show, from left to right, the best-fitting slope, normalisation and intrinsic scatter as a function of redshift. Different colours correspond to different choices for the lower mass threshold, as indicated in the legend. Error bars correspond to the 68 per cent confidence intervals estimated from bootstrap resampling. Open symbols with dashed lines indicate redshift bins containing fewer than ten objects. Horizontal dashed lines show the self-similar slope.
  }
    \label{fig:mass_dependence}
\end{figure*}

\subsection{Size of the lower mass threshold}\label{A:mass_dependence}
Fig.~\ref{fig:mass_dependence} shows the best-fitting parameters as a function of redshift for the gas mass--total mass relation and the total mass--(mass-weighted) temperature relation for different choices of the mass threshold. We discuss the mass dependence of the other scaling relations below.

We demonstrate three choices for the lower mass threshold: the fiducial limit used in the main body of this paper, $3 \times 10^{13} \, E(z)^{-0.5} \, M_{\odot}$, a high mass threshold of $10^{14} \, E(z)^{-0.5} \, M_{\odot}$ and an intermediate threshold of $6 \times 10^{13} \, E(z)^{-0.5} \, M_{\odot}$.
The highest mass threshold corresponds to the same mass limit as the \MACSIS\ \citep{MACSIS} and \cite{Truong2018} (hereafter T18) samples at $z=0$, which is important for the comparisons presented in the main body of this paper. Note however that the mass limits diverge at higher redshift as \MACSIS\ use a redshift-independent mass threshold while T18 scale it by $E(z)^{-1}$. The T18 mass threshold therefore drops to $5.6 \times 10^{13} M_{\odot}$ and $3.3 \times 10^{13} M_{\odot}$ at $z=1$ and $z=2$, respectively, while our highest mass threshold corresponds to $7.5 \times 10^{13} M_{\odot}$ and $5.8 \times 10^{13} M_{\odot}$ at these redshifts.
In Fig.~\ref{fig:mass_dependence} we highlight redshift bins containing fewer than ten objects as open symbols, at which point we can not reliably measure the best-fitting relation.

As the mass threshold increases, the gas mass--total mass relation becomes shallower and its normalisation evolves somewhat more slowly. This is understandable since AGN feedback can expel a larger gas mass fraction from lower mass haloes due to their shallower gravitational potential wells. Similarly, the intrinsic scatter decreases with increasing mass, which can also be attributed to the decreased effectiveness of AGN feedback at the high mass end.
We find no significant change in the redshift evolution of the slope or intrinsic scatter for the mass ranges considered here.

The slope and normalisation of the total mass--(mass-weighted) temperature relation increase slightly towards higher masses, although their redshift evolution is not significantly affected. The difference is small compared with the uncertainties but may reflect an increase in the mean entropy of the ICM due to AGN feedback, particularly in less dense, lower mass haloes. Indeed, we find that the mass-weighted temperature at $M_{500} \sim 3 \times 10^{13} M_{\odot}$ is somewhat lower ($\sim 0.05$ dex) in our full volume simulation without AGN feedback.
Alternatively, the mass dependence may reflect an increasing contribution from non-thermal pressure support in more massive clusters at fixed redshift, as found in \cite{MACSIS}.
The intrinsic scatter is somewhat lower in the higher mass samples, which may reflect the reduced impact of AGN feedback. The scatter may also be biased low due to the decreased sample size, although \cite{LeBrun2017} find a similar mass dependence using a large, volume-limited sample.

In contrast, the total mass--temperature relation based on the spectroscopic temperature is fairly insensitive to the mass threshold (not shown). This is because, as discussed in detail in Section~\ref{subsubsec:M-T}, the spectroscopic temperature is biased high in massive clusters ($T \gtrsim 5$ keV) due to hot, dense gas in the cluster core. The spectroscopic temperature bias introduces additional scatter to the total mass--temperature relation at the high mass end so that the intrinsic scatter is insensitive to the lower mass threshold.

The $Y_{\mathrm{X}}$--total mass relation has a steeper slope and increased intrinsic scatter towards lower masses similar to the gas mass--total mass relation. On the other hand, the normalisation is fairly insensitive to the mass threshold and the redshift evolution of the best-fitting parameters is unchanged.

The X-ray luminosity--total mass relation is insensitive to the mass threshold at $z < 1$, although at $z \gtrsim 1$ the slope and intrinsic scatter are slightly reduced in higher mass samples.
For the high mass sample ($M_{500} > 10^{14} \, E(z)^{-0.5} \, M_{\odot}$) the redshift evolution of the slope is consistent with zero while for lower mass samples it increases mildly with increasing redshift.
The X-ray luminosity--temperature relation shares the same mass dependencies as the X-ray luminosity--total mass relation, although the intrinsic scatter is slightly more sensitive to the mass of the sample due to the added mass dependence of the intrinsic scatter in temperature at fixed total mass (bottom-right panel of Fig.~\ref{fig:mass_dependence}).

The SZ flux--total mass relation has a very similar mass dependence to the gas mass--total mass relation. That is, the slope, normalisation and intrinsic scatter are slightly reduced in a higher mass sample. For the most massive sample the slope of the $Y_{500}-M_{500}$ relation is consistent with the self-similar value and the increase in normalisation between $z=0$ and $z=1$ relative to self-similarity is reduced from $\sim 17$ per cent to $\sim 7$ per cent compared with the fiducial sample.

\subsection{Redshift dependence of the lower mass threshold}\label{A:mass_evol_dependence}
\begin{figure}
  \includegraphics[width=\columnwidth]{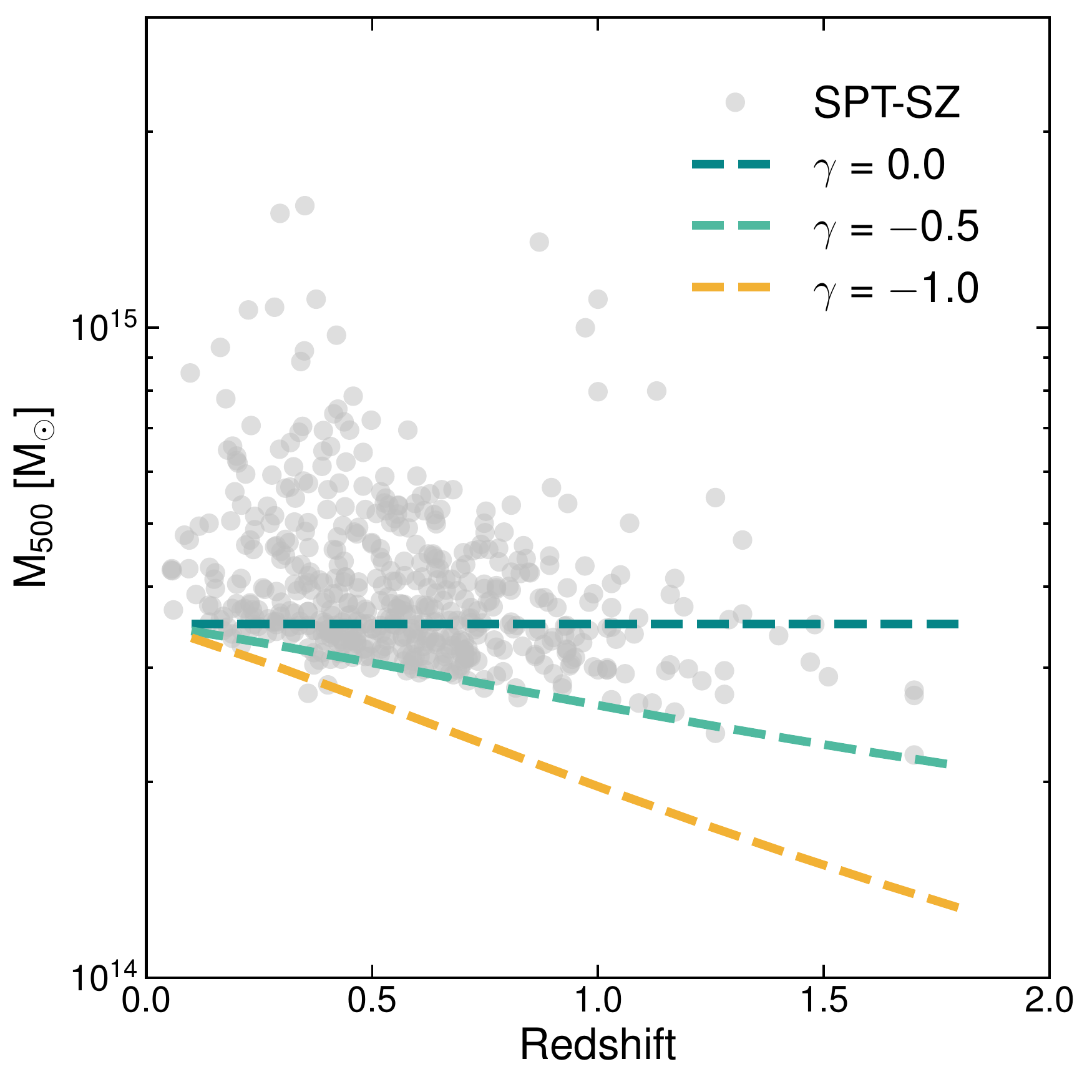}
  \caption{Total mass as a function of redshift for clusters detected in the SPT-SZ 2500 deg$^2$ survey with mass and redshift estimates from \protect\cite{Bocquet2018} (grey circles). Coloured lines indicate different choices for the $E(z)$ dependence of the lower mass limit, $M_{500}\propto E(z)^\gamma$. These are arbitrarily normalised to $M_{500} = 3.5 \times 10^{14} M_{\odot}$ at $z=0.1$.
  }
    \label{fig:SPT_mass_vs_z}
\end{figure}

Our fiducial choice for the redshift scaling of the lower mass limit is $E(z)^{-0.5}$, which is intended to mimic an SZ-selected cluster survey. We demonstrate this using results from the SPT-SZ 2500 deg$^2$ survey \citep{Bleem2015}.
Fig.~\ref{fig:SPT_mass_vs_z} plots total mass as a function of redshift for clusters in the SPT-SZ 2500 deg$^2$ survey using the latest redshift measurements and weak lensing-calibrated mass estimates from \cite{Bocquet2018}. From this figure we can see that the lower mass threshold of the sample varies with redshift as $\sim E(z)^{-0.5}$.
This scaling is consistent with the cosmological analysis of SPT-SZ data presented in \cite{Haan2016} as described in the following. The SPT cluster sample was defined as those detections above a given threshold value for the so-called unbiased detection significance, $\zeta$. In the notation of \cite{Haan2016}, the scaling relation between $\zeta$ and $M_{500}$ implies that, at fixed $\zeta$, the corresponding halo mass varies as $E(z)^{-\mathrm{C_{SZ}/B_{SZ}}}$, where $\mathrm{C_{SZ}}$ and $\mathrm{B_{SZ}}$ are parameters of the scaling relation. The best-fitting values of $\mathrm{C_{SZ}}$ and $\mathrm{B_{SZ}}$ in \cite{Haan2016} depend on the external data sets used in the analyses, however the ratio $-\mathrm{C_{SZ}/B_{SZ}}$ ranges from $-0.33$ to $-0.52$, consistent with our fiducial choice of $\gamma = -0.5$.

\begin{figure*}
  \begin{center}
  {\includegraphics[width=\factorthree\textwidth]{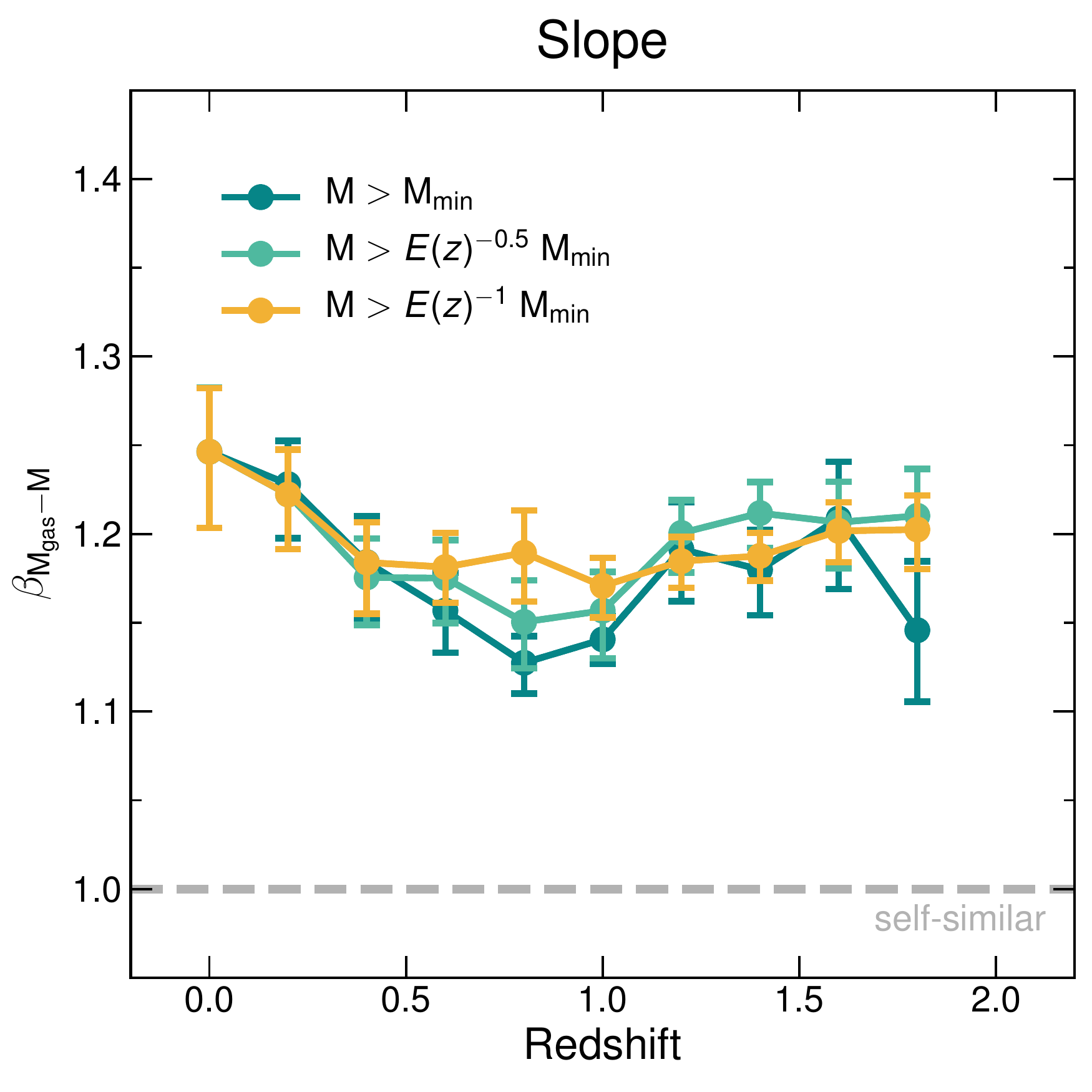}}
  {\includegraphics[width=\factorthree\textwidth]{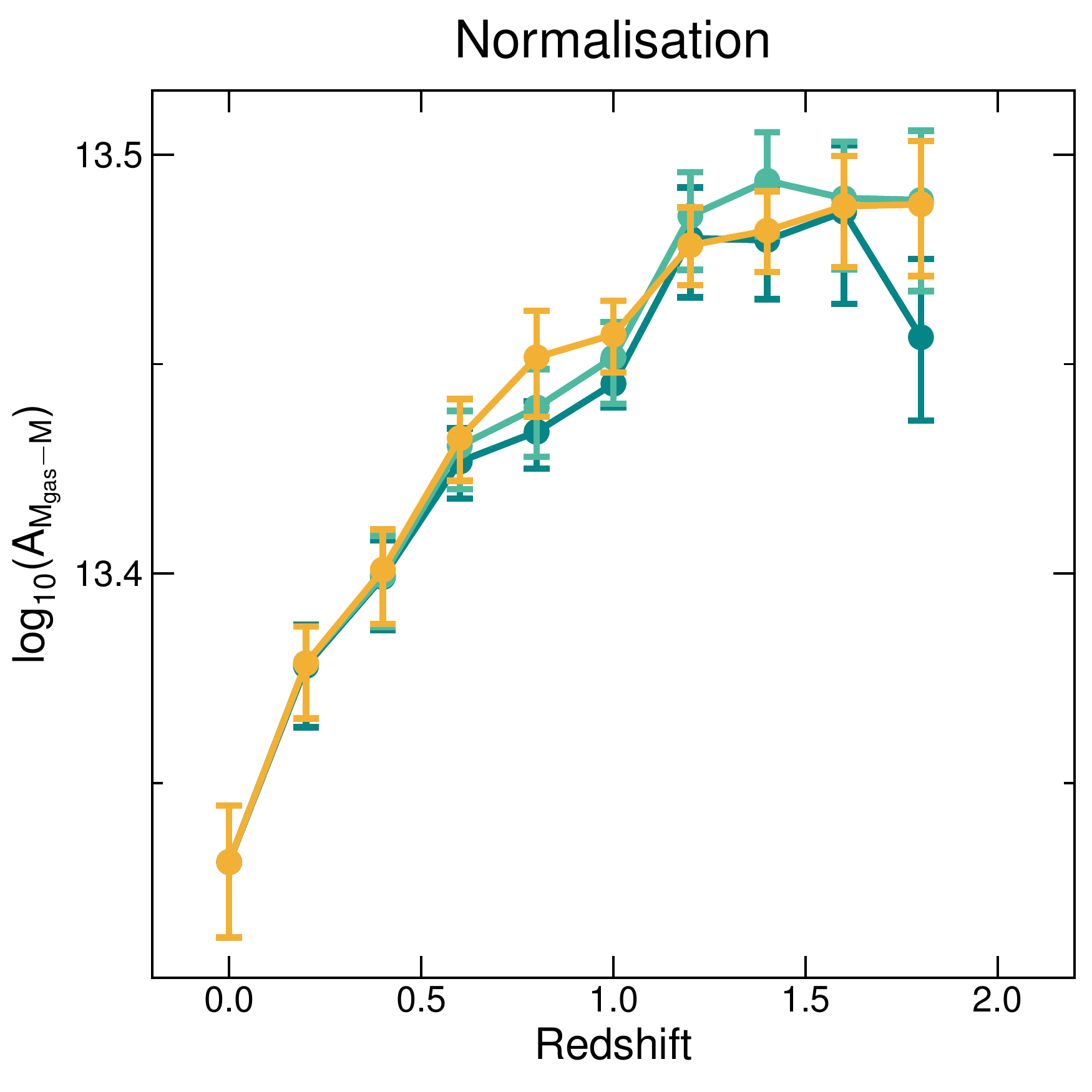}}
  {\includegraphics[width=\factorthree\textwidth]{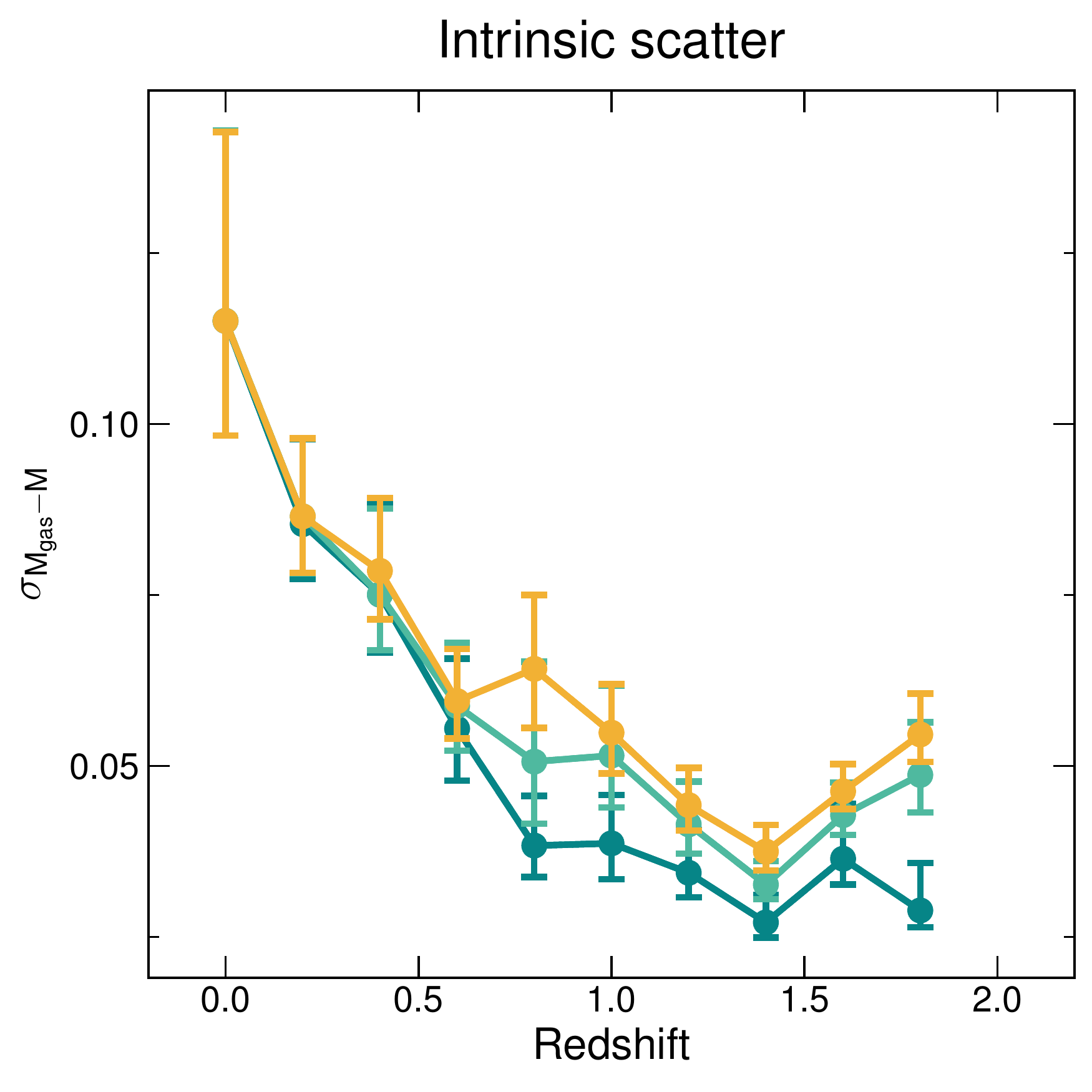}} \\
  {\includegraphics[width=\factorthree\textwidth]{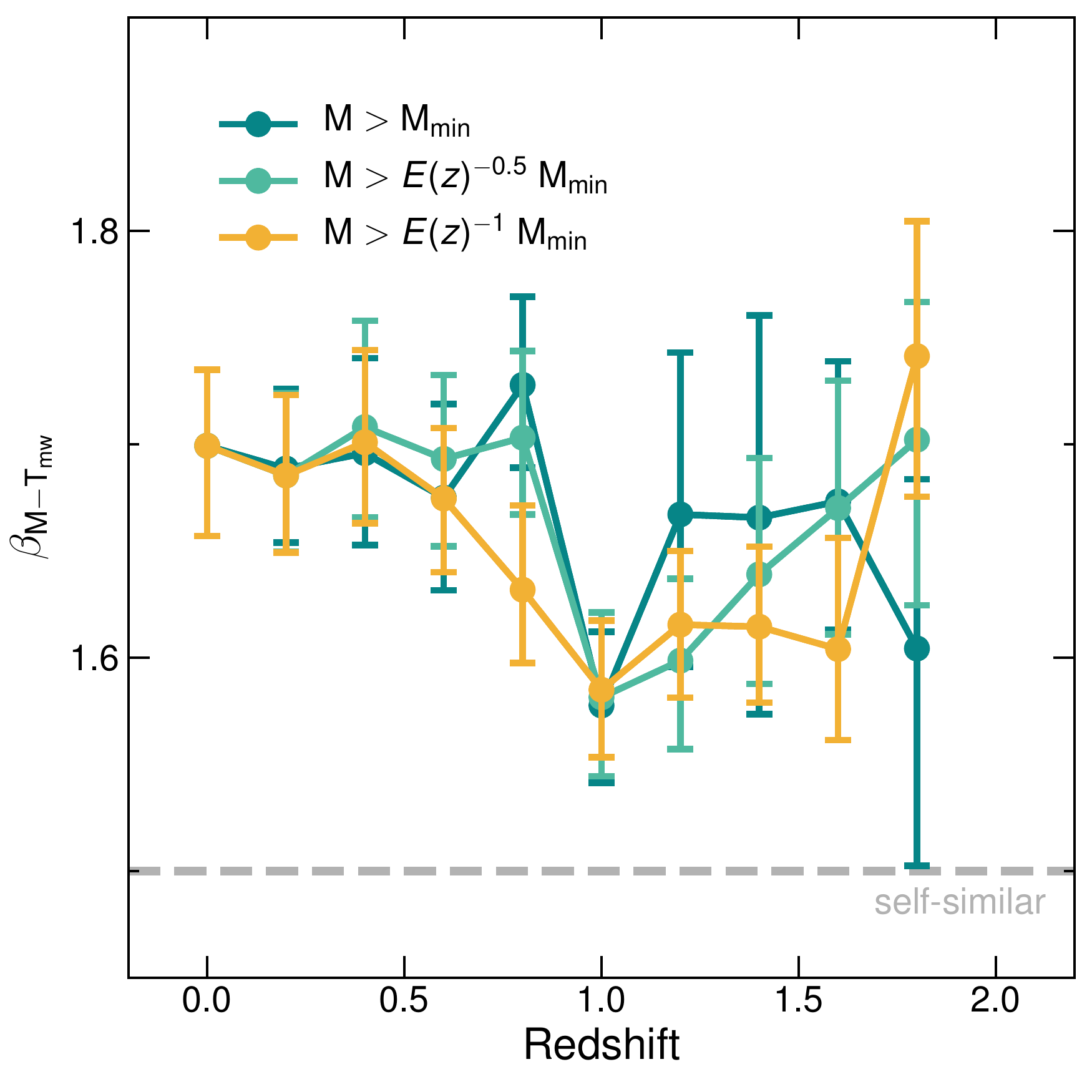}}
  {\includegraphics[width=\factorthree\textwidth]{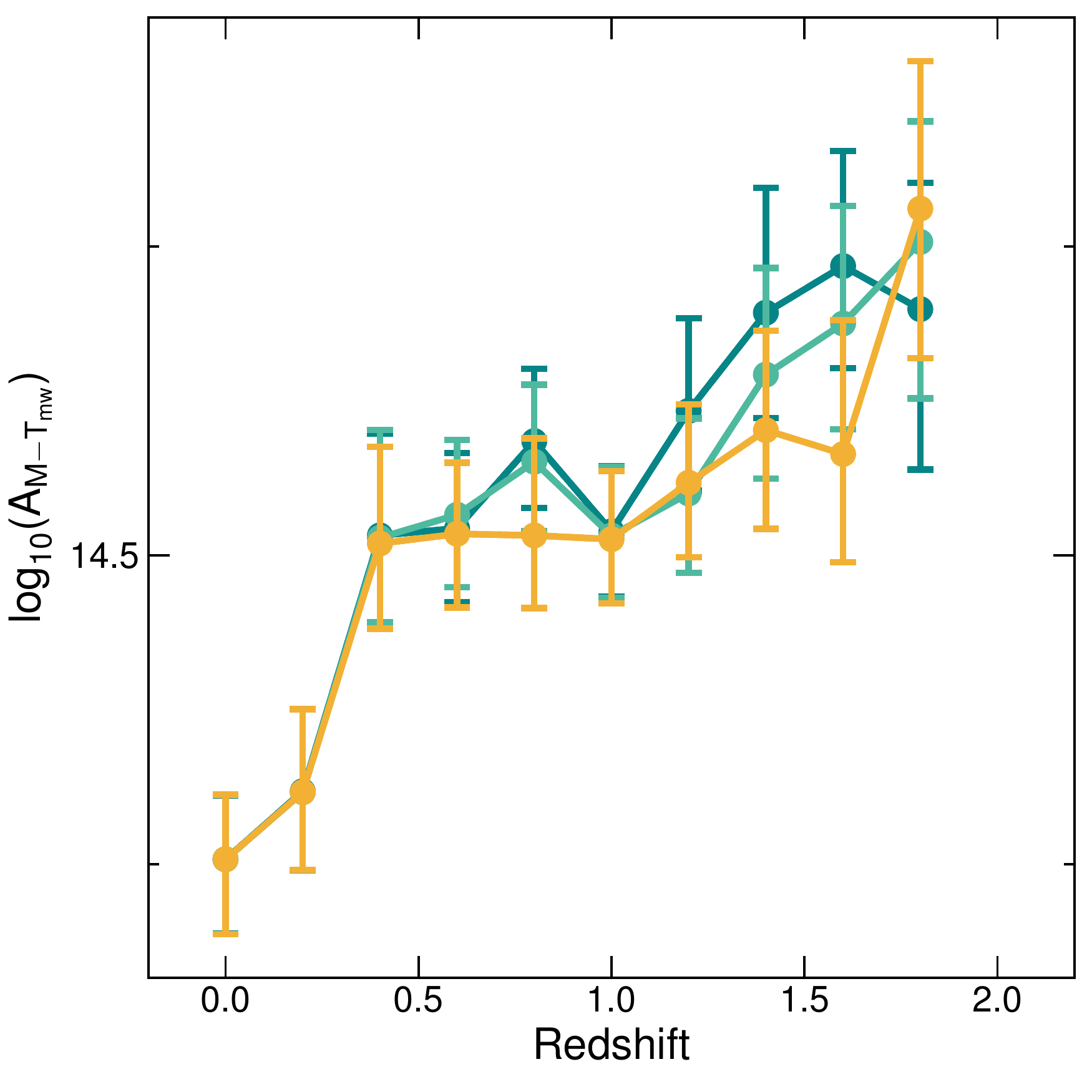}}
  {\includegraphics[width=\factorthree\textwidth]{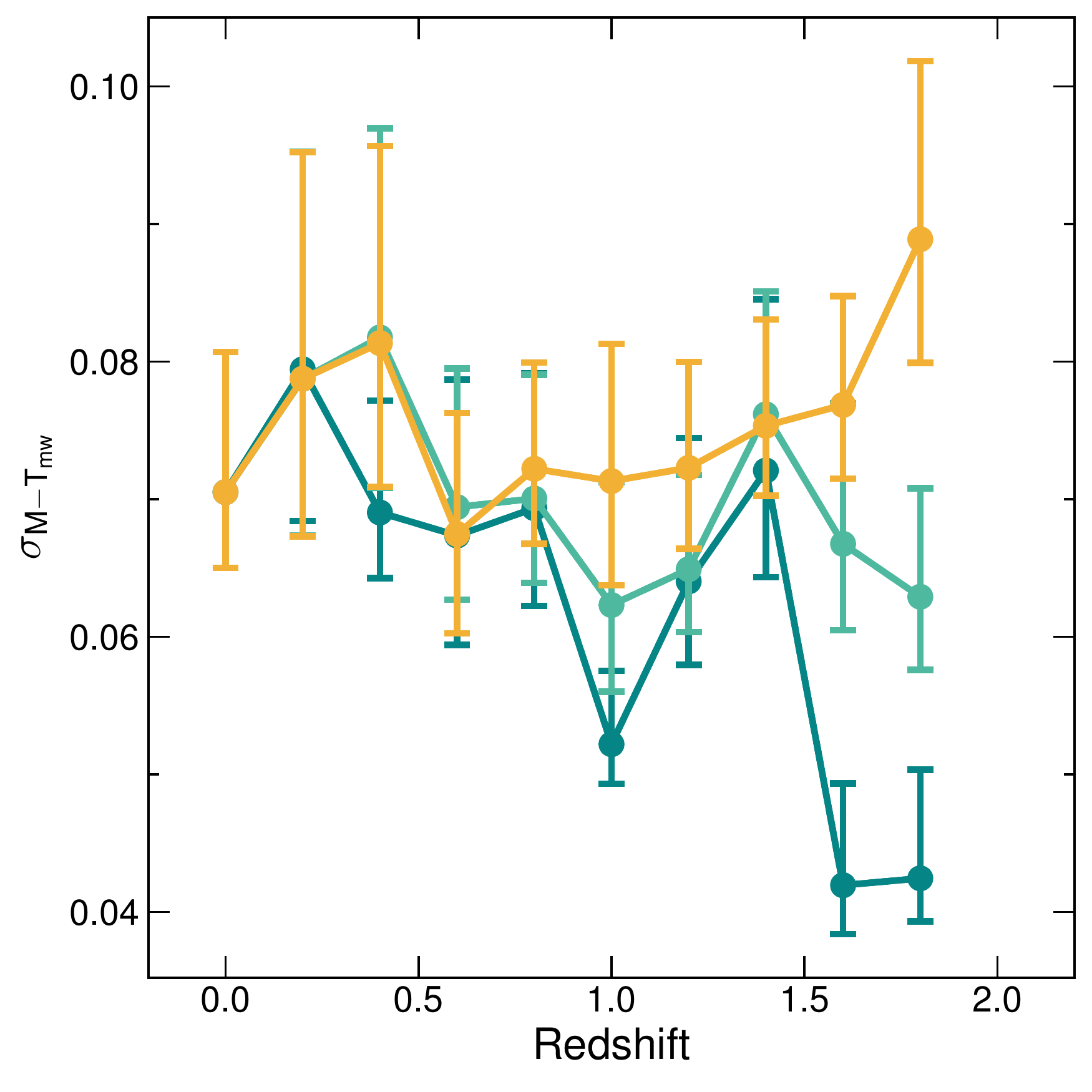}}
  \end{center}
  \caption{The redshift evolution of the gas mass--total mass (top row) and total mass--mass-weighted temperature (bottom row) relations and their dependence on the redshift evolution of the lower mass threshold of the sample. Symbols are the same as Fig.~\ref{fig:mass_dependence}. Different colours indicate different choices for the $E(z)$ dependence of the lower mass limit of the sample, $M_{500} > E(z)^\gamma \, M_{\mathrm{500, min}}$, with $M_{\mathrm{500, min}} = 3 \times 10^{13} M_{\odot}$.
  }
    \label{fig:mass_evol_dependence}
\end{figure*}

Fig.~\ref{fig:mass_evol_dependence} demonstrates the dependence of the gas mass--total mass and total mass--temperature relations on the redshift evolution of the lower mass threshold of the sample.
The redshift evolution is parametrized as $E(z)^{\gamma}$, where $\gamma$ is either zero (i.e. the mass threshold is independent of redshift), $\gamma = -0.5$ (our fiducial choice) or $\gamma = -1.0$. The latter value is used in T18 and corresponds to the redshift evolution of halo mass at fixed temperature in the self-similar scaling model. Hence $\gamma = -1.0$ can be thought of as applying a fixed temperature threshold to the sample.

Changes to the best-fitting parameters are small and limited to $z \gtrsim 0.5$. They tend to act in the direction expected for the changing mass of the sample, the effects of which we have described in the previous section. For example, the intrinsic scatter in the gas mass--total mass relation falls more slowly with increasing redshift for $\gamma = -1$ compared with $\gamma = -0.5$ or $\gamma = 0$ because the sample contains an increasing fraction of low-mass objects, which show increased scatter in gas mass at fixed total mass. Similar reasoning applies to the intrinsic scatter of the total mass--temperature relation, which is approximately redshift-independent for $\gamma = -1$ but decreases mildly with increasing redshift for $\gamma = 0$.
Similar changes to the intrinsic scatter are found for the SZ flux and X-ray luminosity-based relations at $z \gtrsim 0.5$, however these do not affect our conclusions.
Furthermore, we find no significant change to the redshift evolution of the slope or normalisation of the relations by using $\gamma = -1$ or $\gamma = 0$ instead of our fiducial value of $\gamma = -0.5$.

\bsp	
\label{lastpage}
\end{document}